\documentstyle[12pt]{article}
  \def\lsim{\raise0.3ex\hbox{$<$\kern-0.75em\raise-1.1ex\hbox{$\sim$}}}
\def\gsim{\raise0.3ex\hbox{$>$\kern-0.75em\raise-1.1ex\hbox{$\sim$}}}
\def\noi{\noindent}
\def\nn{\nonumber}
\def\bea{\begin{eqnarray}}  \def\eea{\end{eqnarray}}
\def\beq{\begin{equation}}   \def\eeq{\end{equation}}
\def\sq{\hbox {\rlap{$\sqcap$}$\sqcup$}}
\def\beeq{\begin{eqnarray}} \def\eeeq{\end{eqnarray}}
\newcommand\mysection{\setcounter{equation}{0}\section}
\renewcommand{\theequation}{\thesection.\arabic{equation}}
\newcounter{hran} \renewcommand{\thehran}{\thesection.\arabic{hran}}

\def\bmini{\setcounter{hran}{\value{equation}}
  \refstepcounter{hran}\setcounter{equation}{0}
  \renewcommand{\theequation}{\thehran\alph{equation}}\begin{eqnarray}}

\def\bminiG#1{\setcounter{hran}{\value{equation}}
\refstepcounter{hran}\setcounter{equation}{-1}
\renewcommand{\theequation}{\thehran\alph{equation}}
\refstepcounter{equation}\label{#1}\begin{eqnarray}}

\def\emini{\end{eqnarray}\relax\setcounter{equation}{\value{hran}}\renewcommand{\theequation}{\thesection.\arabic{equation}}}

\begin{document}
\begin{center}
{\large \bf The gaussian propagator formalism and} \\ 
{\large \bf the determination of the leading Regge} \\
{\large \bf trajectory for $\phi^3$ field theory} \\
\vskip 8 truemm
{\bf Richard Hong Tuan}\\

Laboratoire de Physique Th\'eorique et Hautes Energies\footnote{Laboratoire associ\'e
au Centre National de la Recherche Scientifique - URA D0063} \\
Universit\'e de Paris XI, B\^atiment 210,
F-91405 Orsay Cedex, France \\
\end{center}

\begin{abstract} 

As the number of loops goes to infinity Feynman $\alpha$-parameters undergo a fixing mechanism
which entails a gaussian representation for propagators in scalar field theories. Here, we describe
this mechanism in the fullest detail. The fixed values are in fact mean-values which can be
determined via consistency conditions. The consistency conditions imply that one $\alpha$-parameter
is integrated in the usual way and the dependence of the mean-values of the other $\alpha$-parameters
on it must be determined. Here we present a method for doing this exactly which requires the
solution of an equation system. We present an analytic solution for this equation system in the
case of the ladder-graph topology. The Regge behaviour is obtained in a simple way as well as an
analytic expression for the leading Regge trajectory. Then, the consistency equations for the two
(in the ladder case) independent $\alpha$-parameters mean-values are solved numerically. Agreement
with previous determinations of the intercept $\alpha (0)$ is obtained for $\alpha (0) \ \gsim$ 0.3.
However, we are able to calculate $\alpha (t/m^2)$ for - 3.6 $\lsim$ $t/m^2$ $\lsim$ 1.8 and find
that it is close to linear. \par

We consider the massless limit of the theory and find that the $\alpha$-parameters mean-values and
the trajectory $\alpha (t)$ have limits which are independent of the mass, a phenomenon which also
occurs for renormalizable theories via the renormalization group equations. 
\end{abstract}

\vskip 1 truecm

\noi LPTHE Orsay 97-38 \par
\noi July 1997 \par
\newpage
\pagestyle{plain}
\mysection{Introduction} 
\hspace*{\parindent}
Methods for calculating in field theory are something of paramount importance, allowing in
particular to understand the dynamical structure of interactions in particle physics. Some are
non-perturbative like instantons \cite{1r} and lattices \cite{2r}. Instantons help to elucidate 
the vacuum structure and lattices up to now are confined to calculating masses and couplings. So we
can consider that, in the near future, those ways of calculating will be restricted to very specific
sectors of field theories. Recently, the holomorphic unraveling \cite{3r} of the properties of
supersymmetric QCD  has aroused interest. But, of course, one of the most important unattained goals
is fin\-ding a way of dealing with QCD itself. Due to its asymptotic freedom, perturbation is useful
when high momenta transfers are involved. However, we would also like to have a control on QCD when
couplings are not small as is the case in soft physics. In some cases, the leading log
approximation has been used, for instance in the calculation of the Pomeron trajectory \cite{4r}.
But we are far from something satisfactory because the sub-leading level, if we want to take it into
account, forces an enormous amount of work to be done, if only possible. Another method has been
advocated, some years ago \cite{5r}, using string theory to derive one-loop multi-gluons amplitudes
taking the limit where $\alpha '$, the Regge slope or the inverse string tension, tends to zero.
One of its most salient features is that this technique can be translated into a set of rules which
requires no knowledge of string theory. Most strikingly, $\phi^3$-like diagrams only \cite{6r} are
required to evaluate the kinematic factors. Indeed, each kinematic factor is the loop-integral that
one would expect from {\it a $\phi^3$ zero mass scalar field theory}, expressed in terms of the
well-known Feynman $\alpha$-parameters, with a factor taking into account the specificity of QCD.
This factor called a reduced kinematic factor takes into account external polarizations and
on-shell external momenta. At the one-loop level it contains only terms linear in the
$\alpha$-parameters or contaning no $\alpha$-parameter. \par

Calculations were made explicitly for one-loop \cite{6r} $n$-gluons amplitudes but the field theory
correspondence with string theory is expected to hold for an arbitrary number of loops with three
gluon vertices. \par

Here, we shall expose a method for calculating Feynman graphs with {\it an infinite number of
loops for scalar massive} $\phi^3$ field theory that we {\it will extend to the massless case}.
Multiplying the integrand by the appropriate reduced kinematic factor then yields an extension of
our method for calculating amplitudes with an infinite number of loops in QCD. In fact, were not
for the additional $\alpha$-parameter dependence of the reduced kinematic factor, all the
calculations made here for the Regge behaviour and the determination of the Regge trajectory
would be valid for the same objects in QCD. \par

We start from an expression \cite{7r} of 1-particle, 1 vertex-irreducible, Euclidean Feynman graphs
amplitudes in terms of the well-known $\alpha$-parameters. We separate an overall scale which is
integrated over separately. This integration controls the divergence of the amplitudes. The
rescaled $\alpha$-parameters are then integrated over by using the mean-value theorem. When the
number of $\alpha$'s tends to infinity, which is precisely the limit we are interested in here,
the mean-value theorem has an important consequence~: the mean-values $\bar{\alpha}_i$ of the
$\alpha$-parameters have to be of order $1/I$ if $I$ is the number of propagators (and
$\alpha$-parameters) of the graph. Moreover, each $\bar{\alpha}_i$ can be determined by a
consistency equation obtained by equating the value of the amplitude obtained by using the
mean-value theorem for all $I$ $\alpha$-parameters and the value of the same amplitude obtained by
integrating with the mean-value theorem over $I-1$ $\alpha_j$-parameters, $j \not= i$, and
integrating in the usual way over the last ($\alpha_i$) $\alpha$-parameter. Of course, in principle
there are as many consistency equations to solve as there are $\alpha$-parameters. However, when
some symmetry exists in the topology of the graph, the number of consistency equations can be
greatly reduced. The simplest topology that one can imagine with an infinite number of loops is the
ladder-graph topology in which only two independent $\bar{\alpha}_i$'s survive. This is, of course,
a very interesting topology because it leads to Reggeisation \cite{8r,9r} when the invariant $s \to
\infty$. In the present article we deal with the two consistency equations for $\bar{\alpha}_-$ and
$\bar{\alpha}_+$, which are respectively the mean-value for the $\alpha$-parameter of the central
rungs of the ladder and the mean-value for the $\alpha$-parameter of the side rungs. The formalism
we develop here gives the amplitude under a compact form \cite{10r} where no integration is
left-over. This compactness is very useful. Knowing the $\bar{\alpha}_i$'s, we simply plug in their
values in the amplitude expression to get the amplitude's value. In this way, we got a compact
expression \cite{11r} for the leading Regge trajectory which is analytic as a function of
$\bar{\alpha}_+$ and $\bar{\alpha}_-$. The consistency equations, when solved, also give for
$\bar{\alpha}_+$ and $\bar{\alpha}_-$ analytic functions of $t$, the other invariant, $\gamma$, the
coupling constant, and $m$ the mass of the theory. \par

The other important consequence of compactness is that we can easily derive the massless limit of
the theory. In the expression for the amplitude, the mass $m$ only appears in one place in the
combination

\beq
Q_G (P, \{\bar{\alpha}\}) + h_0 \ m^2
\label{1.1}
\eeq

\noi where $h_0$ is the sum of the rescaled $\alpha_i$'s, $Q_G$ is the ratio of two homogeneous
polynomials of degree $L+1$ and $L$ in the $\bar{\alpha}_i$'s, $L$ being the number of loops of the
diagram considered. The polynomial in the numerator of $Q_G$ also depends linearly on $s$ and $t$
(or in general any invariant) via a form quadratic in the external momenta $P_j$. So, letting $m
\to 0$ and obtaining the resulting amplitude is rather trivial~! Of course, we have to introduce a
mass in order to avoid infrared problems in the definition of the Feynman graph amplitudes, but
once this is done, one can calculate everything as  a function of $m$ and let $m$ tend to zero. \par

This is what we will do here for the Regge trajectory which has an $m \to 0$ limit independent of
$m$ as well as the consistency equations for $\bar{\alpha}_+$ and $\bar{\alpha}_-$. We remark that
this phenomenon is close in spirit to what happens in the renormalization group \cite{12r} when the
mass dependence also disappears when momenta are taken to go to infinity. This possibility of having
a massless limit independent of the mass was already contained in our first paper \cite{10r} proving
the exactness of the Gaussian representation for propagators and where a compact expression for
amplitudes was already given. (Replacing $\alpha_i$ by $\bar{\alpha}_i$ gives a Gaussian
representation for the propagator $i$). \par
 
In section 2 we give the basics of the $\alpha$-parameter representation for Euclidean scalar
massive $\phi^3$ field theory. The mean-value theorem allows us to give the amplitude for any
diagram in a compact form where no integration is left over. The integration over an overall scale
for the $\alpha_i$'s is done separately and controls the possible ultraviolet divergences via a
gamma function factor $\Gamma (I - dL/2)$ where $d$ is the dimension of space-time. A consistency
equation for $\bar{\alpha}_i$ is obtained by using the mean-value theorem for all $\alpha_j$, $j
\not= i$ and integrating normally on $\alpha_i$, thus giving an expression for the diagram
amplitude where $\bar{\alpha}_i$ does not appear. Consistency requires that this expression is
equal to the expression where the mean-value theorem is used for all $I$ $\alpha$-parameters,
including $\alpha_i$. \par

Because of the constraint $\sum\limits_{\ell} \alpha_{\ell} = h_0$ the phase-space allowed for each
$\alpha_{\ell}$ is of order $h_0/I$. Then, assuming $\bar{\alpha}_j = O(h_0/I)$, the consistency
equation naturally introduces a scale $\Lambda^{-1}$ where $\Lambda$ is proportional to $(I -
dL/2)/h_0$ such that we expect $\bar{\alpha}_i\Lambda$ to be of order one. \par

In the consistency equation a parameter $\mu_i$, defined in (\ref{1.2}) below, which is of degree
one in the $\bar{\alpha}_j$'s, appears added to $\alpha_i$ and $\bar{\alpha}_i$. Because $\mu_i$ is
of degree one in the $\bar{\alpha}_j$'s we also expect $\mu_i$ to be of order $h_0/I$. If we assume
this, then the consistency equation gives $\bar{\alpha}_i = O(h_0/I)$ proving consistency. It is
shown that if $\mu_i\Lambda \to 0$, i.e. if $\mu_i$ is decreasing faster than $(h_0/I)$, the
consistency equation entails that $\bar{\alpha}_i/\mu_i \to \infty$ breaking the expected
proportionality between $\bar{\alpha}_i$ and $\mu_i$. Assuming $\mu_{\ell}\Lambda \to \infty$ for a
non-infinitesimal fraction of the $\mu_{\ell}$'s is also forbidden because this would violate the
constraint $\sum\limits_{\ell} \bar{\alpha}_{\ell} = h_0$. Therefore, only when $\mu_i \Lambda =
O(1)$ is consistency achieved, {\it thereby showing that only $\ \bar{\alpha}_i\Lambda = O(1)$ or
$\bar{\alpha}_i = O(h_0/I)$ is consistent}. It is also proved that even if some $\bar{\alpha}_j$ are
assumed to be decreasing faster than $h_0/I$ the consistency equation still gives that
$\bar{\alpha}_i$ should be $O(h_0/I)$. In the Appendix a proof is given, independent of the
precedent arguments, that $\mu_i$ cannot decrease faster than $h_0/I$ using a continuous fraction
expression for $\mu_i$ as a function of the $\bar{\alpha}_j$'s. \par

In section 3 we specialize to infinite ladder-diagrams and calculate the sum of their amplitudes
through a saddle-point estimate. We find that the value of $L$ at the saddle-point is proportional
to $\ell n (s/s_0)$ where $s_0$ is some finite scale. This is a very natural result expected in any
parton-like or multiperipheral model, giving a multiplicity proportional to the rapidity. An
explicit analytic expression for the leading Regge trajectory is obtained giving a function of
$\gamma$, $m$, $\bar{\alpha}_+$ and $\bar{\alpha}_-$ and $\Lambda$, $\bar{\alpha}_+$ being the
mean-value of the $\alpha$-parameters along the sides of the ladder and $\bar{\alpha}_-$ the
mean-value of the $\alpha$-parameters of the central rungs. The $m \to 0$ limit of the Regge
trajectory is examined and is shown to be independent of $m$ if $\bar{\alpha}_+$, $\bar{\alpha}_-$
and $\Lambda$ also have $m$-independent limits, which will be proved in section 9. \par

We discuss the dependence of $\bar{\alpha}_j$ on $\alpha_i$ when $\alpha_i$ is left free to vary.
$Q_G(P, \{\bar{\alpha}\})$ also varies because $\bar{\alpha}_j \not= \alpha_i$ when $\alpha_i$
varies. These two dependencies on $\alpha_i$ were not taken into account in the consistency
equation written in section 2, which therefore was a simplified version of the true consistency
equation. These dependencies introduce correction factors which will be calculated in the next
sections. \par

In section 4 we derive the explicit dependence of $Q_G(P, \{ \bar{\alpha} \}_{j \not= i},
\alpha_i)$ on $\alpha_i$ distinguishing the two cases where $\alpha_i$ is an $\alpha_+$-parameter
or an $\alpha_-$-parameter. That dependence introduces a factor $H_i(\alpha_i - \bar{\alpha}_i)$ in
the integrand of the consistency equation which is explicitly evaluated in both cases, $+$ and
$-$. \par

In section 5 the dependence of $P_G(\{\bar{\alpha}\})$ on $\alpha_i$ is examined for the factor
$b_i(\{\alpha\}_{j\not= i})$ if $P_G(\{I\})$ is a homogeneous polynomial of degree $L$ in the
$\alpha_i$'s written when $\alpha_i$ varies

\beq
P_G(\{\bar{\alpha}\}_{j\not= i}, \alpha_i) = b_i (\{ \bar{\alpha} \}_{j\not=i}) (\mu_i + \alpha_i) \
\ \ . \label{1.2}
\eeq

The polynomial $P_G(\{ \bar{\alpha}\})$ intervenes explicitly in the compact expression for the
diagram amplitude. In the consistency equation written in section 2 only the $(\mu_i + \alpha_i)$
part of $P_G(\{\bar{\alpha}\})$ depends on $\alpha_i$. As only relative variations are useful the
variable\break \noindent Log $P(\{ \bar{\alpha}\})$ is used and assimilated to Log $b_i(\{
\bar{\alpha} \}_{j\not= i})$ both quantities differing only infinitesimally in their variations. The
$\bar{\alpha}_j(\alpha_i)$ dependence then introduces an additional factor $H_0(\alpha_i -
\bar{\alpha}_i)$ in the integrand of the consistency equation. \par

Then, a general method for finding $\bar{\alpha}_j(\alpha_i)$ is given which relies on the
evaluation of a particular $\bar{\alpha}_{\ell}$ as a function of $\alpha_i$. For that purpose the
$\alpha_{\ell}$ variable must also be left free to vary in order to write a consistency equation
for $\bar{\alpha}_{\ell}$. Then, if terms proportional to $\alpha_i \alpha_{\ell}$ are neglected
the consistency equation written by letting $\alpha_i$ {\it and} $\alpha_{\ell}$ vary together
factorizes into two independent consistency equations which are each identical to the one written by
letting only one variable $\alpha_i$ or $\alpha_{\ell}$ vary. Then, we examine the source of the
$\alpha_i \alpha_{\ell}$ terms and give an explicit expression for a factor containing all of them in
the consistency equation where both $\alpha_i$ and $\alpha_{\ell}$ are free to vary. This factor
implies that the scale $\Lambda$ is ``renormalized'' into a scale $\Lambda_{\ell}^R = \Lambda +
d/2 \ L E_{\ell}^0 + \Lambda_{\ell}(\alpha_i)$ for the $\alpha$-parameter mean-value
$\bar{\alpha}_{\ell}$, where $E^0_{\ell}$ is some constant and $\Lambda_{\ell}(\alpha_i)$ a term
of order one (and therefore infnitesimal with respect to $\Lambda$ and $d/2 LE^0_{\ell}$) containing
the dependence of the scale on $\alpha_i$. The constant $E^0_{\ell}$ comes from the factor
$H_0(\alpha_{\ell}-\bar{\alpha}_{\ell})$ discussed above. If, however, the $\alpha_i
\alpha_{\ell}$ terms were neglected these would be no $\bar{\alpha}_{\ell}(\alpha_i)$ or
$\bar{\alpha}_j(\alpha_i)$ dependence and $E^0_{\ell}$ would be zero. In fact, now, the problem is to
derive $E_{\ell}^0$ knowing $\Lambda_{\ell}(\alpha_i)$. \par

In section 6 we tackle the problem of determining $E_{\ell}^0$ and $\Lambda_{\ell}(\alpha_i)$. As
$\alpha_i$ varies $\Lambda_{\ell}^R$ will vary and so does $\bar{\alpha}_{\ell}$. Differentiating
the consistency condition written for $\bar{\alpha}_{\ell}$ allows to write a relation tying the
variation of $\Lambda_{\ell}^R$ to that of $\bar{\alpha}_{\ell}$

\beq
\delta \bar{\alpha}_{\ell}/ \bar{\alpha}_{\ell} = A_{\ell} \ \delta \Lambda_{\ell}^R/\Lambda^R_{\ell}
\label{1.2a}
\eeq 

\noi where $A_{\ell}$ is calculated as a function of several quantities ($\ell$ can be of the $+$
or $-$ kind)~: $y$, $\delta \Lambda_-^R/\Lambda_-^R$, $\delta\Lambda_+^R/\Lambda_+^R$,
$z_{\pm}$, $\zeta_{\pm}$ and $\varepsilon_{\pm}$ which we now briefly describe. $y$ is
$(2\bar{\alpha}_{\mu}/\bar{\alpha}_-)^{1/2}$, $z_{\ell}$ is proportional to $\partial$ Log
$(J_{\ell}^{-1} H_{\ell} \widetilde{H}_0)/\partial (\bar{\alpha}_{\ell}\Lambda_{\ell}^R)$,
$\zeta_{\ell}$ is proportional to $\partial$ Log $(J_{\ell}^{-1} H_{\ell}
\widetilde{H}_0)/\partial (\mu_{\ell} \Lambda_{\ell}^R)$, $\varepsilon_{\ell}$ is equal to one minus
a weighted ``mean-value'' of $(\mu_{\ell} + \bar{\alpha}_{\ell})/(\mu_{\ell} + \alpha_{\ell})$,
this mean-value being obtained by inserting $(\mu_{\ell} + \bar{\alpha}_{\ell})/(\mu_{\ell} +
\alpha_{\ell})$ into the integrand of the integral of the consistency equation for
$\bar{\alpha}_{\ell}$ (the consistency equation is written as one minus the integral equal zero).
$J_{\ell}$ is a Jacobian allowing to change the integration variable from $\alpha_{\ell}$ to
$\alpha_{\ell} \Lambda_{\ell}^R$. $\widetilde{H}_0$ is the factor in $H_0(\alpha_{\ell} -
\bar{\alpha}_{\ell})$ which does not depend on $\alpha_{\ell}$ and it is equal to $\exp (d/2 \ L
E_{\ell}^0 \bar{\alpha}_{\ell})$. $H_{\ell}$ is the factor described in section 4 which takes
into account the effect on the consistency integral of the variation of $Q_G(P,
\{\bar{\alpha}\}_{j\not= \ell}, \alpha_{\ell})$. Equation (\ref{1.2a}) is essential as it will
allow us to determine the dependence of $\bar{\alpha}_{\ell}$ on $\alpha_i$ and find
$E^0_{\ell}$. \par

The determination of $\Lambda_{\ell}(\alpha_i)$ is made by regrouping all terms in the
consistency equation proportional to $\alpha_i \alpha_{\ell}$ and writing them down as 

\beq
\alpha_{\ell} \Lambda_{\ell} (\alpha_i) + \alpha_i \Lambda_i (\alpha_{\ell}) = F_{i \ell} (\alpha_i,
\alpha_{\ell}) \ \ \ . \label{1.3}
\eeq    

The decomposition (\ref{1.3}) {\it is not unique}, even if we take into account possible symmetry
constraints and demand moreover that $\Lambda_{\ell}(\alpha_i)$ should be the same, $\alpha_i$
taking the value $\bar{\alpha}_+$ or $\bar{\alpha}_-$. We are led to introduce five parameters in
order to take into account this non-uniqueness. However, two constraints are imposed on them by
requiring $\Lambda_{\ell}(\bar{\alpha}_i)$ to be independent of $i$ being $+$ or $-$. \par
We introduce the variables

\beq
x_{\ell i} = \bar{\alpha}_{\ell}^{-1} \  d \bar{\alpha}_{\ell}/d\alpha_i
\label{1.4}
\eeq 

\noi and the equation (\ref{1.2}) can be expressed as a system of equations for these variables
because

\beq
d\Lambda_{\ell}^R/d \alpha_i = d \Lambda/d\alpha_i + d/2\ L \ d \ E_{\ell}^0/d \alpha_i + d
\Lambda_{\ell}(\alpha_i)/d \alpha_i  \label{1.5}
\eeq
 
\noi where $d\Lambda/d\alpha_i$ is shown to be proportional to $x_{+i}$, and $E_{\ell}^0$ is shown
to be a linear combination of $x_{-\ell}$ and $x_{+\ell}$. Then, $d\Lambda_{\ell}(\alpha_i)/d
\alpha_i$ plays the role of the inhomogeneous term determined from the $\alpha_i \alpha_{\ell}$
terms. So we get a system of four equations for the four variables $x_{++}$, $x_{+-}$, $x_{-+}$ and
$x_{--}$. This system is solved and two compatibility conditions emerge,

\beq
dE_+^0/dy = 0 \quad ; \quad dE_-^0/dy = 0 \ \ \ .
\label{1.6}
\eeq

\noi $E_+^0$ and $E_-^0$ are explicitly determined. \par

In section 7, explicit expressions for the quantities $J_{\ell}$, $z_{\ell}$, $\zeta_{\ell}$,
$\delta \Lambda_{\ell}(\alpha_i)$ and $d\varepsilon_{\ell}/dy$ ($\ell = \pm$) are given. They all
enter in the consistency equations determining $\bar{\alpha}_+$ and $\bar{\alpha}_-$. Then, explicit
expressions for $\delta \Lambda_{\ell}^R(\alpha_i)/\Lambda_{\ell}^R(\alpha_i)$ are also given,
these quantities entering in the definition of $A_{\ell}$ in (\ref{1.2a}). \par

In section 8, we study the consistency of the assumption $\bar{\alpha}_j = O$ ($h_0/I$) in the
realm of the complete consistency conditions for $\bar{\alpha}_+$ and $\bar{\alpha}_-$. This
implies replacing $\Lambda$ by $\Lambda_i^R$ and taking into account the effect of the factors
$J_{\ell}^{-1}$, $H_i(\alpha_i - \bar{\alpha}_i)$ and $\widetilde{H}_0(\bar{\alpha}_i)$. This was not
done in section 2 for simplicity and pedagogical reasons. We conclude that this change consists in
introducing in the integrand a bounded function $B_i(x)$, $x$ being the integration variable and
therefore that the consistency arguments developed in section 2 still hold. Then, $\bar{\alpha}_i$
should be $O(h_0/I)$. \par

In section 9, we examine the massless limit of the consistency equations. We find that the
quantities appearing in them have definite $m$ independent limits and therefore that
$\bar{\alpha}_i$ has a massless limit which is independent of $m$. This entails that the whole
scheme is also $m$-independent when $m \to 0$. \par

In section 10, we present numerical solutions of the consistency equations and the compatibility
conditions. The trajectory $\alpha (t/m^2)$ is determined for $- 3.6 < t/m^2 < 1.8$ (in Minkowski
space). The numerical solutions are difficult to obtain, due to chaotic effects because many
quantities are determined through calculations loops. Nevertheless, we obtain results with a
reasonable precision using a very performant minimization algorithm. The results show an agreement
for the trajectory intercept $\alpha (0)$ with previous determinations (which were assuming
massless fields for the central rungs of the ladder) for the range 0.3 $\lsim \ \alpha (0) \
\lsim$ 1.6. Previously no result was, to our knowledge, available at $t \not= 0$. We find, with an
improvement over a previous determination \cite{11r}, that the leading trajectory $\alpha (t/m^2)$ is
compatible with linearity in the range where we could determine it. \par

Section 11 will be the conclusion. 

\mysection{Basics of the $\alpha$-parameter representation} 

\hspace*{\parindent} Let us start with a 1-line irreducible, 1-vertex irreducible graph $G$ with
$L$ loops, $I$ propagators in $d$ dimensions and let us give its amplitude $F_G$ in Euclidean
space when the coupling is equal to $- 1$. We have \cite{7r}

\[ F_G = (4 \pi)^{-dL/2} h_0 \int_0^{h_0} \prod_1^I d \alpha_i \ \delta \left ( h_0 - \sum_i
\alpha_i \right ) \]
 \beq
\left [ P_G(\{\alpha \}) \right ]^{-d/2} \int_0^{\infty} d \lambda \ \lambda^{I - d/2\ L} \exp \left
\{ - \lambda \left [ Q_G (P, \{ \alpha \}) + m^2 h_0 \right ] \right \} \label{2.1}
\eeq  

\noi where $P_G(\alpha )$ is a homogeneous polynomial of degree $L$ defined as

\beq
P_G(\{ \alpha \}) = \sum_{\cal T} \prod_{\ell \not\subset {\cal T}} \alpha_{\ell}
\label{2.2}
\eeq

\noi with a sum over all spanning trees ${\cal T}$ of $G$. We recall that a spanning tree has to
be incident with all vertices of $G$. We have

\beq
Q_G(P, \{ \alpha \}) = \left [ P_G (\{ \alpha \} ) \right ]^{-1} \sum_{\cal C} s_{\cal C}
\prod_{\ell \subset {\cal C}} \alpha_{\ell} \label{2.3}
\eeq

\noi where the sum runs over all cuts ${\cal C}$ of $L + 1$ lines that divide $G$ into two
connected parts $G_1({\cal C})$ and $G_2({\cal C})$ with

\beq
s_{\cal C} = \left ( \sum_{v \in G_1({\cal C})} P_v \right )^2 = \left ( \sum_{v \in G_2({\cal C})}
P_v \right )^2
 \label{2.4}
\eeq  

\noi the external momenta $P_v$ being associated with external lines of $G$. As a cut ${\cal C}$ is
the complement of a spanning tree ${\cal T}$ plus one propagator, $\prod\limits_{C \subset {\cal
C}} \alpha_{\ell}$ in (\ref{2.3}) is of degree $L + 1$. Integrating over $\lambda$ in (\ref{2.1})
gives a convergent integral for $\phi^3$ when $d < 6$. The divergence at $d = 6$ is consistent
with the fact that $\phi^3$ is renormalizable for this value of $d$. In what follows we will
choose $d < 6$ and in fact most of time $d = 4$. Using the mean-value theorem for the $I$
variables $\alpha_i$ in (\ref{2.1}) we obtain, integrating also on $\lambda$,

\[
F_G = (4 \pi )^{-dL/2} \ h_0 \left [ P_G (\{ \bar{\alpha} \}) \right ]^{-d/2} \left [ Q_G (P, \{
\bar{\alpha}\}) + m^2 h_0 \right ]^{-(I - d/2\ L)} \cdot  \]
\beq
\Gamma (I - d/2\ L) \ h_0^{I-1}/(I - 1)~! 
\label{2.5}
\eeq 

\noi the factor $h_0^{I-1}/(I - 1)~!$ representing the phase-space volume for $I$ variables
$\alpha_i$. $F_G$ is then given under a compact form where no integral subsists. We remark that
$h_0$ is a free parameter representing the sum of all $\bar{\alpha}_i$'s because of the constraint
$h_0 = \sum\limits_i \alpha_i$ in (\ref{2.1}) and $F_G$ should not depend on it. \par

Our next step will consist in giving a way of determining $\bar{\alpha}_i$. This will be done
through a consistency equation which is obtained by using the mean-value theorem for $I - 1$
variables $\alpha_j$, letting $\alpha_i$ unintegrated. For that purpose we need to isolate the
dependence on $\alpha_i$ of $P_G(\{ \alpha \})$ and $Q_G(P, \{ \alpha \})$. Therefore we write~:

\bminiG{q-def}
P_G (\{ \alpha \}) = a_i + b_i \alpha_i = b_i \left ( a_i/b_i + \alpha_i \right )
\label{2.6a} 
\eeeq
\beeq 
\sum_{\cal C} s_{\cal C} \prod_{\ell \subset {\cal C}} \alpha_{\ell} = d_i + e_i \alpha_i = e_i
\left ( d_i/e_i + \alpha_i \right )  \label{2.6b}  
\emini

\noi where $a_i$ and $b_i$ are respectively polynomials of degree $L$ and $L - 1$ in the
$\bar{\alpha}_j$'s, $d_i$ and $e_i$ being polynomials of degree $L + 1$ and $L$ in the
$\bar{\alpha}_j$ 's. We demonstrated also that \cite{13r}

\beq
a_i/b_i = d_i/e_i 
\label{2.7}
\eeq

\noi should be equal up to terms vanishing as $I \to \infty$. Therefore we get

\beq
Q_G(P, \{ \bar{\alpha} \}) = e_i/b_i
\label{2.8}
\eeq  

\noi which does not depend anymore explicitly on $\alpha_i$. $Q_G(P, \{ \bar{\alpha}\})$ is then
homogeneous to one power of $\bar{\alpha}_j$. (In $Q_G(P, \{ \bar{\alpha} \})$ every $\alpha_j$ is
replaced by $\bar{\alpha}_j$). \par

The equality (\ref{2.7}) can be understood by saying that the cutting of a tree ${\cal T}$ far from
a propagator $i$ will bring up the same factor for trees going through $i$ and trees not going
through $i$. That is, for infinitely large graphs there is a factorization of spanning trees on
domains which are far apart on $G$ in terms of the minimum number of propagators separating them.
Now, due to the constraint given by $\delta (h_0 - \sum\limits_i \alpha_i)$,

\beq
h_0 - \alpha_i = \sum_{j\not= i} \bar{\alpha}_j
\label{2.9}
\eeq

\noi when taking the mean-values of $I - 1$ variables $\alpha_j$, which gives a rescaling of all
$\bar{\alpha}_j$'s when $\alpha_i$ varies. Then, (\ref{2.9}) let us think that we can write

\beq
\bar{\alpha}_j = O \left [ (h_0 - \alpha_i)/(I - 1) \right ] \ \ \ .
\label{2.10}
\eeq

\noi This is consistent with the fact that the phase-space for $(I - 1)$ variables can be written

\beq
(h_0 - \alpha_i)^{I-2}/(I - 2)~! \sim \left [ e (h_0 - \alpha_i)/I \right ]^{I-1}
\label{2.11}
\eeq

\noi which leaves a phase-space of order $h_0/I$ for each $\bar{\alpha}_j$. Taking into account
(\ref{2.10}) we can express $a_i$, $b_i$ and $Q_G(P, \{ \bar{\alpha}\})$ as 

\bminiG{EDh}
a_i = a_{i0} \left [ (h_0 - \alpha_i)/(I - 1) \right ]^L
\label{2.12a}
\eeeq
\beeq
b_i = b_{i0} \left [ (h_0 - \alpha_i)/(I - 1) \right ]^{L-1}
 \label{2.12b}
\eeeq
\beeq
Q_G(P, \{ \bar{\alpha}\}) = Q_{G0} (h_0 - \alpha_i)/(I - 1) \ \ \ .  
\label{2.12c}
\emini

In a first approximation we will consider $a_{i0}$, $b_{i0}$ and $Q_{G0}$ constant as $\alpha_i$
differs from $\bar{\alpha}_i$. (In section 5 we shall take them varying with $\alpha_i$, but here
we want to present the matter in a first approximation as simply as possible). Consequently, we have
as $I$, $L \to \infty$
\newpage
\[
F_G = (4 \pi )^{-dL/2} \ h_0 \ \int_0^{h_0} d \alpha_i \left [ (h_0 - \alpha_i )^{I-2}/(I-2) ! \right
] b_i^{-d/2} \left ( a_i/b_i + \alpha_i \right )^{-d/2}
 \]
\[
\Gamma (I - dL/2) \left [ Q_{G0}(h_0 - \alpha_i)/(I - 1) + h_0 m^2 \right ]^{-(I - dL/2)}
 \]
\[ 
= (4 \pi )^{-dL/2} \ h_0 \ \Gamma (I - dL/2) \left [ h_0^{I-2}/(I - 2) ! \right ] \left [
b_{i0}(h_0/I)^{L-1} \right ]^{-d/2} \cdot  \]
\[
\left [ Q_{G0}(h_0/I) + h_0 m^2 \right ]^{-(I -dL/2)}
\]
\[
\exp \left \{ - dL/(2I) - (1 - dL/(2I)) \left [ Q_{G0}(h_0/I) /\left [ Q_{G0}(h_0/I) + h_0 m^2
\right ] \right ] \right \}
\]
\beq
\int_0^{h_0} d \alpha_i \left ( a_i/b_i + \alpha_i \right )^{-d/2} \exp \left [ - (I - dL/2)(1 -
\beta ) \alpha_i/h_0 \right ] \label{2.13}
\eeq

\noi where $1 - \beta$ is defined as 

\beq
\label{2.14}
1 - \beta \equiv \left [ 1 + Q_{G0}(h_0/I)/(h_0 m^2) \right ]^{-1} = \left [ 1 + Q_G(P, \{
\bar{\alpha}\})/h_0m^2) \right ]^{-1} \eeq

\noi $P_G(\{ \bar{\alpha} \})$ in (\ref{2.5}) is $P_G(\{ \alpha \})$
with every $\alpha_j$ replaced by $\bar{\alpha}_j$ and

\beq
P_G(\{\bar{\alpha} \}) = \left ( a_i/b_i + \bar{\alpha}_i \right ) b_{i0}(h_0/I)^{L-1}
\label{2.15}
\eeq

\noi because $\bar{\alpha}_j = O(h_0/I)$ when mean-values of $I$ variables are taken. Equating
expression (\ref{2.5}) and (\ref{2.13}) for $F_G$ we get

\[
\left ( \mu_i + \bar{\alpha}_i \right )^{-d/2} = \exp \left [ - dL/(2I) - (1 - dL/(2I)) \beta
\right ] \]
\beq
\left [ (I - 1)/h_0 \right ] \int_0^{h_0} d \alpha_i \left ( \mu_i + \alpha_i \right )^{-d/2} \exp
(- \alpha_i \Lambda ) \label{2.16}
\eeq

\noi with $\Lambda$ and $\mu_i$ defined as

\beq
\Lambda \equiv (I - dL/2) (1 - \beta )/h_0
\label{2.17}
\eeq

\beq
\mu_i \equiv a_i/b_i \ \ \ .
\label{2.18}
\eeq

\noi Now, depending on the size of $\mu_i$, let us see what $\bar{\alpha}_i$ should be. $\mu_i$ is
homogeneous to one power of $\bar{\alpha}_j$ and therefore, according to our guess (\ref{2.10}),
we should have $\mu_i \Lambda = O(1)$ or

\beq
\mu_i = O(h_0/I) \ \ \ .
\label{2.19}
\eeq

\noi In the Appendix we give an argument for $\mu_i$ satisfying (\ref{2.19}) (the
ratio $\mu_i/\bar{\alpha}_i$ is, in fact, the ratio of the sum of weights of spanning trees going
through $i$ to the sum of weights of spanning trees not going through $i$, any propagator $j$ being
weighted by $\bar{\alpha}_j^{-1}$). \par

Then, writing the integral in (\ref{2.16}) as 

\[
\Lambda^{-1} \int_O^{\infty} d(\alpha_i \Lambda ) \exp (- \alpha_i \Lambda ) \left ( \mu_i +
(\alpha_i \Lambda )/\Lambda \right )^{-d/2} \]
\beq
= \Lambda^{-1 + d/2} \exp (\mu_i \Lambda ) \int_{\mu_i \Lambda}^{\infty} dx \exp (- x) x^{-d/2} \ \
\ , \label{2.20}
\eeq

\noi we obtain
\[
\left ( \mu_i + \bar{\alpha}_i \right )^{-d/2} = \exp \left [ - dL/(2I) - \left ( 1 - dL/(2I) \right
) \beta \right ] \cdot
 \]
\beq
\left [ (I - 1)/(h_0 \Lambda ) \right ]  \Lambda^{d/2} \exp (\mu_i \Lambda ) \int_{\mu_i
\Lambda}^{\infty} dx \exp (-x) \ x^{-d/2} \label{2.21}
\eeq

\noi where

\beq
\beta = \left [ Q_G/(h_0m^2) \right ]/ \left [ 1 + Q_G/(h_0m^2) \right ]
\label{2.22}
\eeq

\noi is $O(1)$. Then, the only factor which is not bounded by a constant on the right-hand side of
(\ref{2.21}) is $\Lambda^{d/2}$ and thus

\beq
\mu_i + \bar{\alpha}_i = O(\Lambda^{-1})
\label{2.23}
\eeq

\noi which means that $\bar{\alpha}_i$ should also be $O(\Lambda^{-1})$, demonstrating the
consistency of our assumption (\ref{2.10}). \par

We can go further and ask what happens when we assume that $\mu_i \Lambda \to 0$. Then, the
integral in (\ref{2.21}) is dominated by its contribution near its lower boundary and behaves like

\beq
\left ( \mu_i \Lambda \right )^{1-d/2} \exp (- \mu_i \Lambda )
\label{2.24}
\eeq

\noi which leads to

\beq
\left ( \mu_i + \bar{\alpha}_i \right )^{-d/2} \sim \mu_i I \quad \mu_i^{-d/2}
\label{2.25}
\eeq

\noi and because $\mu_iI \to 0$, we have

\beq
\bar{\alpha}_i/\mu_i \to \infty \ \ \ . 
\label{2.26}
\eeq

\noi This is inconsistent because $\mu_i$ is homogeneous to one power of $\bar{\alpha}_j$. Remark
that {\it our conclusion stays unchanged if we multiply the integrand in (\ref{2.21}) by some
function $B_i(x)$ with a bounded variation}. This remark will be later useful when we will include
the variations of $\bar{\alpha}_j(\alpha_i)$ and $Q_G(P, \{ \bar{\alpha}\}_{j\not= i} \alpha_i)$, not
due to the constraint (\ref{2.9}), this inclusion leading to such a factor. \par

The only case left to see is $\mu_i \Lambda \to \infty$. We eliminate this possibility straight
away if it is to hold true for any $i$ because it would mean $\bar{\alpha}_iI \to \infty$ and
therefore

\beq
\sum_{i=1}^I \bar{\alpha}_i \to \infty
\label{2.27}
\eeq

\noi if all $\bar{\alpha}_i$'s have this behaviour, which is incompatible with the constraint 

\beq
\sum_{i=1}^I \bar{\alpha}_i = h_0 \ \ \ .
\label{2.28}
\eeq

We now consider the most general case where an $\bar{\alpha}_i$ can have the following behaviour

\beq
\label{2.29}
\bar{\alpha}_{i \delta} = C_{\delta} (h_0/I)^{1 + \delta}
\eeq

\noi where $\bar{\alpha}_{i\delta}$ denotes such an $\bar{\alpha}_i$ and $\delta$ is real with
$\delta \geq - 1$. Then, (\ref{2.28}) can be written

\beq
\label{2.30}
\sum_{\delta } n_{\delta} \ C_{\delta} (h_0/I)^{1 + \delta} = h_0 \ \ \ ,
\eeq

\noi and the relation

\beq
\label{2.31}
\sum_{\delta} n_{\delta} = I
\eeq

\noi takes care of the fact that there are $I$ propagators. For $\delta > 0$, $\bar{\alpha}_{i
\delta}$ is then decreasing faster than $(h_0/I)$, $n_{\delta}(h_0/I)^{1 + \delta}$ is tending to
zero, and the {\it sum of the corresponding} $\bar{\alpha}_i$'s {\it contributes infinitesimally to
the sum in (\ref{2.30})}. For $\delta < 0$, $\bar{\alpha}_{i\delta}$ is decreasing slower than
$(h_0/I)$, and therefore this case corresponds to $\bar{\alpha}_i\Lambda$, $\mu_i \Lambda \to
\infty$ and (\ref{2.30}) implies

\beq
\label{2.32}
n_{\delta} \leq (h_0/C_{\delta}) (I/h_0)^{1 + \delta} \ \ \ ,
\eeq

\noi i.e. $n_{\delta}/I$ is infinitesimal. Taking into account the fact that $\delta > 0$ is
inconsistent (we will see below that this is even forbidden if $\delta \not= 0$ is allowed from
the start), the sum in (\ref{2.31}) is saturated by $n_{\delta}$ with $\delta = 0$. \par

Now, let us consider some monomial in (\ref{2.2}) which can be written

\beq
\label{2.33}
\prod_{\ell \not\subset {\cal T}} \bar{\alpha}_{\ell} = \exp \left ( - \sum_{\ell \not\subset {\cal
T}} \ {\rm Log} \ \bar{\alpha}_{\ell} \right ) \ \ \ . \eeq

\noi Looking at the contribution to the sum in (\ref{2.33}) coming from $\bar{\alpha}_{\ell}$'s
with $\delta < 0$, we get, $\delta_-$ corresponding to an $\bar{\alpha}_{\ell 
\delta}$ with $\delta < 0$,

\beq
\label{2.34}
{\rm Log} \ \left ( \prod_{\delta_-} \bar{\alpha}_{i \delta} \right ) = - n_{\delta_-} <1 +
\delta_- > \ {\rm Log} \ I  \eeq

\noi neglecting non-leading terms, $<1 + \delta_- >$ being the average of $(1 + \delta )$ for $\delta
< 0$ and $n_{\delta_-}$ the total number of them. We see that (\ref{2.34}) is infinitesimal with
respect to the contribution of $\bar{\alpha}_i$'s with $\delta = 0$, which is

\beq
\label{2.35}
- \left ( L - n_{\delta_-} \right ) \ {\rm Log} \ I
\eeq

\noi with $L$ being a constant times $I$. We therefore conclude that Log $P(\{ \bar{\alpha}\})$ is
built up from $\bar{\alpha}_i$'s with $\delta = 0$ up to a possible relatively infinitesimal
contribution from $\bar{\alpha}_i$'s with $\delta < 0$. \par

Starting from (\ref{2.29}) we have assumed that $\bar{\alpha}_j$ can have a different behaviour
than that of (\ref{2.10}) where only the case $\delta = 0$ was considered at first. Now, we would
like to extend the argument made for excluding $\delta < 0$ to $\delta > 0$. \par

Repeating an argument made before \cite{10r} we will conclude that consistency is achieved only for
all $\bar{\alpha}_i$'s having the behaviour (\ref{2.10}). \par

Let us restate this argument (in a somewhat modified form). \par 

I - If some $\bar{\alpha}_j$'s decrease faster than $O(h_0/I)$, i.e. if $\delta > 0$ for some of
them, this would either \par
i) not change the behaviour

\[ b_i = b_{i0} \left [ (h_0 - \alpha_i )/I \right )^{L-1} \]

\noi and then, as we have seen previously, $\bar{\alpha}_i$ is $O(h_0/I)$ for any $i$.  \par

ii) or change $b_i$ such that $b_i$ would decrease faster than $[(h_0 - \alpha_i)/I]^{L-1}$. Then,
two cases are to be evaluated \par

a) $(h_0 - \alpha_i)^{(I-d/2L)(1 - \beta)}$ is replaced by $(h_0 - \alpha_i)^{M_a}$ with $M_a =
O(I)$. Then, $\bar{\alpha}_i$ has still to be $O(h_0/I)$, for any $i$ and this contradicts our
assumption $\bar{\alpha}_j I \to 0$. \par

b) $(h_0 - \alpha_i)^{(I - d/2L)(1 - \beta)}$ is replaced by $(h_0 - \alpha_i)^{M_b}$ with $M_b <
O(I)$. Then, $\Lambda$ is replaced by $M_b/h_0$ in (\ref{2.21}). If we assume $\mu_iM_b = O(1)$,
then (\ref{2.21}) gives us

\beq
\label{2.36}
\left [ \left ( \mu_i + \bar{\alpha}_i \right ) M_b \right ]^{-d/2} \sim I/M_b
\eeq

\noi which leads to a contradiction because the left-hand side is $O(1)$ and the right-hand side is
infinite. If we assume $\mu_i M_b \to \infty$, then
(\ref{2.21}) gives us

\beq
\label{2.37}
\left [ \left ( \mu_i + \bar{\alpha}_i \right ) M_b \right ]^{-d/2} \sim (I/M_b) (\mu_i M_b)^{-d/2}
\eeq
\noi which, again, is inconsistent with $(I/M_b) \to \infty$. \par

If we assume $\mu_i M_b \to 0$, we get the relation (\ref{2.25}) and because $\mu_i$ cannot
decrease faster than $1/I$, the only possibility left is $\mu_i = O(1/I)$ which entails
$\bar{\alpha}_i = O(1/I)$ and (\ref{2.10}) is recovered. \par

Of course, we also have to take into account the change on $\beta$ produced by the possible
altering of some $\bar{\alpha}_i$'s decreasing faster than $O(h_0/I)$. We could have instead of
(\ref{2.12c})

\beq
\label{2.38}
Q_G (P, \{ \bar{\alpha} \}) = Q_{G_0} \left [ (h_0 - \alpha_i )/(I - 1) \right ]^{1 + \delta_Q}
\eeq 

\noi with $\delta_Q > 0$. However, we see that this will make $Q_G(P, \{ \bar{\alpha}\})$ tend to
zero and $\beta$ will be tending to zero too. So the reasoning made above remains unchanged and the
decrease of $\bar{\alpha}_i$ faster than $1/I$ remains forbidden. \par

Finally, we can rewrite (\ref{2.21}) under the following form

\[ 
\left ( \mu_i \Lambda + \bar{\alpha}_i \Lambda \right )^{-d/2} = \exp \left [ - dL/(2I) - (1 -
dL(2I))\beta \right ] \]
\beq
\left [ (I - 1)/(h_0 \Lambda ) \right ] \int_0^{\infty} dx \left ( \mu_i \Lambda + x \right )^{-d/2}
\exp (-x) \label{2.39}
\eeq

\noi which enables a numerical resolution, $\mu_i \Lambda$, $\bar{\alpha}_i \Lambda$, $dL/(2I)$,
$(I-1)/\Lambda$ and $\beta$ being $O(1)$. For each independent $\bar{\alpha}_i$ then exists such
an equation. So, in principle, we have {\it a complete perturbative solution of scalar massive
$\phi^3$ field theory by solving such equations}. As we will see in the end of the next section and
section 9 the massless case can also be solved in the same way.

\mysection{Regge behaviour, the leading Regge trajectory and its $m \to 0$ limit}

\noi {\large \bf A - Regge behaviour} \\

One may wonder how the formalism we have described in the last section can yield the Regge
behaviour for appropriate graphs. Previous work made some thirty years ago \cite{8r,9r,14r} will
serve as a check of ours ideas. In fact we will see that using the ladder graphs we can get the
Regge behaviour in an easy but curious way. The leading Regge trajectory will also be easy to write
down, even for arbitrary argument $t/m^2$ which was not the case in the approaches using the
Bethe-Salpeter equation or the multiperipheral model. \par

So let us take ladder graphs of massive scalar $\phi^3$ field theory. For those graphs, due to the
symmetry existing for central propagators and side propagators of the ladder, we only have
\cite{15r} two independent $\bar{\alpha}_i$ left that we call $\bar{\alpha}_+$ for the mean-value of
the $\alpha$-parameters on the sides and $\bar{\alpha}_-$ for the mean-value of the
$\alpha$-parameters in the center. (See Fig. 1). Defining

\beq
y \equiv \left ( 2 \bar{\alpha}_+/ \bar{\alpha}_- \right )^{1/2}
\label{3.1}
\eeq 

\noi we get \cite{15r}, neglecting all terms which vanish as $L \to \infty$,

\bminiG{EDh}
P_G(\{ \bar{\alpha}\}) = (\bar{\alpha}_-)^L \exp (yL) f(g)
\label{3.2a}
\eeeq
\beeq
f(y) = 1/2 \quad y (1 + y^{-1})^2
\label{3.2b}
\emini

\noi and

\beq
Q_G(P, \{ \bar{\alpha}\}) = t/2 \ L \ \bar{\alpha}_+ + \bar{\alpha}_- \ s \exp (-yL) [f(g)]^{-1} \ \
\ .  \label{3.3}
\eeq

\noi Reinstating the coupling constant dependence through the factor $(- \gamma )^{2L+2}$, we get,
for $\phi^3$ at $d = 4$ and therefore replacing $I - d/2L$ by $L + 1$,

\[
F_G = (e^2/\sqrt{3}) \left [ - \gamma/(m f(y)) \right ]^2 \left [ - \gamma e /(m 4\pi 3 \sqrt{3})
\right ]^{2L} \]
\beq
\left [ \exp (-y)/g_- \right ]^{2L} (1 - \beta )^{3L+1}
\label{3.4}
\eeq

\noi with $g_- \equiv \bar{\alpha}_- \Lambda$. Now, we can sum over $L$ the ladder amplitudes
(\ref{3.4}) and find the saddle-point equation

\beq
2 \ell n \ {\rm C}^{\rm st} + 3 \ell n (1 - \beta ) + (3L + 1) \left [ y + 1/(L+1) \right ]
bs/(1 - \beta ) = 0 \label{3.5}
\eeq

\noi with the following definitions

\bminiG{EDh}
{\rm C}^{\rm st} \equiv \left [ - \gamma e/(m 4 \pi 3 \sqrt{3}) \right ] \left [ \exp (-y)/g_-
\right ] \label{3.6a}
\eeeq
\beeq
\beta = a + bs
\label{3.6b}
\eeeq
\beeq
a = t/2 \ \bar{\alpha}_+ \ \Lambda/m^2 
\label{3.6c}
\eeeq
\beeq
bs = s \left [ g_-/(m^2 f(y)) \right ] \exp (-y L)/(L + 1) \ \ \ .
\label{3.6d}
\emini

We see immediately that as $s \to \infty$, there is no solution of (\ref{3.5}) for finite $L$,
which is satisfactory. $bs$ being constant also brings no solution and $bs \to \infty$ gives $F_G$
behaving like $(-bs)^L$ which is exploding. The only possibility left is $bs$ tending to zero.
Then, (\ref{3.5}) becomes 

\beq
(1 - a) \left [ 2/3 \ \ell n \ {\rm C}^{\rm st} + \ell n (1 - a) \right ] + L \ y \ bs = 0
\label{3.7}
\eeq

\noi which has the solution $L_{sp}$ at the saddle point with

\bminiG{EDh}
L_{sp} = (1/y) \ \ell n \ (s/s_0)
\label{3.8a}
\eeeq

\noi with

\beeq
- (1 - a) \left [2/3 \ \ell n \ {\rm C}^{\rm st} + \ell n (1 - a) \right ] = (s_0/m^2)y \ g_-/f(y) \
\ \ .  \label{3.8b}
\emini

The relation (\ref{3.8a}) is a very natural one meaning that the dominant ladders have a length
proportional to the ``rapidity'' $\ell n (s/s_0)$. The same phenomenon appears in the multiperipheral
model \cite{9r} and in the parton model \cite{16r}. Putting this value of $L$ in (\ref{3.4}) we
immediately obtain the leading Regge trajectory

\beq
\alpha (t) = y^{-1} \left [ 2 \ \ell n \ {\rm C}^{\rm st} + 3 \ \ell n (1 - \beta ) \right ] \ \ \ , 
\label{3.9}
\eeq

\noi which is a simple analytic expression. \\

\noi {\large \bf B - The $m \to 0$ limit} \\

Let us now examine the case where $m^2 \to 0$ in order to obtain {\it massless} $\phi^3$ results.
The first thing to note is that $1 - \beta \to 0$ as $m^2 \to 0$ due to the fact that by definition

\beq
1 - \beta \equiv \left [ 1 + Q_G(P, \{ \bar{\alpha} \})/h_0m^2 \right ]^{-1} \ \ \ . 
\label{3.10}
\eeq

We expect, looking at (\ref{2.5}), {\it that $F_G$ becomes independent of $m$ as $m \to 0$}. Of
course the introduction of a mass is necessary in order to avoid infrared problems in the
definition of Feynman integrals but the final result for the amplitude $F_G$ and all physical
quantities should be that they are well defined and independent of $m$ as the mass $m$ tends to
zero. \par

We will first verify that the trajectory $\alpha (t)$ obtained in (\ref{3.9}) is indeed independent
of $m$ as $m$ tends to zero. Looking at the definition (\ref{3.6a}) of the ${\rm C}^{\rm st}$ we
get 

\beq
\alpha (t) = y^{-1} \left \{ 2 \left [ \ell n \left ( e/(4 \pi 3 \sqrt{3}) \right ) - y + \ell n (-
\gamma /m) - \ell n (\bar{\alpha}_- \Lambda ) \right ] + 3 \ell n (1 - \beta ) \right \} \label{3.11}
\eeq

\noi Because $\Lambda = (I - d/2\ L) (1 - \beta )/h_0$ (see (\ref{2.17})), we have

\beq
\ell n (\bar{\alpha}_- \Lambda ) = \ell n (1 - \beta ) + \hbox{terms independent of $m$} \ \ \ ,
\label{3.12}
\eeq

\noi so the sum of terms dependent on $m$ in the bracket of (\ref{3.11}) is

\beq
2 \left [ - \ell n \ m - \ell n (1 - \beta ) \right ] + 3 \ell n (1 - \beta ) = - \ell n \ m^2 + \ell
n (1 - \beta ) \label{3.13}
\eeq

\noi which is independent of $m$ in the limit $m \to 0$ as can be readily deduced from
(\ref{3.10}). Of course, there could be some indirect dependence on $m$ through $\bar{\alpha}_-$ or
$y = (2 \bar{\alpha}_+/ \bar{\alpha}_-)^{1/2}$ but we will see later on in section 9 that the
equations determining $\bar{\alpha}_+$ and $\bar{\alpha}_-$ are indeed independent of $m$ as $m$
tends to zero, making the transition to QCD possible for what concerns this limit. \\

\noi {\large \bf C - Introduction to the determination of $\bar{\alpha}_+$ and $\bar{\alpha}_-$} \\

We now turn to the task of determining the values of $\bar{\alpha}_+$ and $\bar{\alpha}_-$ in the
case of the ladder graphs. Some refinements of the consistency equations (\ref{2.16}) or
(\ref{2.21}) are necessary in order to really be able to calculate the leading Regge trajectory. In
fact, until now only the dependence on $\alpha_i$ induced by the $\delta$-function $\delta (h_0 -
\sum\limits_i \alpha_i)$ has been taken into account. Taking a careful look at quantities which may
have additional dependence on $\alpha_i$ when it is varied as in (\ref{2.16}) and (\ref{2.21}) we
find two kinds of dependencies. \par

The first one is due to the fact that $Q_G(P, \{\bar{\alpha} \})$, where all $\alpha_{\ell}$'s have
been replaced by their mean-values is not exactly equal to $Q_G(P, \{ \bar{\alpha}\})_i$ which is
$Q_G(P, \{ \alpha \})$ where $(I - 1)$ variables $\alpha_{\ell}$ have been replaced by
$\bar{\alpha}_{\ell}$ and where $\alpha_i$ is left free to vary. We will examine in the next
section the corrections which must be added to (\ref{2.16}) and (\ref{2.20}) in order to take into
account this phenomenon. \par

The second one is due to the dependency of the mean-values $\bar{\alpha}_{\ell}$ on $\alpha_i$. The
relative variation $\delta \bar{\alpha}_{\ell}/\bar{\alpha}_{\ell}$ of one mean-value will be
found to be of order $1/L$. However, as we are dealing with powers $(\bar{\alpha}_{\ell})^L$, the
final outcome will be a sort of ``renormalization'' of the constants $\bar{\alpha}_+ \Lambda$,
$\bar{\alpha}_- \Lambda$. The definition of the scale $\Lambda$ will also be affected and $\Lambda$
itself will also be renormalized. The details of this renormalization will be exposed in section 5.

\mysection{The $Q_G(P, \{\bar{\alpha}\})_i$ dependence on $\alpha_i$} 
\hspace*{\parindent} First, let us give the way the expression for $P_G(\{\bar{\alpha}\})$ is
obtained. In Fig. 2a we displayed a ladder with $L$ loops where $p$ central propagators are
removed and $L - p$ side propagators removed giving rise to a monomial $(\bar{\alpha}_-)^p
(\bar{\alpha}_+)^{L-p}$. In the topology displayed in Fig. 2a we have $L - p$ ``cells'' of lengths
$\ell_1, \ell_2, \cdots , \ell_{L-p}$ where a cell of length $\ell_i$ is obtained by removing
$\ell_i - 1$ center propagators. For each cell a side-propagator has to be removed in order to
obtain ``open cells'', i.e. no loop left. The remaining propagators then form a spanning tree on the
ladder graph. The two other topologies displayed in Fig. 2b and Fig. 2c correspond to one and two
end-cells being opened by removing a center-propagator instead of a side-propagator. \par

Then, the expression for $P_G^{(a)}(\{ \bar{\alpha}\})$ which is the part of
$P_G(\{\bar{\alpha}\})$ corresponding to the topology displayed in Fig. 2a is

\beq
P_G^{(a)}(\{ \bar{\alpha}\}) = \sum_{p=0}^{L-1} (\bar{\alpha}_-)^p (\bar{\alpha}_+)^{L-p}
\sum_{\ell_1 + \cdots + \ell_{L-p} = L} 2^{L-p} \ \ell_1 \cdots \ell_{L-p} \label{4.1}
\eeq  

\noi where the factors $2\ell_i$ comes from the fact that there are $2 \ell_i$ side-propagators
along a cell which can be removed in order to open it. \par

In order to obtain $\sum\limits_{\cal C} s_{\cal C} \prod\limits_{\ell = {\cal C}}
\bar{\alpha}_{\ell}$ in (\ref{2.3}) from $P_G(\{\bar{\alpha}\})$ one has to remove one more
propagator. To select the part with $s_{\cal C} = t$ one has to remove a second side-propagator on
the opposite side of the first one along a cell. The $s_{\cal C} = s$ part is obtained by removing
the $L + 1$ center-propagators and no side-propagator. So, we have

\beq
Q_G^{(a)}(P, \{ \bar{\alpha}\}) = \left [ (1/2) \bar{\alpha}_+ \ t \ L \ P_G^{(a)}(\{ \bar{\alpha}\})
+ s \ \bar{\alpha}_-^{L+1} \right ] P_G^{-1} (\{ \bar{\alpha}\}) \ \ \ , \label{4.2}
\eeq 

\noi the factor $L$ coming from $L$ side-propagators which can be removed, and the factor 1/2 in
order not to double-count the cuts obtained. Of course $Q_G(P, \{ \bar{\alpha}\})$ is the sum of
the three contributions obtained from cutting $P_G^{(a)}(\{ \bar{\alpha}\})$, $P_G^{(b)}(\{
\bar{\alpha}\})$ and $P_G^{(c)}(\{ \bar{\alpha}\})$. (We have integrated in $Q_G^{(a)}(P, \{
\bar{\alpha}\})$ the term proportional to $s$). We now have to replace one $\bar{\alpha}_+$ by
$\alpha_{i_+}$ or one $\bar{\alpha}_-$ by $\alpha_{i_-}$ to compute the dependence of $Q_G(P, \{
\bar{\alpha}\})_i$ on either $\alpha_{i_+}$ or $\alpha_{i_-}$ when one of them is left free to
vary. \\

\noi {\large \bf A - The $\alpha_{i_+}$ dependence} \\

So let us consider

\beq
P_G^{(a)}(\{ \bar{\alpha}\})_{i_+} = a_{i_+} + b_{i_+} \alpha_{i_+}
\label{4.3}
\eeq

\noi where $a_{i_+}$ and $b_{i_+}$ are polynomials of mean-values $\bar{\alpha}_j$, $j \not= i_+$.
We know that the pro\-pa\-ga\-tor $i_+$ is along one cell which has a length $\ell$. Therefore, the
cutting of all other cells is unaffected and we get for these cells a term contributing to
$Q_G^{(a)}(P, \{ \bar{\alpha}\})_{i_+}$ which is

\beq
t/2 \ \bar{\alpha}_+ (L - \ell ) \ P_G^{(a)}(\{ \bar{\alpha}\})_{i_+} 
\label{4.4}
\eeq

\noi reminiscent of the first term in (\ref{4.2}). Let us now consider the cutting of the cell which
contains $i_+$ (see fig. 3). The term $b_{i_+} \alpha_{i_+}$ has $i_+$ as the propagator removed
(which is not the case of fig. 3 where $i_+$ has been chosen on the opposite side of the removed
propagator) and therefore cutting the cell on the other side, we have $\ell$ possibilities which
gives the term

\beq
t/2 \cdot \bar{\alpha}_+ \ \ell \ b_{i_+} \alpha_{i+} \ \ \ .
\label{4.5}
\eeq

\noi However when considering $a_{i_+}$, $i_+$ can be any of the $2 \ell - 1$ side-propagators
which are still part of the open cell in Fig. 3. If $i_+$ is on the same side as the cut propagator
it cannot be removed because this would isolate a part of the cell where no external line is
attached (more exactly this gives a zero contribution because $\sum P_v = 0$ is that case in
(\ref{2.4})). Then, we get two terms

\bminiG{EDh}
t/2 \ \bar{\alpha}_+ \ \ell \ \left [ (\ell - 1)/(2 \ell - 1) \right ] a_{i_+}
\label{4.6a}
\eeeq
\beeq
t/2 \left [ (\ell - 1) \bar{\alpha}_+ + \alpha_{i_+} \right ] \left [ \ell/(2 \ell -
1) \right ] a_{i_+} 
\label{4.6b}
\emini

\noi the first one corresponding to $i_+$ on the side of the already cut propagator and the second
one (corresponding to Fig. 3) where $i_+$ is on the opposite side. Summing (\ref{4.4}),
(\ref{4.5}), (\ref{4.6a}) and (\ref{4.6b}) we get 

\beq
t/2 \ L \ \bar{\alpha}_+ \ P_G^{(a)} (\{ \bar{\alpha}\}) + t/2 \left [ \ell/(2 \ell - 1) \right ]
(\alpha_{i_+} - \bar{\alpha}_+) a_{i_+}  \ \ \ . \label{4.7}
\eeq

\noi Defining~:

\beq
\mu_+ \equiv a_{i_+}/b_{i_+} 
\label{4.8}
\eeq

\noi we get

\bminiG{EDh}
Q_G^{(a)}(P, \{ \bar{\alpha}\})_{i_+} = Q_G^{(a)} (P, \{ \bar{\alpha}\}) + \varepsilon_+^{(a)}
\label{4.9a}
\eeeq
\beeq
&&\varepsilon_+^{(a)} = t/2  < \ell/(2\ell - 1)> \left [ \mu_+/(\mu_+ +
\alpha_{i_+}) \right ] 
\nn \\
&&\left ( \alpha_{i_+} - \bar{\alpha}_+ \right ) P_G^{(a)} (\{ \bar{\alpha}
\})_{i_+}/P_G(\{ \bar{\alpha}\})_{i_+} \ \ \ . \label{4.9b} 
\emini

\noi For the other topologies in Fig. 2b and Fig. 2c the same reasoning provides us with the same
expressions replacing (a) by (b) and (c). $\mu_+$ keeps the same value up to terms of order $1/L$
because the topologies differ only on one or two cells and because there are an infinite number of
them. $<\ell/(2 \ell - 1)>$ is the mean-value of $\ell /(2 \ell - 1)$, integrating over all
$\ell$'s. Then,

 \bminiG{EDh}
Q_G(P, \{ \bar{\alpha}\})_{i_+} = Q_G (P, \{ \bar{\alpha} \}) + \varepsilon_+
\label{4.10a}
\eeeq
\beeq
\varepsilon_+ = t/2  <\ell /(2 \ell - 1)> (\alpha_{i_+} - \bar{\alpha}_+)
\mu_+/(\mu_+ + \alpha_{i_+}) \ \ \ .  \label{4.10b}
\emini

In a recent letter \cite{15r} we gave an estimate for $<1/\ell>$.

\bminiG{EDh}
<1/\ell> = \mu_-/\left ( \mu_- + \bar{\alpha}_- \right ) = y
\label{4.11a}
\eeeq
\beeq
< 1/(2 \ell )> = \bar{\alpha}_+/\left ( \mu_+ + \bar{\alpha}_+ \right ) = y/2 \ \ \ .
\label{4.11b}
\emini

Writing $\ell/(2 \ell - 1)$ as $(2 - 1/\ell )^{-1}$ and replacing $<(2 - 1/\ell )^{-1}>$ with $(2
-\break \noindent  <1/\ell >)^{-1}$ we get a crude estimate for this average which is $(2 - y)^{-1}$.
We used such an estimation in our numerical solutions of equations. \par

Now, $\varepsilon_+$ is of order $1/L$ relative to $Q_G(P, \{ \bar{\alpha}\}) + h_0m^2$, but as
this quantity is elevated to a power $- (I - d/2\ L)$ in (\ref{2.5}) and in (\ref{2.13}) we get a
finite correction factor. With $\Lambda$ given in (\ref{2.17}) we get a factor

\[
H_{i_+} \left ( \alpha_{i_+} - \bar{\alpha}_+ \right ) = \exp \left ( - \varepsilon_+ \Lambda/m^2
\right )
\]
\beq
 = \exp \left \{ - t/2 \ <\ell /(2 \ell - 1)> \mu_+(\Lambda/m^2) \left ( \alpha_{i_+} -
\bar{\alpha}_+ \right ) / \left ( \mu_+ + \alpha_{i_+} \right ) \right \} 
\label{4.12}
\eeq

\noi which must be inserted in the integrand of the right-hand side of (\ref{2.16}) when
calculating $\bar{\alpha}_+$. \\

\noi {\large \bf B - The $\alpha_{i_-}$ dependence} \\

We are now interested in the case where one $\alpha_{i_-}$ Feynman parameter of a central
propagator is left free to vary. Then,

\beq
P_G(\{ \bar{\alpha}\})_{i_-} = a_{i_-} + b_{i_-} \alpha_{i_-}
\label{4.13}
\eeq

\noi where $a_{i_-}$ and $b_{i_-}$ are polynomials of mean-values $\bar{\alpha}_j$, $j\not= i_-$.
Correspondingly, $Q_G(P, \{ \bar{\alpha}\})_{i_-}$ being the expression for $Q_G(P, \{ \bar{\alpha}
\})$ when $\alpha_{i_-}$ is free, we have

\beq
Q_G(P, \{\bar{\alpha}\})_{i_-} = (t/2) \ L \ \bar{\alpha}_+ + s \ \bar{\alpha}_-^L \ \alpha_{i_-}
\ P_G^{-1}(\{ \bar{\alpha} \})_{i_-} \label{4.14}
\eeq

\noi and we see that its $\alpha_{i_-}$ dependence is confined to the term containing $s$. We can
therefore write

\beq 
Q_G(P, \{ \bar{\alpha}\})_{i_-} = Q_G (P, \{ \bar{\alpha} \}) + s \ \bar{\alpha}_-^L \ P_G^{-1}(\{
\bar{\alpha} \}) \left [ \alpha_{i_-} \left ( P_G (\left \{ \bar{\alpha} \right
\})_{i_-}/P_G(\{\bar{\alpha}\}) \right )^{-1} - \bar{\alpha}_- \right ] \ . \label{4.15}
\eeq 

\noi Defining

\beq
\mu_- \equiv a_{i_-}/b_{i_-} \ \ \ , 
\label{4.16}
\eeq

\[
\alpha_{i_-} \left ( P_G (\{ \bar{\alpha}\})_{i_-}/P_G(\{ \bar{\alpha}\} \right )^{-1} -
\bar{\alpha}_- = \alpha_{i_-} \left ( \mu_- + \bar{\alpha}_- \right )/\left ( \mu_- + \alpha_{i_-}
\right ) - \bar{\alpha}_- \]
\beq
= \mu_- \left ( \alpha_{i_-} - \bar{\alpha}_- \right )/ \left ( \mu_- + \alpha_{i_-}
\right ) \ \ \ .\label{4.17} \eeq

\noi On the other hand

\beq
s \ \bar{\alpha}_-^L \ P_G^{-1}(\{ \bar{\alpha}\}) = s \exp (- y L) [f(y)]^{-1}
\label{4.18}
\eeq

\noi and if we are at the saddle point $L = L_{sp}$, $s \exp(-yL) = s_0$ and (\ref{3.8b}) gives

\beq
s_0 [f(y)]^{-1} = \left [ m^2/(y g_-) \right ] \left \{ -(1 - a) \left [ 2/3 \ \ell n \ {\rm C}^{\rm
st} + \ell n (1 - a)\right ] \right \}   \ \ \ . \label{4.19}
\eeq

\noi So,

\bminiG{EDh}
\label{4.20a}
Q_G(P, \{ \bar{\alpha}\})_{i_-} = Q_G (P, \{ \bar{\alpha}\}) + \varepsilon_-
\eeeq
\noi with
\beeq
\varepsilon_- = s_0[f(y)]^{-1} \left ( \alpha_{i_-} - \bar{\alpha}_- \right ) \mu_-/\left ( \mu_- +
\alpha_{i_-} \right ) \ \ \ . \label{4.20b}
\emini

\noi Again, $Q_G(P, \{ \bar{\alpha}\})_{i_-}$ appears in 

\beq
\left [ Q_G(P, \{ \bar{\alpha}\})_{i_-} + h_0 m^2 \right )^{-(I - d/2L)}
\label{4.21}
\eeq

\noi with for $d = 4$ and $\phi^3$, $I - d/2 \ L = L + 1$. So with $\Lambda = (I - d/2\ L) (1 - \beta
)/h_0$ we get a factor 

\[
H_{i_-}(\alpha_{i_-} - \bar{\alpha}_- ) = \exp (- \varepsilon_- \Lambda/m^2 ) 
\]
\beq
= \exp \left \{ y^{-1}(1 - a) \left [ 2/3 \ \ell n \ {\rm C}^{\rm st} + \ell n (1 - a) \right ]
(\mu_-/\bar{\alpha}_-) (\alpha_{i_-} - \bar{\alpha}_-)/(\mu_- + \alpha_{i_-}) \right \}
\label{4.22}
\eeq 

\noi which must be inserted in the integrand of the right-hand side of (\ref{2.16}) when calculating
$\bar{\alpha}_-$. In the following the notation $H_i(\alpha_i - \bar{\alpha}_i)$ will designate
either $H_{i_+}(\alpha_{i_+} - \bar{\alpha}_+)$ or $H_{i_-} (\alpha_{i_-} - \bar{\alpha}_-)$
depending on $i$ being a side- or center-propagator.

\mysection{The $\bar{\alpha}_{\ell}(\alpha_i)$ dependence : introduction} 

\hspace*{\parindent} $P_G(\{ \bar{\alpha}\})$ is a polynomial of degree $L$ in the mean-values
$\bar{\alpha}_{\ell}$. Therefore, an infinitesimal variation $\delta \bar{\alpha}_{\ell}(\alpha_i)$
as $\alpha_i$ varies can result in a finite correction factor. In the same way, the variation of
$\bar{\alpha}_+$ in $Q_G(P, \{ \bar{\alpha}\}) = t/2 \ L \ \bar{\alpha}_+$ (up to terms vanishing as
$1/L$) could matter because $Q_G(P, \bar{\alpha}\} + h_0 m^2$ is elevated to the power $- (I -
d/2L) = - (L + 1)$. In this section and the following ones we will be trying to evaluate the
dependence of $\bar{\alpha}_j$ on $\alpha_{i_+}$ or $\alpha_{i_-}$ ($\alpha_{i_-}$ being a
center-propagator variable and $\alpha_{i_+}$ a side-propagator variable) when one of them is free
to vary, with of course $j \not= \alpha_{i_+}$ or $\alpha_{i_-}$. This will be a somewhat lengthy
task, at least to be exposed clearly, so we divided it into several steps. \par

The first one is, given $\bar{\alpha}_{j_-}(\alpha_i)$ and $\bar{\alpha}_{j_+}(\alpha_i)$ (we will
thereafter use the notation $\bar{\alpha}_-(\alpha_i)$ for $\bar{\alpha}_{j_-}(\alpha_i)$ and
$\bar{\alpha}_+(\alpha_i)$ for $\bar{\alpha}_{j_+}(\alpha_i))$, determinating what is the factor
which modifies the integrand of the right-hand side of the consistency equation (\ref{2.16}). \par

The second step is exposing the method we followed to calculate $\bar{\alpha}_-(\alpha_i)$ and
$\bar{\alpha}_+(\alpha_i)$. In this section we deal with these two first steps. The actual
determination of $\bar{\alpha}_+(\alpha_i)$ and $\bar{\alpha}_-(\alpha_i)$ will be exposed in the
next section. \\

\noi {\large \bf A - General form of the correction factor} \\

We start with the factor

\beq
\left [ b_i(\{ \bar{\alpha}\}) \right ]^{-d/2} \equiv \left [ P_G(\{ \bar{\alpha}\})/ \left ( \mu_i
+ \bar{\alpha}_i \right ) \right ]^{-d/2} \label{5.1}
\eeq

\noi which appears as $[b_{i0}(h_0/I)^{L-1}]^{-d/2}$ in (\ref{2.13}). In the notation used in
(\ref{2.13}) $b_{i0}$ is a constant. Here $b_i(\{ \bar{\alpha}\})$ is $b_{i0}$ multiplied by
$(h_0/I)^{L-1}$. Because $\delta \bar{\alpha}_j/\bar{\alpha}_j$ will be infinitesimal when the
$\alpha_i$ variation is $O(h_0/I)$, we will consider that Log $b_i(\{\bar{\alpha}\})$ and Log
$P_G(\{ \bar{\alpha}\})$ have the same variation. Then, as $L \to \infty$ (see (\ref{3.2a}))

\beq
{\rm Log} \ P(\{\bar{\alpha}\}) = L \left [ {\rm Log} \ \bar{\alpha}_- + (2
\bar{\alpha}_+/\bar{\alpha}_-)^{1/2} \right ] + \ {\rm Log} \ f(y) \ \ . \label{5.2}
\eeq

\noi Taking the variation and neglecting $d f(y)$ because it is of order $(1/L)$ with respect to
the other terms 

\[
d \ {\rm Log} \  P_G(\{\bar{\alpha}\}) = L \left [ d \bar{\alpha}_-/\bar{\alpha}_- + 1/2 \ (2
\bar{\alpha}_+/\bar{\alpha}_-)^{1/2} d \bar{\alpha}_+/\bar{\alpha}_+ -  1/2 \ (2
\bar{\alpha}_+/\bar{\alpha}_-)^{1/2} d \bar{\alpha}_-/\bar{\alpha}_- \right ] \]
\beq
= L \left [ (1 - y/2) d \bar{\alpha}_-/\alpha_- + y/2 \ d \bar{\alpha}_+/\bar{\alpha}_+ \right ]
\label{5.3}
\eeq

\noi which is equal to $db_{i}(\{\bar{\alpha}\})$ up to $O(1/L)$ terms. Next comes the variation
of\break \noindent Log $[Q_G(P, \{ \bar{\alpha}\}) + h_0 m^2]^{-(I-d/2L)}$ which is (the term
proportional to $\bar{\alpha}_-$ in $Q_G(P, \{ \bar{\alpha} \})$ is $O(h_0/I)$ at the saddle-point,
see (\ref{3.3}))

\beq
d \ {\rm Log} \ \left [ t/2 \ L \ \bar{\alpha}_+ + h_0 m^2 \right ]^{-(I-d/2\ L)} = - (I-d/2\ L)\ 
t/2 \ (\Lambda /m^2) d \bar{\alpha}_+ \ \ \ .  \label{5.4}
\eeq

\noi Because $I - d/2L = L + 1$, we can define 

\beq
\eta \equiv (2/d) \ t/2 \ \bar{\alpha}_+ \Lambda/m^2
\label{5.5}
\eeq

\noi so that we get the correction factor

\bea
H_0(\alpha_i - \bar{\alpha}_i) &=& \exp \left \{ - d/2 L \left [ (1 - y/2)
d\bar{\alpha}_-/\bar{\alpha}_- + (y/2 + \eta ) d \bar{\alpha}_+/\bar{\alpha}_+ \right ] \right \}
\nn \\ 
&\equiv& \exp \left [ - d/2 \ L \ E_0(\alpha_i - \bar{\alpha}_i ) \right ] \label{5.6} 
\eea

\noi where terms of order $(1/L)$ relative to Log $H_0(\alpha_i - \bar{\alpha}_i)$ have been
neglected in the argument of the exponential. \\

\noi{\large \bf B - General method for determining $\bar{\alpha}_{\ell}(\alpha_i)$} \\

The first thing we will do is to leave unintegrated two variables $\alpha_{\ell}$ and $\alpha_i$
instead of zero or one until now. Then, we will be able to see how $\alpha_i$ influences the
consistency equation for $\bar{\alpha}_{\ell}$. With the two variables $\alpha_i$ and
$\alpha_{\ell}$ left free we will have to take into account the $\delta$-function constraint $\delta
(h_0 - \sum\limits_i \alpha_i)$ through the replacement of $h_0$ (in the case when no variable was
left free) by $h_0 - (\alpha_i + \alpha_{\ell})$. Then, we have to evaluate $P_G(\{
\bar{\alpha}\})_{i, \ell}$ when the mean-values have been taken on $I - 2$ variables $j \not= i,
\ell$. We write

\bminiG{EDh}
P_G (\{ \bar{\alpha}\})_{i, \ell} = a_i (\alpha_{\ell}) + b_i (\alpha_{\ell}) \alpha_i
\label{5.7a}
\eeeq
\beeq
a_i(\alpha_{\ell}) = a_{i1} + a_{i2} \ \alpha_{\ell}
\label{5.7b}
\eeeq
\beeq
b_i(\alpha_{\ell}) = b_{i1} + b_{i2} \ \alpha_{\ell}
\label{5.7c}
\emini

\noi giving

\beq
P_G(\{\bar{\alpha}\})_{i, \ell} = \left ( a_{i2} + b_{i2} \alpha_i \right ) \left [ \left ( a_{i1}
+ b_{i1} \alpha_i \right )/ \left ( a_{i2} + b_{i2} \alpha_i \right ) + \alpha_{\ell} \right ] \ \ .
\label{5.8} \eeq

\noi Now, if $i$ and $\ell$ are infinitely far away along the ladder, the value of the ratio
$a_i/b_i$ will be independent of $\ell$. This is the factorization phenomenon
\cite{13r}, \cite{15r} for infinite graphs. So, we will be able to write

\beq
\left ( a_{i2}/b_{i2} \right )_{\pm} = \left ( a_{i1}/b_{i1} \right )_{\pm} = \mu_{i \pm}
\label{5.9}
\eeq 

\noi where $+$ or $-$ refers to $i$ being a side- or center-propagator. (Here, of course $i_1$ and
$i_2$ refer to the same propagator $i$ but for trees going through $\ell$ or not). So we get

\beq
P_G(\{ \bar{\alpha}\})_{i, \ell} = \left ( a_{i2} + b_{i2} \alpha_i \right ) \left ( b_{i1}/b_{i2}
+ \alpha_{\ell} \right ) \ \ \ . \label{5.10}
\eeq

\noi Invoking again the factorization phenomenon

\beq
\left ( a_{i1}/a_{i2} \right )_{\pm} = \left ( b_{i1} /b_{i2} \right )_{\pm} = \mu_{\ell_{\pm}}
\label{5.11}
\eeq
 
\noi but with the role of $i$ and $\ell$ interchanged (here $+$ or $-$ refers to $\ell$ being a
side- or center-propagator) we finally get 

\beq
P_G(\{\bar{\alpha}\})_{i, \ell} = b_{i2} \left ( \mu_i + \alpha_i \right ) \left ( \mu_{\ell} +
\alpha_{\ell } \right ) \label{5.12}
\eeq

\noi provided $i$ and $\ell$ are infinitely far away along the ladder. However, as this the case
which dominates when $i$ and $\ell$ are arbitrary, we will take (\ref{5.12}) to be valid in
general, making so an error of order $1/L$. As for $\bar{\alpha}_j(\alpha_i)$ we will also simplify
our notation and write

\beq
\mu_{i_{\pm}} = \mu_{\ell_{\pm}} = \mu_{\pm}
\label{5.13}
\eeq

\noi which is justified for symmetry reasons. $\mu_i$ or $\mu_{\ell}$ will be used if the nature of
$i$ or $\ell$ is not specified. Now, keeping only terms proportional to a constant as $L \to \infty$,
it is easy to proceed as for the obtention of (\ref{2.16}) and get 

\newpage
\bminiG{EDh}
\label{5.14a}
&&1 = \kappa^2 (I - 1) (I - 2)/h_0^2 \nn \\
&&\int_0^{h_0} d \alpha_i \left [ \left ( \mu_i + \bar{\alpha}_i \right )/ \left ( \mu_i +
\alpha_i \right ) \right ]^{-d/2} \exp \left  ( - \alpha_i \Lambda \right ) H_i\left ( \alpha_i -
\bar{\alpha}_i \right ) H_0 \left ( \alpha_i - \bar{\alpha}_i \right ) \nn \\
&&\int_0^{h_0} d \alpha_{\ell} \left [ \left ( \mu_{\ell} + \bar{\alpha}_{\ell} \right )/ \left (
\mu_{\ell} + \alpha_{\ell} \right ) \right ]^{-d/2} \exp (- \alpha_{\ell} \Lambda ) H_{\ell}
\left ( \alpha_{\ell} - \bar{\alpha}_{\ell} \right ) H_0 \left ( \alpha_{\ell} -
\bar{\alpha}_{\ell} \right ) \nn \\
\eeeq
\beeq
\kappa = \exp \left [ - dL/(2I) - (1 - dL/(2I)) \beta \right ] \ \ \ .
\label{5.14b}
\emini

Using (\ref{2.16}), with the factor $H_i (\alpha_i - \bar{\alpha}_i) H_0(\alpha_i -
\bar{\alpha}_i)$ included (taking into account the corrections evaluated in section 4 and in A of
this section) we see that (\ref{5.14a}) would then simply imply two independent consistency
equations for $\alpha_i$ and $\alpha_{\ell}$. So we can say that at the ``leading order'' there is
a decoupling of $\alpha_i$ and $\alpha_{\ell}$. The coupling of $\alpha_i$ and $\alpha_{\ell}$
which arises when $\alpha_i \alpha_{\ell}$ terms appear are only visible when we go to the next
order, i.e. when we take into account terms of order $1/L$. This $(\alpha_i, \alpha_{\ell})$
coupling only at the next-to-leading order is of course welcome because it means that the
derivative $d \bar{\alpha}_{\ell}(\alpha_i)/d\alpha_i$ is of order $1/L$, so that powers
$\bar{\alpha}_{\ell}^L$ will only give rise to a finite correction factor. Then, to leading order,
because of $d\bar{\alpha}_j (\alpha_i)/d \alpha_i = 0$ we have $H_0(\alpha_i - \bar{\alpha}_i)
= 1$. Next-to-leading order will give $d\bar{\alpha}_j(\alpha_i)$ of order $1/L$, i.e.
$d\bar{\alpha}_-/\bar{\alpha}_-$ and $d \bar{\alpha}_+/\bar{\alpha}_+$ in (\ref{5.6}) will be
$O(1)$ and $H_0(\alpha_i - \bar{\alpha}_i)$ will contribute to a ``renormalization'' of the scale
$\Lambda$. \par

Let us now examine the source of these $\alpha_i \alpha_{\ell}$ terms. We recall that the
$\delta$-function. $\delta (h_0 - \sum\limits_i \alpha_i)$ implies a replacement of
$\bar{\alpha}_j$ by $\bar{\alpha}_j [1 - (\alpha_i + \alpha_{\ell})/h_0]$ in $(b_{i2})^{-d/2}$ and as
Log $(1 - x) = - x - x^2/2$ we get the replacements

\bminiG{EDh}
\label{5.15a}
&&(b_{i2})^{-d/2} \to (b_{i2})^{-d/2} \exp \left [ d/2(L - 2) (\alpha_i + \alpha_{\ell})/h_0 \right ]
\exp \left [ d/2 (L - 2) \alpha_i \alpha_{\ell}/h_0^2 \right ] \nn \\
\eeeq
\beeq
h_0^{I-3} \to h_0^{I-3} \exp \left [ - (I - 3) (\alpha_i + \alpha_{\ell})/h_0 \right ] \exp \left
[ - (I-3) \alpha_i \alpha_{\ell}/h_0^2 \right ]  \label{5.15b}
\eeeq
\beeq
&&\left \{ 1 + \left [ Q_G (P, \{ \bar{\alpha}\})/h_0m^2 \right ] \left ( 1 - (\alpha_i +
\alpha_{\ell})/h_0 \right ) \right \}^{-(I -d/2\ L)} \to \nn \\
&&\left [ 1 + Q_G(P, \{ \bar{\alpha} \})/h_0m^2 \right ]^{-(I - d/2\ L)} \exp \left [ (I - d/2\ L )
\beta (\alpha_i + \alpha_{\ell})/h_0 \right ]
\nn \\
&&\exp \left [ (I - d/2 \ L ) \beta^2 \alpha_i \alpha_{\ell}/h_0^2 \right ] \ \ \ .
\label{5.15c}
\emini

\noi Also, because (remember that $\mu_i$ is of degree 1 in $\bar{\alpha}_j$)

\beq
\mu_i \left ( 1 - (\alpha_i + \alpha_{\ell})/h_0 \right ) + \alpha_i = (\mu_i + \alpha_i) \left \{ 1
- \left [ (\alpha_i + \alpha_{\ell} )/h_0 \right ] \mu_i/(\mu_i + \alpha_i ) \right \} \label{5.16}
\eeq

\noi we get the replacement

\[ \left [ \left ( \mu_i + \alpha_i \right ) \left ( \mu_{\ell} + \alpha_{\ell} \right ) \right
]^{-d/2} \to \]
\beq 
\left [ \left ( \mu_i + \alpha_i \right ) \left ( \mu_{\ell} + \alpha_{\ell} \right ) \right
]^{-d/2} \exp \left \{ d/2 \left [ (\alpha_i + \alpha_{\ell})/h_0 \right ] \left [ \mu_i/\left (
\mu_i + \alpha_i \right ) + \mu_{\ell}/ \left ( \mu_{\ell} + \alpha_{\ell} \right ) \right ]
\right \} \ .
\label{5.17}
\eeq

\noi The replacement $\bar{\alpha}_j \to \bar{\alpha}_j (1 - 2/I)$ implied because we take the
mean-value over $(I - 2)$ variables instead of $I$ provides us with the factor $\kappa^2$ in
(\ref{5.14a}). Gathering all terms containing $\alpha_i \alpha_{\ell}$ we get the factor

\[
\exp \left \{ - (I - dL/2)(1 - \beta^2 ) \alpha_i \alpha_{\ell}/h_0^2 \right .
\]
\beq
\left . + d/2 \left [ (\alpha_i/h_0) \mu_{\ell}/ \left ( \mu_{\ell} + \alpha_{\ell} \right ) + \left
( \alpha_{\ell}/h_0 \right ) \mu_i / \left ( \mu_i + \alpha_i \right ) \right ] \right \}
\ \ \ . \label{5.18} \eeq

\noi The terms linear in $\alpha_i$ or $\alpha_{\ell}$ give $\exp (- \alpha_{\ell} \Lambda )$ and
$\exp (- \alpha_i \Lambda )$ in (\ref{5.14a}). We will interpret the terms in the exponential in
(\ref{5.17}) as additional contributions to $\alpha_{\ell} \Lambda$ and $\alpha_i \Lambda$, that we
define as $\alpha_{\ell} \Lambda_{\ell}(\alpha_i)$ and $\alpha_i \Lambda_i (\alpha_{\ell})$ such
that~:

\[
\alpha_{\ell} \Lambda_{\ell} (\alpha_i) + \alpha_i \Lambda_i (\alpha_{\ell}) = (I - d/2\ L) (1 -
\beta ^2) \alpha_i \alpha_{\ell}/h_0^2 \]
\beq
- d/2 \left [ \left ( \alpha_{\ell}/h_0 \right ) \mu_i/\left ( \mu_i + \alpha_i \right ) + \left (
\alpha_i/h_0 \right ) \mu_{\ell}/\left ( \mu_{\ell} + \alpha_{\ell} \right ) \right ] \
\ \ . \label{5.19} \eeq

We therefore get a ``renormalization'' of the scale $\Lambda$ which is dependent on $\alpha_i$, the
integration variable. We note that another contribution to this renormalization comes from
$H_0(\alpha_i - \bar{\alpha}_i)$. To the extent where $E_0(\alpha_i - \bar{\alpha}_i)$ can be
linearized, {\it which happens to be justified by the fact that $\alpha_i - \bar{\alpha}_i$ is
infinitesimal}, writing

\beq
E_0(\alpha_i - \bar{\alpha}_i) = E_i^0 \alpha_i - E_i^0 \bar{\alpha}_i \ \ \ , 
\label{5.20}
\eeq

\noi $E^0_i$ being some constant, we can introduce a renormalized $\Lambda$,

\beq
\Lambda_i^R = \Lambda + d/2 \ L \ E_i^0 + \Lambda_i (\alpha_{\ell}) \ \ \ .
\label{5.21}
\eeq

\noi Then, defining

\beq
\widetilde{H}_0(\bar{\alpha}_i) \equiv \exp \left ( d/2 \ L \ E_i^0 \ \bar{\alpha}_i \right ) \ \ \ ,
\label{5.22}
\eeq

\noi the consistency equation for $\bar{\alpha}_i$ now reads

\beq
1 = (I/h_0) \kappa \int_0^{h_0} d \alpha_i \left [ \left ( \mu_i + \bar{\alpha}_i \right )/\left (
\mu_i + \alpha_i \right ) \right ]^{d/2} \exp \left (- \alpha_i \Lambda_i^R \right ) H_i \left (
\alpha_i - \bar{\alpha}_i \right ) \widetilde{H}_0(\bar{\alpha}_i) \ \ . \label{5.23} 
\eeq

And, of course, the same equation holds for $\bar{\alpha}_{\ell}$ interchanging $i$ and $\ell$. Our
following task will be the determination of $\Lambda_i(\alpha_{\ell})$ and $E_i^0$. This is done in
the next section.

\mysection{Determination of $\bar{\alpha}_{\ell}(\alpha_i)$} 

\hspace*{\parindent} In order to get $\bar{\alpha}_{\ell}(\alpha_i)$ we have several steps to
accomplish. First, we have to write down an equation relating the variation of
$\bar{\alpha}_{\ell}(\alpha_i)$ with that of $\Lambda_{\ell}^R(\alpha_i)$ as $\alpha_i$
varies. We use the consistency equation for $\bar{\alpha}_{\ell}$ under the form

\bminiG{EDh}
&&\left [ \left ( \mu_{\ell} + \bar{\alpha}_{\ell} \right ) \Lambda_{\ell}^R \right ]^{-d/2} =
(I/h_0) \Lambda_{\ell}^{R^{-1}} \kappa \cdot 
\nn \\
&&\int_0^{\infty} dx \left ( \mu_{\ell} \Lambda_{\ell}^R + x \right )^{-d/2} \exp (-x) J_{\ell}^{-1}
H_{\ell} \left ( \alpha_{\ell} - \bar{\alpha}_{\ell} \right ) \widetilde{H}_0 (\bar{\alpha}_{\ell})
\label{6.1a} \eeeq
\noi with
\beeq
J_{\ell} = 1 + \left ( \alpha_{\ell}/ \Lambda_{\ell}^R \right ) \partial \Lambda_{\ell}^R/\partial
\alpha_{\ell} \label{6.1b}
\emini

\noi and obtain a relation between $\delta \bar{\alpha}_{\ell}$, $\delta \mu_{\ell}$ and $\delta
\Lambda_{\ell}^R$ by differentiating (\ref{6.1a}). This will give us the two coefficients $A_{\ell}$
defined by

\beq
\delta \bar{\alpha}_{\ell}/\bar{\alpha}_{\ell} \equiv A_{\ell} \ \delta
\Lambda_{\ell}^R/\Lambda_{\ell}^R \ \ \ .  \label{6.2}
\eeq

\noi Secondo, we determine $\Lambda_{\ell}(\alpha_i)$ with the help of (\ref{5.19}). $\delta
\Lambda_{\ell}^R$ will then be expressed as a function of $\delta \bar{\alpha}_+$ and $\delta
\bar{\alpha}_-$ using (\ref{5.21}) and (\ref{5.6}). Then,
(\ref{6.2}) will give a system of equations for $d\bar{\alpha}_-/d\alpha_{\ell}$ and
$d\bar{\alpha}_+/d\alpha_{\ell}$. Finally, this system will be solved and
$\bar{\alpha}_-(\alpha_{\ell})$ and $\bar{\alpha}_+(\alpha_{\ell})$ determined as well as
$H_0(\bar{\alpha}_{\ell})$, $\alpha_{\ell}$ being either $\alpha_{\ell_+}$ or $\alpha_{\ell_-}$,
i.e. a side- or center-variable ($\bar{\alpha}_{\ell} (\alpha_{i_+})$ and
$\bar{\alpha}_{\ell}(\alpha_{i_-})$ will be different functions). We start with the determination of
$A_{\ell}$. \\

\noi {\large \bf A - Determination of $A_{\ell}$} \\

Let us first define $z_{\ell}$, $\zeta_{\ell}$ and $\varepsilon_{\ell}$ such that 

 \bminiG{EDh}
\label{6.3a}
z_{\ell} = (2/d) \left ( \mu_{\ell} + \bar{\alpha}_{\ell} \right ) \Lambda_{\ell}^R \ \partial \ {\rm
Log} \ \left ( J_{\ell}^{-1} H_{\ell} \widetilde{H}_0 \right )/\partial (\bar{\alpha}_{\ell}
\Lambda_{\ell}^R)  \eeeq
\beeq
\label{6.3b}
\zeta_{\ell} = (2/d) \left ( \mu_{\ell} + \bar{\alpha}_{\ell} \right ) \Lambda_{\ell}^R \ \partial \
{\rm Log} \ \left ( J_{\ell}^{-1} H_{\ell} \widetilde{H}_0 \right )/\partial (\mu_{\ell}
\Lambda_{\ell}^R)  \eeeq

\beeq
&&\varepsilon_{\ell} = 1 - (I/h_0) \Lambda_{\ell}^{R^{-1}} \exp \left [ - dL/(2I) - (1 - dL/(2I))
\beta \right ] \cdot \nn \\
&&\left [ \left ( \mu_{\ell} + \bar{\alpha}_{\ell} \right ) \Lambda_{\ell}^R \right ]^{d/2+1}
\int_0^{\infty} dx \left ( \mu_{\ell} \Lambda_{\ell}^R + x \right )^{-d/2-1} \exp (-x)
J_{\ell}^{-1} H_{\ell} (\alpha_{\ell} - \bar{\alpha}_{\ell}) \widetilde{H}_0 (\bar{\alpha}_{\ell})\ .
\nn \\
  \label{6.3c} 
\emini

\noi We remind us that $\beta = t/2 \ \bar{\alpha}_+ \ \Lambda/m^2$ from (\ref{3.6b}) and $bs \to
0$. If (\ref{6.1a}) is written

\beq
\left [ \left ( \mu_{\ell} + \bar{\alpha}_{\ell} \right ) \Lambda_{\ell}^R \right ]^{-d/2} =
I_{\ell}
\label{6.4}
\eeq

\noi we obtain, taking variations,

\[ - d/2 \left [ \left ( \mu_{\ell} + \bar{\alpha}_{\ell} \right ) \Lambda_{\ell}^R \right ]^{-d/2
- 1} \delta \left [ \left ( \mu_{\ell} + \bar{\alpha}_{\ell} \right ) \Lambda_{\ell}^R \right ] =
- \left [ \left ( 1 - dL/2I) \right ) \beta \delta \bar{\alpha}_+/\bar{\alpha}_+ + \delta
\Lambda_{\ell}^R/\Lambda_{\ell}^R \right ] I_{\ell} \]
\[
+ d/2 \left [ z_{\ell} \delta \left ( \bar{\alpha}_{\ell} \Lambda_{\ell}^R \right ) + \zeta_{\ell}
\delta \left ( \mu_{\ell} \Lambda_{\ell}^R \right ) \right ] \left [ \left ( \mu_{\ell} +
\bar{\alpha}_{\ell} \right ) \Lambda_{\ell}^R \right ]^{-1} I_{\ell}
\]
\beq
- d/2 \ \delta \left (\mu_{\ell} \Lambda_{\ell}^R \right ) (1 - \varepsilon_{\ell}) \left [ \left (
\mu_{\ell} + \bar{\alpha}_{\ell} \right ) \Lambda_{\ell}^R \right ]^{-d/2-1}
 \label{6.5}
\eeq

\noi or replacing $dL/(2I)$ by $2/3$, $d/2$ by 2 and multiplying by $(2/d) [(\mu_{\ell} +
\bar{\alpha}_{\ell} ) \Lambda_{\ell}^R ]^{d/2 + 1}$,

\[ ( 1 + z_{\ell}) \delta \left ( \bar{\alpha}_{\ell} \Lambda_{\ell}^R \right ) + \left (
\varepsilon_{\ell} + \zeta_{\ell} \right ) \delta \left ( \mu_{\ell} \Lambda_{\ell}^R \right ) = 
\]
\beq
1/2 \left [ \left ( \mu_{\ell} + \bar{\alpha}_{\ell} \right ) \Lambda_{\ell}^R \right ] \left (
\beta /3 \ \delta \bar{\alpha}_+/\bar{\alpha}_+ + \delta \Lambda_{\ell}^R/\Lambda_{\ell}^R \right )
\label{6.6} \eeq

\noi or
\newpage

\[ (1 + z_{\ell}) \left [ \bar{\alpha}_{\ell} / \left ( \mu_{\ell} + \bar{\alpha}_{\ell} \right
)\right ] \delta \bar{\alpha}_{\ell}/\bar{\alpha}_{\ell} - \beta /6 \ \delta
\bar{\alpha}_+/\bar{\alpha}_+ + \left ( \varepsilon_{\ell} + \zeta_{\ell} \right ) \left [
\mu_{\ell}/ \left ( \mu_{\ell} + \bar{\alpha}_{\ell} \right ) \right ] \delta \mu_{\ell}/\mu_{\ell}
=  \]
\beq
\left \{ 1/2 - (1 + z_{\ell}) \left [ \bar{\alpha}_{\ell}/\left ( \mu_{\ell} +
\bar{\alpha}_{\ell} \right ) \right ] - \left ( \varepsilon_{\ell} + \zeta_{\ell} \right ) \left [
\mu_{\ell}/\left ( \mu_{\ell} + \bar{\alpha}_{\ell} \right ) \right ] \right \} \delta
\Lambda_{\ell}^R/\Lambda_{\ell}^R \ \ . \label{6.7} \eeq

\noi This provides us with two equations. Two more equations come from the relations (derived in
\cite{15r})

\bminiG{EDh}
\label{6.8a}
1 + \bar{\alpha}_-/\mu_- = y^{-1}
\eeeq
\beeq
1 + \mu_+/\bar{\alpha}_+ = 2 y^{-1}
\label{6.8b}
\emini

\noi and a fifth equation can be derived from $y = (2 \bar{\alpha}_+/\bar{\alpha}_-)^{1/2}$. So
that five equations arise for $\delta \bar{\alpha}_+$, $\delta \mu_+$, $\delta \bar{\alpha}_-$,
$\delta \mu_-$ and $\delta y$ which can be expressed as functions of $\delta \Lambda_+^R$ and
$\delta \Lambda_-^R$. The solution of this five equations system is easy to obtain. Defining $a$,
$b$, $c$, $d$, $\Delta_-$, $\Delta_+$ as 

\bminiG{EDh}
\label{6.9a}
a \equiv 1/2 \ \left ( \varepsilon_- + \zeta_- \right ) \left [ y/(1 - y) \right ] - \beta /6
\eeeq
\beeq
b \equiv (1 + z_-) (1 - y) + \left ( \varepsilon_- + \zeta_- \right ) y \left [ 1 - 1/2\ /(1 - y)
\right ] \label{6.9b}
\eeeq
\beeq
\label{6.9c}
c \equiv (1 + z_+) y/2 - \beta /6 + \left ( \varepsilon_+ + \zeta_+ \right ) (1 - y/2) \left [ 1 -
1/(2 - y) \right ] \eeeq
\beeq
\label{6.9d}
d \equiv \left ( \varepsilon_+ + \zeta_+ \right ) (1 - y/2)/(2 - y)
\eeeq
\beeq
\Delta_- \equiv \left [ 1/2 - (1 + z_-)(1 - y) - \left ( \varepsilon_- + \zeta_- \right ) y \right ]
\delta \Lambda_-^R/\Lambda_-^R \label{6.9e}
\eeeq
\beeq
\Delta_+ \equiv \left [ 1/2 - (1 + z_+) y/2 - \left ( \varepsilon_+ + \zeta_+ \right ) (1 - y/2)
\right ] \delta \Lambda_+^R/\Lambda_+^R \ \ \ , \label{6.9f}
\emini

\noi the result is
\bminiG{EDh}
A_+ = \left [ 1/2 - (1 + z_+) y/2 - \left ( \varepsilon_+ + \zeta_+ \right ) (1 - y/2) \right ]
\left [ d \left ( \Delta_-/\Delta_+ \right ) - b \right ] /(ad - bc) \nn \\
\label{6.10a}
\eeeq
\beeq
A_- = \left [ 1/2 - (1 + z_-)(1 - y) - \left ( \varepsilon_- + \zeta_- \right ) y \right ] \left [
a(\Delta_+/\Delta_-) - c \right ]/(ad -bc) \ \ \ .  \label{6.10b}
\emini

\noi The next step is the calculation of $\Lambda_{\ell}(\alpha_i)$. \\

\noi {\large \bf B - Determination of $\Lambda_{\ell}(\alpha_i)$} \\

Let us rewrite (\ref{5.19}) under the form 

\bminiG{EDh}
\label{6.11a}
\alpha_{\ell} \Lambda_{\ell}(\alpha_i) + \alpha_i \Lambda_i (\alpha_{\ell}) = F_{i \ell} (\alpha_i,
\alpha_{\ell}) \eeeq
\beeq
&&F_{i \ell} \equiv (I - d/2\ L)(1 - \beta^2) \alpha_i \alpha_{\ell}/h_0^2
\nn \\
&&- d/2 \left [ ( \alpha_{\ell}/h_0) \mu_i/\left ( \mu_i + \alpha_i \right ) + \left ( \alpha_i/h_0
\right ) \mu_{\ell}/\left ( \mu_{\ell} + \alpha_{\ell} \right ) \right ] \ \ . \label{6.11b}
\emini

\noi If $i$ and $\ell$ are both side-propagators or center-propagators, a symmetry exists
exchanging $i$ and $\ell$, i.e. $\Lambda_{\ell}$ and $\Lambda_i$ should be the
same function. In
particular taking $\alpha_i$ and $\alpha_{\ell}$ at their common mean-value $\bar{\alpha}_+$ or
$\bar{\alpha}_-$ we get

\bminiG{EDh}
\label{6.12a}
\Lambda_{\ell_+}(\bar{\alpha}_{i_+}) = \Lambda_{i_+}(\bar{\alpha}_{\ell_+})
\eeeq
\beeq
\Lambda_{\ell_-} (\bar{\alpha}_{i_-}) = \Lambda_{i_-}(\bar{\alpha}_{\ell_-})
\label{6.12b}
\eeeq
\noi with
\beeq
\bar{\alpha}_{i_+} = \bar{\alpha}_{\ell_+} = \bar{\alpha}_+ \quad ; \quad \bar{\alpha}_{i_-} =
\bar{\alpha}_{\ell_-} = \bar{\alpha}_- \ \ \ .  \label{6.12c}
\emini

\noi The second constraint we shall impose is that {\it we want the equation for
$\bar{\alpha}_{\ell}$ to be the same whatever $i$ is}, i.e. $i_+$ or $i_-$. This can be
approximately realized by demanding

\bminiG{EDh}
\Lambda_{\ell_+} (\bar{\alpha}_{i_+}) = \Lambda_{\ell_+}(\bar{\alpha}_{i_-})
\label{6.13a}
\eeeq
\beeq
\Lambda_{\ell_-}(\bar{\alpha}_{i_+}) = \Lambda_{\ell_-}(\bar{\alpha}_{i_-})
\label{6.13b}
\emini

\noi the same being also true exchanging $i$ and $\ell$. Of course, a priori, it could be that
$\bar{\alpha}_{\ell}$ would be the same using two different equations for it, but that would be some
sort of a miracle. So we feel safer imposing (6.13), which moreover will be easily implementable.
\par

Let us start by writing $\mu_{\ell}/(\mu_{\ell} + \alpha_{\ell})$ under the form

\beq
\mu_{\ell}/(\mu_{\ell} + \alpha_{\ell}) = \left ( \mu_{\ell}/(\mu_{\ell} + \bar{\alpha}_{\ell})
\right ) \left [ 1 - \left ( \alpha_{\ell} - \bar{\alpha}_{\ell} \right ) / \left ( \mu_{\ell} +
\alpha_{\ell} \right ) \right ] \ \ \ , \label{6.14} \eeq

\noi $\mu_i/(\mu_i + \alpha_i)$ being also treated in the same way. Plugging this into
(\ref{6.11b}) we get $h_0 F_{i \ell}$ under the form 

\[ h_0 F_{i \ell} = \Lambda (1 + \beta ) \alpha_i \alpha_{\ell} - (d/2) \left \{ \alpha_i
\left [ 1 - \alpha_{\ell}/\left ( \mu_{\ell} + \alpha_{\ell} \right ) \right ] \mu_{\ell}/\left (
\mu_{\ell} + \bar{\alpha}_{\ell} \right ) \right . \]
\[
+ \alpha_i \left [ \bar{\alpha}_{\ell}/\left ( \mu_{\ell} + \alpha_{\ell} \right ) \right ]
\mu_{\ell}/ \left ( \mu_{\ell} + \bar{\alpha}_{\ell} \right ) + \alpha_{\ell} \left [ 1 -
\alpha_i/\left ( \mu_i + \alpha_i \right ) \right ] \mu_i / \left ( \mu_i + \bar{\alpha}_i \right )
\]
\[
\left . + \alpha_{\ell} \left [ \bar{\alpha}_i/ \left ( \mu_i + \alpha_i \right ) \right ]
\mu_i/\left ( \mu_i + \bar{\alpha}_i \right ) \right \}
\]
\[
= \alpha_i \alpha_{\ell} \left \{ \Lambda (1 + \beta ) + d/2 \left [ \left ( \mu_{\ell} +
\alpha_{\ell} \right )^{-1} \mu_{\ell}/ \left ( \mu_{\ell} + \bar{\alpha}_{\ell} \right ) + \left (
\mu_i + \alpha_i \right )^{-1} \mu_i/ \left ( \mu_i + \bar{\alpha}_i \right ) \right ] \right .
\]
\[ - \alpha_i(d/2) \left [ 1 + \bar{\alpha}_{\ell}/\left ( \mu_{\ell} + \alpha_{\ell} \right )
\right ] \mu_{\ell}/\left ( \mu_{\ell} + \bar{\alpha}_{\ell} \right )
\]
\beq
- \alpha_{\ell} (d/2) \left [ 1 + \bar{\alpha}_i/\left ( \mu_i + \alpha_i \right ) \right ]
\mu_i/\left ( \mu_i + \bar{\alpha}_i \right )
\label{6.15}
\eeq

\noi which has the advantage that in every term a factor $\alpha_i$ or $\alpha_{\ell}$ exists and
that every factor $(\mu_{\ell} + \alpha_{\ell})^{-1}$ is multiplied by $\mu_{\ell}/\mu_{\ell} +
\bar{\alpha}_{\ell}$ which is less than one, thus minimizing the $\alpha_{\ell}$ variation of such
a term (the same being of course true for $(\mu_i + \alpha_i)^{-1})$. \par

Using (\ref{6.11a}) we are now able to write

\bminiG{EDh}
&& h_0 \Lambda_i(\alpha_{\ell}) = x_{i\ell} \Lambda (1 + \beta ) \alpha_{\ell} + a_{i\ell}(d/2)
\left [ \alpha_{\ell}/(\mu_{\ell} + \alpha_{\ell} ) \right ] \mu_{\ell}/(\mu_{\ell} +
\bar{\alpha}_{\ell}) \nn \\
&&+ b_{i \ell} (d/2) \left [ \alpha_{\ell}/(\mu_i + \alpha_i) \right ] \mu_i/(\mu_i +
\bar{\alpha}_i) \nn \\ 
&&- (d/2) \left [ 1 + \bar{\alpha}_{\ell}/(\mu_{\ell} + \alpha_{\ell}) \right ]
\mu_{\ell}/(\mu_{\ell} + \bar{\alpha}_{\ell}) \label{6.16a}
\eeeq
\beeq
&& h_0 \Lambda_{\ell}(\alpha_i) = (1 - x_{i\ell}) \Lambda (1 + \beta) \alpha_i + (1 - a_{i \ell})
(d/2) \left [ \alpha_i/(\mu_{\ell} + \alpha_{\ell}) \right ] \mu_{\ell}/(\mu_{\ell} +
\bar{\alpha}_{\ell}) \nn \\
&&+ (1 - b_{i \ell}) (d/2) \left [ \alpha_i /(\mu_i + \alpha_i) \right ] \mu_i/(\mu_i +
\bar{\alpha}_i) \nn \\ 
&&- (d/2) \left [ 1 + \bar{\alpha}_i/(\mu_i + \alpha_i) \right ] \mu_i/(\mu_i + \bar{\alpha}_i) \ \ \
.  \label{6.16b} \emini

\noi If $i$ and $\ell$ are both $+$ or $-$, the $i \leftrightarrow \ell$ symmetry imposes

\bminiG{EDh}
1 - x_{i \ell} = x_{i \ell}
\label{6.17a}
\eeeq
\beeq
a_{i \ell} = 1 - b_{i\ell}
\label{6.17b}
\eeeq
\beeq
b_{i \ell} = 1 - a_{i\ell}
\label{6.17c}
\emini

\noi which leads to (we recall that $\mu_-/(\mu_- + \bar{\alpha}_-) = y$, $\mu_+/(\mu_+ +
\bar{\alpha}_+) = 1 - y/2$, $y = (2 \bar{\alpha}_+ \bar{\alpha}_-)^{1/2}$) 

\bminiG{EDh}
&&h_0 \Lambda_{\ell_+}(\alpha_{i_+}) = 1/2 \Lambda (1 + \beta) \alpha_{i_+} + (d/2) \left [ a_+
\alpha_{i_+}/(\mu_+ + \alpha_{\ell_+}) \right . \nn \\ 
&&\left . + b_+ \alpha_{i_+}/(\mu_+ + \alpha_{i_+}) - (1 + \bar{\alpha}_+/(\mu_+ + \alpha_{i_+}))
\right ] (1 - y/2) \label{6.18a}
\eeeq
\beeq
&& h_0 \Lambda_{i_+}(\alpha_{\ell_+}) = 1/2 \ \Lambda (1 + \beta ) \alpha_{\ell_+} + (d/2) \left [ (1
- a_+) \alpha_{\ell_+}/(\mu_+ + \alpha_{\ell_+}) \right . \nn \\
&&\left . + (1 - b_+) \alpha_{\ell_+}/(\mu_+ + \alpha_{i_+}) - (1 + \bar{\alpha}_+/(\mu_+ +
\alpha_{\ell_+})) \right ] (1 - y/2) \label{6.18b}
\emini

\noi for the $\ell = i = +$ case and

\bminiG{EDh}
&&h_0 \Lambda_{\ell_-}(\alpha_{i_-}) = 1/2 \ \Lambda (1 + \beta ) \alpha_{i_-} + (d/2) \left [ a_-
\alpha_{i_-}/(\mu_- + \alpha_{\ell_-}) \right . \nn \\
&&\left . + b_- \alpha_{i_-}/(\mu_- + \alpha_{i_-}) - (1 + \bar{\alpha}_-/(\mu_- + \alpha_{i_-}))
\right ] y \label{6.19a}
\eeeq
\beeq
&& h_0 \Lambda_{i_-}(\alpha_{\ell_-}) = 1/2 \ \Lambda (1 + \beta ) \alpha_{\ell_-} + (d/2) \left [ (1
- a_-) \alpha_{\ell_-}/(\mu_- + \alpha_{\ell_-}) \right . \nn \\
&&\left . + (1 - b_-) \alpha_{\ell_-}/(\mu_- + \alpha_{i_-}) - (1 + \bar{\alpha}_-/(\mu_- +
\alpha_{\ell_-})) \right ] y \label{6.19b}
\emini

\noi for the $\ell = i = -$ case. We have taken the notation $a_+ = 1 - a_{++}$, $b_+ = 1 -
b_{++}$, $a_- = 1 - a_{--}$, $b_- = 1 - b_{--}$. Then, (\ref{6.17b}) and (\ref{6.17c}) give 

 \bminiG{EDh}
a_+ + b_+ = 1
\label{6.20a}
\eeeq
\beeq
a_- + b_- = 1 \ \ \ .
\label{6.20b}
\emini

\noindent For $\ell = +$ and $i = -$, we get

\bminiG{EDh} 
&&h_0 \Lambda_{\ell_+}(\alpha_{i_-}) = x \Lambda (1 + \beta ) \alpha_{i_-} + (d/2) \left [ u (1 -
y/2) \alpha_{i_-} /(\mu_+ + \alpha_{\ell_+}) \right . \nn \\ 
&&\left . + z \ y \ \alpha_{i_-}/(\mu_- + \alpha_{i_-}) - y (1 + \bar{\alpha}_-/(\mu_- +
\alpha_{i_-})) \right ] 
\label{6.21a}
\eeeq
\beeq
&& h_0 \Lambda_{i_-}(\alpha_{\ell_+}) = (1 - x) \Lambda (1 + \beta ) \alpha_{\ell_+} + (d/2) \left
[ (1 - u) (1 - y/2) \alpha_{\ell_+}/(\mu_+ + \alpha_{\ell_+}) \right . \nn \\
&&\left . + (1 - z) y \ \alpha_{\ell_+}/(\mu_- + \alpha_{i_-}) - (1 - y/2) (1 + \bar{\alpha}_+ /
(\mu_+ + \alpha_{\ell_+})) \right ]  \label{6.21b}
\emini

\noi and also the same equations exchanging $i$ and $\ell$ for $\ell = -$ and $i = +$. Here we have
taken the notations $x = 1 - x_{-+} = x_{+-}$, $u = 1 - a_{-+} = a_{+-}$, $z = 1 -  b_{-+} =
b_{+-}$. \par

Now, if we want to satisfy the constraint (\ref{6.13a}), $\Lambda_{\ell_+} (\bar{\alpha}_{i_+}) =
\Lambda_{\ell_+}(\bar{\alpha}_{i_-})$, we have

\bminiG{EDh}
&&1/2 \ \Lambda (1 + \beta ) \bar{\alpha}_{i_+} +
(d/2) (1 - y/2) \left [ a_+ \bar{\alpha}_{i_+} /(\mu_+ + \alpha_{\ell_+}) + b_+ y/2  - (1 + y/2)
\right ] \nn \\
&&= x \Lambda (1 + \beta ) \bar{\alpha}_{i_-} + (d/2) \left [ u(1 - y/2) \bar{\alpha}_{i_-}/(\mu_+
+ \alpha_{\ell_+}) + z \  y(1 - y) \right . \nn \\
&&\left . - y(1 + 1 - y) \right ] 
\label{6.22a}
\eeeq
\noi and for the constraint (\ref{6.13b}), $\Lambda_{\ell_-}(\bar{\alpha}_{i_+}) = \Lambda_{\ell_-}
(\bar{\alpha}_{i_-})$
\beeq
&& 1/2 \ \Lambda (1 + \beta ) \bar{\alpha}_{i_-} + (d/2) y \left [ a_- \bar{\alpha}_{i_-}/(\mu_- +
\alpha_{\ell_-}) + b_-(1 - y) - (1 + 1 - y )\right ] \nn \\
&& = (1 - x) \Lambda (1 + \beta ) \bar{\alpha}_{i_+} + (d/2) \left [ (1 - u) (1 - y/2) y/2 + (1 -
z) y \bar{\alpha}_{i_+}/(\mu_- + \alpha_{\ell_-}) \right . \nn \\
&&\left . - (1 - y/2) (1 + y/2) \right ] \ \ \ .
\label{6.22b}
\emini

\noi The effect of integrating over $\alpha_{\ell_+}$ or $\alpha_{\ell_-}$ will be taken into
account by replacing $(\mu_+ + \alpha_{\ell_+})^{-1}$ by $(1 - x_+) (\mu_+ +
\bar{\alpha}_{\ell_+})^{-1}$ and $(\mu_- + \alpha_{\ell_-})^{-1}$ by $(1 - x_-) (\mu_- +
\bar{\alpha}_{e_-})^{-1}$, $x_+$ and $x_-$ being two parameters to be determined. Then, the
constraints (\ref{6.22a}) and (\ref{6.22b}) read

\bminiG{EDh}
&& (2/d) (x - 1/2 \ \bar{\alpha}_+/\bar{\alpha}_-) \bar{\alpha}_- \Lambda (1 + \beta ) + (1 - y/2) 
(1 + a_+ x_+ y/2) \nn \\
&&+ u (\bar{\alpha}_-/ \bar{\alpha}_+) (1 - x_+) (1 - y/2) y/2 - zy - (1 - z) (2 - y) = 0
\label{6.23a}
\eeeq
\beeq
&&(2/d) \left [ (1 - x) \bar{\alpha}_+/\bar{\alpha}_- - 1/2 \right ] \bar{\alpha}_- \Lambda (1 +
\beta) + y \left [ 1 + a_- x_- (1 - y) \right ] \nn \\
&&+ (1 - z) (\bar{\alpha}_+/\bar{\alpha}_-) (1 - x_-) y(1 - y) - (1 - u) (1 - y/2) - u(1 - y/2)(1 +
y/2) = 0 \ \ . \nn \\
\label{6.23b}
\emini 

\noi Solving the constraints we can express $u$ and $z$ as a function of $x$, $a_+$, $a_-$, $x_+$
and $x_-$. In practice, we keep the constraints under the form (6.23) because, then, numerical
calculations are much more stable. Having explicit expressions for $\Lambda_{\ell}(\alpha_i)$ we
can proceed further and calculate $\delta \bar{\alpha}_{\ell}(\alpha_i)$. \\

\noi {\large \bf C - Solving equations for $\delta \bar{\alpha}_{\ell} (\alpha_i)$} \\

Let us recall that we were looking for the variation of $\Lambda_{\ell}^R$ given by 

\beq
\Lambda_{\ell}^R = \Lambda + d/2 \ L \ E_{\ell}^0 + \Lambda_{\ell} (\alpha_i)
\label{6.24}
\eeq

\noi which is merely rewriting (\ref{5.21}) with an interchange of $i$ and $\ell$. Let us now, as a
preliminary task, evaluate the variation of $\Lambda$, the unrenormalized scale, with respect to
$\bar{\alpha}_+(\alpha_i)$. (\ref{2.17}), (\ref{3.6b}), (\ref{3.6c}) (and the limit $bs \to 0$) give

\bea
\label{6.25}
d\Lambda /d\alpha_i &=& - \left [ (I - dL/2)/h_0 \right ] d\beta /d \alpha_i \nn \\
&=& - \left [ (I - dL/2)/h_0 \right ] \left ( \Lambda d \bar{\alpha}_+/d\alpha_i + \bar{\alpha}_+ d
\Lambda/d \alpha_i \right ) t/(2m^2)
\eea

\noi which leads to

\beq
\Lambda^{-1} d \Lambda / d \alpha_i = - \beta \ \bar{\alpha}_+^{-1} \ d \bar{\alpha}_+/d\alpha_i \
\ \ . \label{6.26}
\eeq

\noi Defining the variable $x_{\ell i}$ through

\beq
\label{6.27}
x_{\ell i} \equiv \bar{\alpha}_{\ell}^{-1} \ d\bar{\alpha}_{\ell}/d\alpha_i
\eeq

\noi we get (see (\ref{6.2}))

\beq
\label{6.28}
x_{\ell i} = A_{\ell i} \ \Lambda_{\ell}^{R^{-1}} \left ( d \Lambda /d \alpha_i + (d/2)L \
dE_{\ell}^0/d \alpha_i + d \Lambda_{\ell} (\alpha_i)/d \alpha_i \right ) \eeq

\noi where $A_{\ell i}$ is $A_{\ell}(\alpha_i)$, the dependence on $\alpha_i$ coming from the
dependence of $\Delta_-$ and $\Delta_+$ on $\delta \Lambda_-^R/\Lambda_-^R$ and $\delta
\Lambda_+^R/\Lambda_+^R$ respectively (see (\ref{6.9e}) and (\ref{6.9f})). Defining ($c_{\ell i}$
and $d_{\ell i}$ being read on (6.18), (6.19) and (6.21))

\bminiG{EDh} 
\label{6.29a}
c_{\ell i} + d_{\ell i}/ \left [ (\mu_i + \alpha_i ) L \right ] \equiv \left ( \Lambda_{\ell}^R
\right )^{-1} d \Lambda_{\ell} (\alpha_i )/d \alpha_i  \eeeq 
\beeq
z_{R_{\ell}} \equiv \Lambda / \Lambda_{\ell}^R
\label{6.29b}
\emini

\noi and using (\ref{5.6}), (\ref{5.20}) through

\beq
E_{\ell}^0 = \left [ (1 - y/2) \bar{\alpha}_-^{-1} d\bar{\alpha}_-/d \alpha_{\ell} + (y/2 + \eta )
\bar{\alpha}_+^{-1}  d\bar{\alpha}_+/d\alpha_{\ell} \right ] \ \ \ , \label{6.30}
\eeq

\noi we get from (\ref{6.28}) the system of equations

\bminiG{EDh}
&&\left ( \beta z_{R_+} + A_{++}^{-1} \right ) x_{++} = c_{++} + d_{++}/ \left [ (\mu_+ +
\alpha_+)L \right ] \nn \\
&&+ d/2 \ (L/\Lambda_+^R) d/d \alpha_+ \left [ (1 - y/2) x_{-+} + (y/2 + \eta ) x_{++} \right ] 
\label{6.31a}  
\eeeq 
\beeq
&&\left ( \beta z_{R_+} + A_{+-}^{-1} \right ) x_{+-} = c_{+-} + d_{+-}/ \left [ (\mu_- +
\alpha_-)L \right ] \nn \\
&&+ d/2 \ (L/\Lambda_+^R) d/d \alpha_- \left [ (1 - y/2) x_{-+} + (y/2 + \eta ) x_{++} \right ]
\label{6.31b}
\eeeq
\beeq
&&A_{-+}^{-1} \ x_{-+} + \beta z_{R_-} x_{++} = c_{-+} + d_{-+}/ \left [ (\mu_+ + \alpha_+ )L \right
] \nn \\
 &&+ d/2 \ (L/\Lambda_-^R) d/d \alpha_+ \left [ (1 - y/2) x_{--} + (y/2 + \eta ) x_{+-} \right
] \label{6.31c}
\eeeq
\beeq
&&A_{--}^{-1} \ x_{--} + \beta z_{R_-} x_{+-} = c_{--} + d_{--}/ \left [ (\mu_- + \alpha_- )L \right
] \nn \\
 &&+ d/2 \ (L/\Lambda_-^R) d/d \alpha_- \left [ (1 - y/2) x_{--} + (y/2 + \eta ) x_{+-} \right
] \ \ .
\label{6.31d}
\emini

\noi We will now write $x_{\ell i}$ as a power series in $[(\mu_i + \alpha_i)L]^{-1}$, coefficients
depending only on $y$. This is because all dimensionless quantities may be written as a function of
$y$ and $(\mu_i + \alpha_i)L$. Limiting ourselves to the constant term and the first power in
$[(\mu_i + \alpha_i )L]^{-1}$ (which will be sufficient as we show next) we write

\bminiG{EDh} 
\label{6.32a}
x_{\pm \ell} = a_{\pm \ell}^0 + a_{\pm \ell}/\left [ ( \mu_+ + \alpha_+ )L \right ]
\eeeq
\beeq
x_{\pm \ell} = b_{\pm \ell}^0 + b_{\pm \ell}/\left [ ( \mu_- + \alpha_- )L \right ]
\label{6.32b}
\emini

\noi where the first expression occurs when $\alpha_i$ is a $+$ variable and the second when
$\alpha_i$ is a $-$ variable. We also have two useful relations stemming from $y = (2
\bar{\alpha}_+/\bar{\alpha}_-)^{1/2}$

\bminiG{EDh} 
\label{6.33a}
dy/d\alpha_+ = \left ( x_{++} - x_{-+} \right ) y/2
\eeeq
\beeq
dy/d \alpha_- = \left ( x_{+-} - x_{--} \right ) y/2 \ \ \ .
\label{6.33b}
\emini

\noi We now make the observation that 

\beq
d/d\alpha_i \left [ (\mu_i + \alpha_i)L \right ]^{-1} = - L \left [ (\mu_i + \alpha_i ) L \right
]^{-2} \label{6.34}
\eeq

\noi diverges as $L \to \infty$. Therefore, the coefficents of diverging terms should be $O(1/L)$
in order to cancel the divergence. This entails the constraints

\bminiG{EDh} 
\label{6.35a}
(1 - y/2) a_{- \ell} + (y/2 + \eta ) a_{+ \ell} = O(1/L)
\eeeq
\beeq
(1 - y/2) b_{-\ell} + (y/2 + \eta ) b_{+ \ell} = O(1/L)
\label{6.35b}
\emini

\noi easily derived from (6.31). \par

Let us take the derivative in (\ref{6.31a})

\[ d/d\alpha_+ \left [ (1 - y/2) x_{-+} + (y/2 + \eta ) x_{++} \right ] = \left ( x_{++} - x_{-+}
\right ) y/2 \cdot \] 
\[ 
\left \{ \left [ (\mu_+ + \alpha_+) L \right ]^{-1} d/dy \left [ (1 - y/2))a_{-+} + (y/2 +
\eta ) a_{++} \right ]\right .\] 
\beq
\left . + d/dy \left [ (1 - y/2) a_{-+}^0 + (y/2 + \eta ) a_{++}^0 \right ] \right \} + \hbox{h.o.}
\label{6.36}
\eeq

\noi where h.o. means higher order terms $\sim [L(\mu_+ + \alpha_+)]^{-n}$, $n > 1$. Now, due to
the constraints (6.35), we see that the first term in the bracket of the right-hand side of
(\ref{6.36}) vanishes. This property also holds for higher-order coefficients because of
constraints similar to (6.35) acting for higher-orders as it is easy to verify. Keeping
track of powers of $[L(\mu_+ + \alpha_+)]^{-1}$ equal to zero and one, one gets the following
constraints obtained by using (\ref{6.36}) in (6.31), identifying coefficients for a
given power of $[L(\mu_+ + \alpha_+)]^{-1}$ and introducing $E_{\ell i}^0 \equiv
E_{\ell}^0(\alpha_i )$,

\bminiG{EDh} \label{6.37a}
\left ( \beta z_{R_+} + A_{++}^{-1} \right ) a_{++}^0 = c_{++} + \left [ dy L/(4 \Lambda_+^R)
\right ] \left ( a_{++}^0 - a_{-+}^0 \right ) dE_{++}^0/dy \eeeq 
\beeq
\left ( \beta z_{R_+} + A_{+-}^{-1} \right ) b_{+-}^0 = c_{+-} + \left [ dy L/(4 \Lambda_+^R)
\right ] \left ( b_{+-}^0 - b_{--}^0 \right ) dE_{+-}^0/dy
\label{6.37b}
\eeeq
\beeq
A_{-+}^{-1} a_{-+}^0 + \beta z_{R_-} a_{++}^0 = c_{-+} + \left [ dy L/(4 \Lambda_-^R ) \right ]
\left ( a_{++}^0 - a_{-+}^0 \right ) dE_{-+}^0/dy \label{6.37c}
\eeeq
\beeq
A_{--}^{-1} b_{--}^0 + \beta z_{R_-} b_{+-}^0 = c_{--} + \left [ dy L/(4 \Lambda_-^R ) \right ]
\left ( b_{+-}^0 - b_{--}^0 \right ) dE_{--}^0/dy
\label{6.37d}
\emini

\noi i.e. four equations determining $a_{++}^0$, $a_{-+}^0$, $b_{+-}^0$, $b_{--}^0$ with
 
\bminiG{EDh} \label{6.38a}
E_{++}^0 = (1 - y/2) a_{-+}^0 + (y/2 + \eta ) a_{++}^0 
\eeeq 
\beeq
\label{6.38b}
E_{+-}^0 = (1 - y/2) b_{-+}^0 + (y/2 + \eta ) b_{++}^0 
\eeeq
\beeq
E_{-+}^0 = (1 - y/2) a_{--}^0 + (y/2 + \eta ) a_{+-}^0 
\label{6.38c}
\eeeq
\beeq
\label{6.38d}
E_{--}^0 = (1 - y/2) b_{--}^0 + (y/2 + \eta ) b_{+-}^0 \ \ \ .
\emini

\noi $b_{-+}^0$, $b_{++}^0$, $a_{--}^0$, $a_{+-}^0$ are undetermined but this is welcome because
one should have   

\bminiG{EDh} 
\label{6.39a}
E_{++}^0 = E_{+-}^0 = E_+^0
\eeeq
\beeq
E_{--}^0 = E_{-+}^0 = E_-^0
\label{6.39b}
\emini

\noi i.e. $E_{\ell}^0$ {\it should be independent of $\alpha_i$}. ($\Lambda_{\ell}^R$ does not
depend on any other $\alpha_i$ when all $\alpha_i$'s, $i \not= \ell$, are integrated over). 
Therefore, $E_+^0$ and $E_-^0$ are given by (\ref{6.38a}) and (\ref{6.38d}) respectively, which
express $E_+^0$ and $E_-^0$ as a function of $(a_{-+}^0$, $a_{++}^0$) and ($b_{--}^0$, $b_{+-}^0$)
respectively. One remarkable thing is that {\it only zero-order coefficients enter in the expression
of $E_{\ell}^0$} because it is proportional to the linear combination $(1 - y/2) x_{- \ell} + (y/2 +
\eta ) x_{+ \ell}$. The knowledge of higher order coefficients is then unnecessary for our purpose.
However, a constraint is visible, looking at (\ref{6.31b}) and (\ref{6.31c}), which demands that the
order of derivation should not matter, namely

\[
d/d\alpha_- \left [(1 - y/2) x_{-+} + (y/2 + \eta ) x_{++} \right ] = \]
\beq
\label{6.40}
d/d\alpha_+ \left [ (1 - y/2) x_{--} + (y/2 + \eta ) x_{+-} \right ]
\eeq

\noi or
\newpage
\[ \Lambda_+^R \left \{ \left ( \beta z_{R_+} + A^{-1}_{+-} \right ) \left ( b_{+-}^0 +
b_{+-}/\left [ (\mu_- + \alpha_- )L\right ] \right ) - c_{+-} - d_{+-}/\left [ (\mu_- + \alpha_-
)L\right ] \right \}
\]
\[ = \Lambda_-^R \left \{ \left ( A_{-+}^{-1} a_{-+}^0 + \beta z_{R_-} a_{++}^0 \right ) + \left (
A_{-+}^{-1} a_{-+} + \beta z_{R_-} a_{++} \right ) / \left [ (\mu_+ + \alpha_+)L \right ] \right .
\]
\beq
\label{6.41}
\left . - c_{-+} - d_{-+}/\left [ (\mu_- + \alpha_- )L \right ] \right \} \ \ \ . 
\eeq

\noi Identifying to zero the coefficients of $[(\mu_- + \alpha_-)L]^{-1}$ and $[(\mu_+ + \alpha_+
)L ]^{-1}$ and ba\-lan\-cing constant coefficients we get

\bminiG{EDh} \label{6.42a}
&&\Lambda_+^R \left [ \left ( \beta z_{R_+} + A_{+-}^{-1} \right ) b_{+-}^0 - c_{+-} \right ] \nn
\\
&&= \Lambda_-^R \left [ \left ( \beta z_{R_-} a_{++}^0 + A_{-+}^{-1} a_{-+}^0 \right ) - c_{-+}
\right ]  \eeeq  \beeq
\label{6.42b}
\left ( \beta z_{R_+} + A_{+-}^{-1} \right ) b_{+-} = d_{+-}
\eeeq
\beeq
\label{6.42c}
\beta z_{R_-} a_{++} + A_{-+}^{-1} a_{-+} = d_{-+} \ \ \ .
\emini

\noi Looking at (\ref{6.37b}) and (\ref{6.37c}), we see that (\ref{6.42a}) is equivalent to

\beq
\left ( b_{+-}^0 - b_{--}^0 \right ) dE^0_+/dy = \left ( a_{++}^0 - a_{-+}^0 \right ) dE^0_-/dy \ \
\ . \label{6.43}
\eeq

\noi On the other hand, (\ref{6.31b}) and (\ref{6.31c}) give, looking at coefficients of $[(\mu_- +
\alpha_-)L]^{-1}$ and $[(\mu_- + \alpha_-)L ]^{-1}$ respectively,

\bminiG{EDh} 
\label{6.44a}
\left ( \beta z_{R_+} + A_{+-}^{-1} \right ) b_{+-} = d_{+-} + \left [ dyL/(4\Lambda R_+)\right ]
\left ( b_{+-} - b_{--} \right ) dE_+^0/dy  \eeeq
\beeq
\beta z_{R_-} a_{++} + A^{-1}_{-+} a_{-+} = d_{-+} + \left [ dyL/(4 \Lambda_-^R) \right ] \left (
a_{++} - a_{-+} \right ) dE_-^0/dy  \label{6.44b}
\emini

\noi which combined with (\ref{6.42b} and (\ref{6.42c}) respectively give

\bminiG{EDh} \label{6.45a}
\left ( b_{+-} - b_{--} \right ) dE^0_+/dy = 0 
\eeeq
\beeq
\label{6.45b}
\left ( a_{++} - a_{-+} \right ) dE^0_-/dy = 0 \ \ \ .
\emini

\noi In general (\ref{6.43}) and (\ref{6.45a}), (\ref{6.45b}) give three constraints. However, two
cons\-traints only are sufficients if

\bminiG{EDh} 
\label{6.46a}
dE^0_+/dy = 0
\eeeq
\beeq
dE^0_-/dy = 0 \ \ \ .
\label{6.46b}
\emini

As this is the minimum-constraint choice, we will stick to (\ref{6.46a}) and (\ref{6.46b}) as the
two constraints to be imposed. Using them in (6.37) gives the solution 

\bminiG{EDh} \label{6.47a}
a_{++}^0 = c_{++} \left ( \beta z_{R_+} + A_{++}^{-1} \right )^{-1} 
\eeeq 
\beeq
\label{6.47b}
a_{-+}^0 = A_{-+} \left ( c_{-+} - \beta z_{R_-} a_{++}^0 \right )
\eeeq
\beeq
\label{6.47c}
b_{+-}^0 = c_{+-} \left ( \beta z_{R_+} + A_{+-}^{-1} \right )^{-1}
\eeeq
\beeq
\label{6.47d}
b_{--}^0 = A_{--} \left ( c_{--} - \beta z_{R_-} b_{+-}^0 \right ) \ \ \ .
\emini

\noi So, finally, we have two equations for $\bar{\alpha}_-$ and $\bar{\alpha}_+$, four constraints
(\ref{6.23a}), (\ref{6.23b}), (\ref{6.46a}), (\ref{6.46b}) and the seven parameters $x$, $u$, $z$,
$a_+$, $a_-$, $x_+$, $x_-$. Another fifth constraint will come from the fact that
$\sum\limits_{i=1}^I \bar{\alpha}_i = h_0$. This will be discussed in section 10 where the
numerical solutions will be presented. We still need some work to be done before tackling the
numerical solution of our equations. In particular we have to give explicit expressions for
$J_{\ell}$, $z_{\ell}$, $\zeta_{\ell}$, $\delta \Lambda_{\ell}^R(\alpha_i)$ and the derivative
$d\varepsilon_{\ell}/dy$ which enters in the expression of $dE_{\ell}/dy$. All this will be done
in the next section.

\mysection{Explicit expressions for $J_{\ell}$, $z_{\ell}$, $\zeta_{\ell}$, $\delta
\Lambda_{\ell}$, $d\varepsilon_{\ell}/dy$} 

\hspace*{\parindent} We begin with the calculation of the Jacobian $J_{\ell} = 1 + (\alpha_{\ell} /
\Lambda_{\ell}^R)d\Lambda_{\ell}^R/ d\alpha_{\ell}$ and therefore we need

\beq
\label{7.1}
\Lambda_{\ell}^R = \left [ (I - dL/2)/h_0 \right ] (1 - \beta) + d/2 \ L \ E_{\ell}^0
\eeq

\noi which is the expression for $\Lambda_{\ell}^R$ when all $\alpha_j$ ($j \not= \ell$) have been
integrated with the mean-value theorem and therefore where the $\Lambda_{\ell}(\alpha_i)$ term
present in (\ref{6.24}) is omitted. Now, the expression for $E^0_{\ell}$ given by (6.38) is a
function of $a_{-\ell}^0$ and $a_{+\ell}^0$, which themselves are functions of $c_{j \ell}$. We
therefore need explicit expressions for $c_{j \ell}$ which will be obtained through the definition
(\ref{6.29a}) by (6.18), (6.19) and (6.21). We get

\bminiG{EDh} \label{7.2a}
h_0 c_{++} = 1/2 \ (1 + \beta ) z_{R_+} + d/2 \ a_+ (1 - y/2)/\left [ (\mu_+ +
\alpha_{\ell_+})\Lambda_+^R \right ]
\eeeq 
\beeq
\label{7.2b}
h_0 c_{+-} = x (1 + \beta ) z_{R_+} + d/2 \ u (1 - y/2)/\left [ (\mu_+ +
\alpha_{\ell_+})\Lambda_+^R \right ]
\eeeq
\beeq
h_0 c_{-+} = (1 - x) (1 + \beta ) z_{R_-} + d/2 \ (1 - z) y /\left [ (\mu_- +
\alpha_{\ell_-})\Lambda_-^R \right ]
\label{7.2c}
\eeeq
\beeq
h_0 c_{--} = 1/2 \ (1 + \beta ) z_{R_-} + d/2 \ a_- y/\left [ (\mu_- +
\alpha_{\ell_-})\Lambda_-^R \right ] \ \ \ .
\label{7.2d}
\emini

\noi We are now ready to calculate $J_{\ell}$. \\

\noi {\large \bf A - Explicit expressions for $J_{\ell}$ and $\zeta_{\ell}$} \\

Taking derivatives and neglecting in a first approximation $\partial
z_{R_{\ell}}/\partial \alpha_{\ell}$ we get 

\bminiG{EDh}
\partial \left ( h_0 c_{++} \right ) /\partial \alpha_+ = - d/2 \ \Lambda_+^R \ a_+(1 - y/2)/\left [
(\mu_+ + \alpha_{\ell_+} ) \Lambda_+^R \right ]^2  \label{7.3a} 
\eeeq 
\beeq
\partial \left ( h_0 c_{-+} \right )/\partial \alpha_+ = 0
\label{7.3b}
\eeeq
\beeq
\partial \left ( h_0 c_{+-} \right ) / \partial \alpha_- = 0
\label{7.3c}
\eeeq
\beeq
\partial \left ( h_0 c_{--} \right )/\partial \alpha_- = - d/2 \ \Lambda_-^R \ a_- y/\left [ (\mu_- +
\alpha_{\ell_-} \Lambda_-^R \right ]^2 \ \ . \label{7.3d}
\emini

\noi Using (6.38), (6.47) and

\beq
L/h_0 = \left [ L/(I - dL/2) \right ] z_{R_{\ell}} 
\ \Lambda_{\ell}^R/(1 - \beta ) \label{7.4}
\eeq

\noi we obtain

\bminiG{EDh} 
\label{7.5a}
&&J_+ = \left | 1 - (d/2)^2 \left [ L/(I - dL/2) \right ] \left [ z_{R_+}/(1 - \beta ) \right ]
\Big [ (y/2 + \eta ) \right . \cdot \nn \\
&& \left . \left ( \beta z_{R_+} + A_{++}^{-1} \right )^{-1} - (1 - y/2) A_{-+} \beta z_{R_-}
\Big ] a_+ (1 - y/2) \alpha_+ \Lambda_+^R / \left [ (\mu_+ + \alpha_+ ) \Lambda_+^R \right
]^2 \right |   \nn \\
\eeeq 
\beeq 
\label{7.5b}
J_- &=& \left | 1 - (d/2)^2 \left [ L/(I - dL/2) \right ] \left [ z_{R_-}/(1 - \beta ) \right ]
(1 - y/2) \right .  \cdot \nn \\
&&\left . A_{--} \ a_- \ y \ \alpha_-  \Lambda_-^R / \left [ (\mu_- +
\alpha_- ) \Lambda_-^R \right ]^2 \right | \ \ .
\emini

\noi We remark that $J_{\ell}$ is a function of $\mu_{\ell} \Lambda_{\ell}^R$ and not of
$\bar{\alpha}_{\ell} \Lambda_{\ell}^R$. Because (neglecting $\partial z_{R_{\ell}}/\partial
(\mu_{\ell} \Lambda_{\ell}^R)$) we will see that $\widetilde{H}_0 (\bar{\alpha}_{\ell})$ only
depends on $\bar{\alpha}_{\ell} \Lambda_{\ell}^R$, we will get $\zeta_{\ell}$ from

\beq
\zeta_{\ell} = 2/d \ \left ( \mu_{\ell} + \bar{\alpha}_{\ell} \right ) \Lambda_{\ell}^R \ \partial \
{\rm Log} \ \left [ J_{\ell}^{-1} H_{\ell} (\alpha_{\ell} - \bar{\alpha}_{\ell} ) \right ]/\partial
\left ( \mu_{\ell} \Lambda_{\ell}^R \right ) \ \ . \label{7.6} \eeq

\noi However, looking at the expressions (\ref{4.12}) and (\ref{4.22}) for $H_{i_+}(\alpha_{i_+} -
\bar{\alpha}_{i_+})$ and $H_{i_-} (\alpha_{i_-} - \bar{\alpha}_-)$ respectively we see that

\bminiG{EDh} 
\label{7.7a}
\partial \ {\rm Log} \ \left [ H_{\ell_+} \left ( \alpha_+ - \bar{\alpha}_+ \right ) \right ]  /
\partial \left ( \mu_+ \Lambda_+^R \right ) \sim \alpha_+ - \bar{\alpha}_+  \eeeq 
\beeq
\label{7.7b}
\partial \ {\rm Log} \ \left [ H_{\ell_-} \left ( \alpha_- - \bar{\alpha}_- \right ) \right ]  /
\partial \left ( \mu_- \Lambda_-^R \right ) \sim \alpha_- - \bar{\alpha}_- \ \ \ .
\emini
 
\noi In a first approximation where the mean-values of the left-hand side is taken, we will neglect
these contributions to $\zeta_{\ell}$, so that finally we will have

\beq
\label{7.8}
\zeta_{\ell} = 2/d \ \left ( \mu_{\ell} + \bar{\alpha}_{\ell} \right ) \Lambda_{\ell}^R \ \partial \
{\rm Log} \ \left [ J_{\ell}^{-1} \right ] / \partial \left ( \mu_{\ell} \Lambda_{\ell}^R \right
) \eeq

\noi which is easily obtained from (7.5). \\

\noi {\large \bf B - Explicit expression for $z_{\ell}$} \\

Looking at (\ref{5.22}), we have (replacing $i$ by $\ell$)

\bea
{\rm Log} \ \widetilde{H}_0(\bar{\alpha}_{\ell}) &=& d/2 \ L \ E_{\ell}^0 \ \bar{\alpha}_{\ell} \nn
\\ &=& \left ( d/2 \ L \ E_{\ell}^0/ \Lambda_{\ell}^R \right ) \bar{\alpha}_{\ell} \Lambda_{\ell}^R
\nn \\
&=& \left ( 1 - z_{R_{\ell}} \right ) \bar{\alpha}_{\ell} \Lambda_{\ell}^R
\label{7.9}
\eea

\noi and therefore, neglecting $\partial z_{R_{\ell}}/\partial (\bar{\alpha}_{\ell}
\Lambda_{\ell}^R)$ in a	first approximation,

\beq
\label{7.10}
\partial \ {\rm Log} \ \widetilde{H}_0 (\bar{\alpha}_{\ell})/\partial \left ( \bar{\alpha_{\ell}}
\Lambda_{\ell}^R \right ) = 1 - z_{R_{\ell}} \ \ \ . 
\eeq

On the other hand looking at (7.5) we see that $J_{\ell}$ does not depend on
$\bar{\alpha}_{\ell}\Lambda_{\ell}^R$ either than through $z_{R_{\ell}}$. Neglecting again
$\partial z_{R_{\ell}}/\partial (\bar{\alpha}_{\ell} \Lambda_{\ell}^R)$ we conclude that the factor
$J_{\ell}^{-1}$ does not contribute to $z_{\ell}$. Then, we have contributions from $H_{\ell}
(\alpha_{\ell} - \bar{\alpha}_{\ell})$. First,

\beq
\label{7.11}
\partial \ {\rm Log} \ H_{\ell_+}/\partial (\bar{\alpha}_+ \Lambda_+^R ) = -t/2m^2 (2 - y)^{-1}
\mu_+ \Lambda / \left [ \left ( \mu_+ + \alpha_{\ell_+} \right ) \Lambda_+^R \right ]  \eeq

\noi where we have used $<(2 - 1/\ell )^{-1}> = (2 - y)^{-1}$. Taking the mean-value of
(\ref{7.11}) we get for our numerical resolution

\bea
\partial \ {\rm Log}\ H_{\ell_+}/\partial \left ( \bar{\alpha_+} \Lambda_+^R \right ) &=& - t/2m^2 (2
- y)^{-1} z_{R_+} (1 - x_+) \nn \\
&=& - t/4m^2 z_{R_+} (1 - x_+)  \ \ \ . \label{7.12}
\eea

\noi Now for the contribution of $H_{\ell_-} (\alpha_- - \bar{\alpha}_-)$, we get

\bea
\label{7.13}
&&\partial \ {\rm Log} \ H_{\ell_-}/\partial (\bar{\alpha}_- \Lambda_-^R ) = y^{-1} (1 - \beta )
\left [ 2/3 \ \ell n \ {\rm C}^{\rm st} + \ell n (1 - \beta ) \right ] \cdot \nn \\
 &&\left ( - \alpha_-
\Lambda_-^R \right )^{-1} \left \{ \left ( \mu_-/\bar{\alpha}_-) \left [ (\alpha_- - \bar{\alpha}_-
\right )/ \left ( \mu_- + \bar{\alpha}_- \right ) \right ] - y \right \} \eea

\noi and, again, taking the mean-value to facilitate the numerical resolution

\beq
\partial \ {\rm Log} \ H_{\ell_-}/\partial \left ( \bar{\alpha}_- \Lambda_-^R \right ) = - (1 -
\beta ) \left [ 2/3 \ \ell n \ {\rm C}^{\rm st} + \ell n (1 - \beta ) \right ] / \left (
\bar{\alpha}_- \Lambda^R_- \right )  \ \ \ . \label{7.14} \eeq

\noi We now have $z_{\ell}$ by adding the contribution obtained from $\widetilde{H}_0
(\bar{\alpha}_{\ell})$ and $H_{\ell} (\alpha_{\ell} - \bar{\alpha}_{\ell})$ and multiplying by
$(2/d)(\mu_{\ell} + \bar{\alpha}_{\ell})\Lambda_{\ell}^R$. \\

\noi {\large \bf C - Explicit expressions for $\delta \Lambda_{\ell}^R(\alpha_i) /\Lambda_{\ell}^R
(\alpha_i )$} \\

$A_{\ell i}$ being a function of

\beq
\label{7.15}
\left [ \delta \Lambda_+^R (\alpha_i)/\Lambda_+^R (\alpha_i) \right ]/\left [ \delta \Lambda_-^R
(\alpha_i)/\Lambda^R_- (\alpha_i ) \right ] \eeq 

\noi (see (\ref{6.9e}) and (\ref{6.9f})) through the ratio $(\Delta_+/\Delta_-)$ (see
(\ref{6.10a}) and (\ref{6.10b})), we provide explicit expressions for $(\Lambda_{\ell}^R)^{-1}
\delta \Lambda_{\ell}^R/\delta \alpha_i$ which are read from (6.18), (6.19) and
(6.21). We get for $\alpha_i = \bar{\alpha}_i$, starting with $\Lambda_{\ell}^{R^{-1}}
\delta \Lambda_{\ell}/\delta \alpha_i$

\bminiG{EDh} 
&&h_0 \Lambda_+^{R^{-1}} \delta \Lambda_+ (\alpha_+) /\delta \alpha_+ = 1/2 \ z_{R_+}
(1 + \beta ) + \nn \\
&&(d/2) \left \{ a_+ (1 - x_+) + \left [ 1 - a_+ (1 - y/2) \right ] \right \} (1 -
y/2) / \left [ \left ( \mu_+ + \bar{\alpha}_+ \right ) \Lambda_+^R \right ] \nn \\
\label{7.16a}  \eeeq 
\beeq 
&&h_0 \Lambda_-^{R^{-1}} \delta \Lambda_- (\alpha_+)/\delta \alpha_+ = (1 - x) z_{R_-}
(1 + \beta ) + \nn \\
&&(d/2) \left \{ (1 - z) (1 - x_-) y/\left [ (\mu_- + \bar{\alpha}_-) \Lambda_-^R \right ] \right
.\nn \\ &&+ (1 - y/2) \left ( z_{R_-}/z_{R_+} \right ) \left [ 1 - u(1 - y/2) \right ] /\left [
(\mu_+ + \bar{\alpha}_+ ) \Lambda_+^R \right ]  \label{7.16b}  \eeeq 
\beeq
&&h_0 \Lambda_+^{R^{-1}} \delta \Lambda_+(\alpha_-)/\delta \alpha_- = x \ z_{R_+} (1 + \beta ) +
\nn \\
&&(d/2) \left \{ u (1 - x_+) (1 - y/2) / \left [ \left ( \mu_+ + \bar{\alpha}_+ \right )
\Lambda_+^R \right ] \right .\nn \\
&&\left .+ y \left ( z_{R_+}/z_{R_-} \right ) \left [ 1 - (1 - z) y \right ] / \left [ \left ( \mu_-
+ \bar{\alpha}_- \right ) \Lambda_-^R \right ] \right \}
 \label{7.16c} 
\eeeq
\beeq
&&h_0 \Lambda_-^{R^{-1}} \delta \Lambda_-(\alpha_-)/\delta \alpha_- = 1/2 \ z_{R_-} (1 + \beta ) +
\nn \\
&&(d/2) \left \{ a_-(1 - x_-) + (1 - a_- y) \right \} y/\left [ \left ( \mu_- + \bar{\alpha}_-
\right ) \Lambda_-^R \right ] \ \ \ .   \label{7.17d} \emini

\noi Here the meaning of $\delta$ is a difference operator,

\bminiG{EDh} 
\label{7.18a}
\delta \Lambda_{\ell} (\alpha_i) = \Lambda_{\ell} (\alpha_i) - \Lambda_{\ell} (\bar{\alpha}_i) 
\eeeq 
\beeq
\delta \alpha_i = \alpha_i - \bar{\alpha}_i
\label{7.18b}
\emini

\noi and $(\mu_i + \alpha_i)^{-1}$ is replaced by its average $(\mu_i + \bar{\alpha}_i)^{-1}$
in order to take into account the integration through the mean-value theorem of the variable
$\alpha_i$. We have not finished our calculation of $\delta
\Lambda_{\ell}^R(\alpha_i)/\Lambda_{\ell}^R(\alpha_i)$ because we need to add $\delta
\Lambda/\Lambda_{\ell}^R$ in

\beq
\Lambda_{\ell}^{R^{-1}} \delta \Lambda_{\ell}^R/\delta \alpha_i = \Lambda_{\ell}^{R^{-1}} \left (
\delta \Lambda / \delta \alpha_i + \delta \Lambda_{\ell}(\alpha_i)/\delta \alpha_i \right ) \ \ \ .
\label{7.19} \eeq

\noi In fact we know from (\ref{6.26}) that

\beq
\Lambda_{\ell}^{R^{-1}} \delta \Lambda /\delta \alpha_i = - \beta z_{R_{\ell}} \ x_{+i} \ \ \ .
\label{7.20}
\eeq 

\noi Looking at (\ref{6.32a}) and (\ref{6.32b}) we note that we need $a_{++}$ and $b_{+-}$ in order
to know the values of $x_{++}$ and $x_{+-}$ in (\ref{7.20}). The equations (\ref{6.31a}) and
(\ref{6.31b}) give, identifying first power coefficients and with (6.45)

\bminiG{EDh} 
\label{7.21a}
a_{++} = d_{++} \left ( \beta z_{R_+} + A_{++}^{-1} \right )^{-1} 
\eeeq 
\beeq
b_{+-} = d_{+-} \left ( \beta z_{R_+} + A_{+-}^{-1} \right )^{-1} \ \ \ .
\label{7.21b}
\emini
 
We have already $a_{++}^0$ and $b_{+-}^0$ from (6.47a) and (6.47c) and so we have, using
(\ref{6.29a}),

\bminiG{EDh} 
\label{7.22a}
\Lambda_+^{R^{-1}} \delta \Lambda / \delta \alpha_+ = - \beta z_{R_+} \left ( \beta z_{R_+} +
A_{++}^{-1} \right ) \cdot 
\Lambda_+^{R^{-1}} \delta \Lambda_+ (\alpha_+) / \delta \alpha_+  \eeeq 
\beeq
\label{7.22b}
\Lambda_-^{R^{-1}} \delta \Lambda / \delta \alpha_- = - \beta z_{R_-} \left ( \beta z_{R_+} +
A_{+-}^{-1} \right ) \cdot 
 \Lambda_+^{R^{-1}} \delta \Lambda_+ (\alpha_-) / \delta \alpha_- \ \ \ .
\emini

\noi Finally, defining (obtained from (7.16))

\beq
\label{7.23}
R_{\pm} \equiv \left [ \delta \Lambda_+ (\alpha_{\pm}) / \Lambda_+^R(\alpha_{\pm})\right ] / \left [
\delta \Lambda_- (\alpha_{\pm})/ \Lambda_-^R(\alpha_{\pm}) \right ]
 \eeq

\noi we get

 \bminiG{EDh} 
\label{7.24a}
&&\left ( \Lambda_+^{R^{-1}} \delta \Lambda_+^R/ \delta \alpha_+ \right )/\left (
\Lambda_-^{R^{-1}} \delta \Lambda_-^R/ \delta \alpha_+ \right ) = 
\left [ - \beta z_{R_-} A_{++} + \left ( 1 + \beta z_{R_+} A_{++} \right ) R_+^{-1} \right ]^{-1}
\nn \\
 \eeeq  
\beeq
&&\left ( \Lambda_+^{R^{-1}} \delta \Lambda_+^R/ \delta \alpha_- \right )/\left (
\Lambda_-^{R^{-1}} \delta \Lambda_-^R/ \delta \alpha_- \right ) = 
\left [ - \beta z_{R_-} A_{+-} + \left ( 1 + \beta z_{R_+} A_{+-} \right ) R_-^{-1} \right ]^{-1}
\nn \\
 \label{7.24b}
\emini

\noi ratios which are needed in order to obtain $A_{\ell i}$ through (6.10) and (6.9).
\\

\noi {\large \bf D - Explicit expression for $d \varepsilon_{\ell}/dy$} \\

>From the definition (\ref{6.3c}) of $\varepsilon_{\ell}$ we deduce the partial derivatives
$\partial \varepsilon_{\ell}/\partial \mu_{\ell}$ and $\partial \varepsilon_{\ell}/\partial
\bar{\alpha}_{\ell}$, denoting $A_{\ell}$ and $B_{\ell}$ by 

 \bminiG{EDh}
B_{\ell} &\equiv& ( \Lambda_{\ell}^R)^{-1} \partial \varepsilon_{\ell}/\partial \mu_{\ell} =
(\varepsilon_{\ell} - 1) \left [ d/2 + 1 + d/2 \ \bar{\zeta}_{\ell} \right ]/ \left [ \left (
\mu_{\ell} + \bar{\alpha}_{\ell} \right ) \Lambda_{\ell}^R \right ] \nn \\
&&- (d/2 + 1) (\eta_{\ell} - 1)/ \left [ \left ( \mu_{\ell} + \bar{\alpha}_{\ell} \right )
\Lambda_{\ell}^R \right ]  
\label{7.25a}  
\eeeq  
\beeq
A_{\ell} \equiv ( \Lambda_{\ell}^R)^{-1} \partial \varepsilon_{\ell}/\partial \bar{\alpha}_{\ell} =
(\varepsilon_{\ell} - 1) \left [ d/2 + 1 + d/2 \ \bar{z}_{\ell} \right ]/ \left [ \left (
\mu_{\ell} + \bar{\alpha}_{\ell} \right ) \Lambda_{\ell}^R \right ]
\label{7.25b}
\emini

\noi with $\bar{\zeta}_{\ell}$ and $\bar{z}_{\ell}$ being the mean-values of $\zeta_{\ell}$ and
$z_{\ell}$ and $\eta_{\ell}$ being defined through

\bea
\eta_{\ell} &=& 1 - (I/h_0) (\Lambda_{\ell}^R)^{-1} \exp \left [ -dL/(2I) - (1 - dL/(2I)) \beta
\right ] \nn \\
&&\left [ \left ( \mu_{\ell} + \bar{\alpha}_{\ell} \right ) \Lambda_{\ell}^R \right ]^{d/2 + 2}
\int_0^{\infty} dx \left ( \mu_{\ell} \Lambda_{\ell}^R + x \right )^{-d/2-2} \exp (-x ) \cdot \nn \\
&&J_{\ell}^{-1} H_{\ell} \left ( \alpha_{\ell} - \bar{\alpha}_{\ell} \right )
\widetilde{H}_0 (\bar{\alpha}_{\ell}) \ \ \ .  \label{7.26} 
\eea

\noi Then, we get $d\varepsilon_{\ell}/dy$ through 

\beq
\label{7.26a}
(\Lambda_{\ell}^R )^{-1} d \varepsilon_{\ell}/dy = \left ( A_{\ell} + B_{\ell}
\ d \mu_{\ell}/d \bar{\alpha}_{\ell} \right ) d \bar{\alpha}_{\ell}/dy \ \ \ .
\eeq

\noi The next step consists in obtaining $d \mu_{\ell}/d \bar{\alpha}_{\ell}$ and
$d\bar{\alpha}_{\ell}/dy$. This can be done by first writing (\ref{6.7}) as

\bea
&&d \mu_{\ell} \left ( \varepsilon_{\ell} + \zeta_{\ell} \right ) + d \bar{\alpha}_{\ell} (1 -
z_{\ell}) - (\beta / 6) \left ( \mu_{\ell} + \bar{\alpha}_{\ell} \right ) d
\bar{\alpha}_+/\bar{\alpha}_+ = \nn \\
&&\left [ 1/2 \left ( \mu_{\ell} + \bar{\alpha}_{\ell} \right ) - (1 + z_{\ell} )
\bar{\alpha}_{\ell} - \left ( \varepsilon_{\ell} + \zeta_{\ell} \right ) \mu_{\ell} \right ] d
\Lambda_{\ell}^R/\Lambda_{\ell}^R \ \ \ .  
 \label{7.27} \eea

>From (\ref{6.26}) we know that 

\beq
\Lambda^{-1} d\Lambda = - \beta \ \bar{\alpha}_+^{-1} \ d \bar{\alpha}_+ \ \ \ .
\label{7.28}
\eeq

\noi Because $\Lambda_{\ell}^R = \Lambda + d/2 \ L E_{\ell}^0$ and $dE_{\ell}^0/dy = 0$ (see
(\ref{5.21}) and (6.46)), we get

\beq
d\Lambda_{\ell}^R = d \Lambda
\label{7.29}
\eeq

\noi and, therefore,

\bea
\Lambda_{\ell}^{R^{-1}} \ d \Lambda_{\ell}^R/d \bar{\alpha}_+ &=& - \left ( \Lambda /
\Lambda_{\ell}^R \right ) \beta/\bar{\alpha}_+ \nn \\
&=& - z_{R_{\ell}} \beta \bar{\alpha}_+^{-1} \ \ \ . \label{7.30}
\eea

\noi Then, putting (\ref{7.30}) into the right-hand side of (\ref{7.27}), one gets two equations

\bea
&&d\mu_{\ell} \left [ \varepsilon_{\ell} + \zeta_{\ell} \right ] + d \bar{\alpha}_{\ell} (1 +
z_{\ell}) - (\beta /6) \left ( \mu_{\ell} + \bar{\alpha}_{\ell} \right ) d \bar{\alpha}_{\ell}
/\bar{\alpha}_+ = \nn \\
&& - z_{R_{\ell}} \beta \left ( d \bar{\alpha}_+/\bar{\alpha}_+ \right ) \left [ 1/2 \left (
\mu_{\ell} + \bar{\alpha}_{\ell} \right ) - (1 + \bar{z}_{\ell}) \bar{\alpha}_{\ell} - \left (
\varepsilon_{\ell} + \zeta_{\ell} \right ) \mu_{\ell} \right ] \label{7.31} \eea

\noi which with the equations $y^{-1} = 1 + \bar{\alpha}_-/\mu_-$ and $2y^{-1} = 1 + \mu_+
/\bar{\alpha}_+$ will allow to determine $d \mu_{\ell}/d \bar{\alpha}_{\ell}$ and $d
\mu_{\ell}/d \bar{\alpha}_{\ell}$. We have 

\beq
\label{7.32}
dy = \left ( a_{\ell} + b_{\ell} \ d \mu_{\ell} /d \bar{\alpha}_{\ell} \right ) d \bar{\alpha}_{\ell}
\eeq

\noi with 

\bminiG{EDh} 
\label{7.33a}
a_- = - y^2/ \mu_- 
\eeeq 
\beeq
\label{7.33b}
b_- = y^2 \ \bar{\alpha}_-/\mu_-^2
\eeeq
\beeq
\label{7.33c}
a_+ = 1/2 \ y^2 \ \mu_+ /\bar{\alpha}_+^2
\eeeq
\beeq
\label{7.33d}
b_+ = - 1/2 \ y^2/\bar{\alpha}_+
\emini

\noi and therefore :

\beq
d\varepsilon_{\ell}/dy = \Lambda_{\ell}^R \left ( A_{\ell} + B_{\ell} \ d
\mu_{\ell}/d\bar{\alpha}_{\ell} \right ) / \left ( a_{\ell} + b_{\ell} \ d \mu_{\ell} / d
\bar{\alpha}_{\ell} \right ) \ \ \ . \label{7.34} \eeq

\noi From (\ref{7.31}) we get

\bea
d\mu_+/d \bar{\alpha}_+ &=& \left ( \varepsilon_+ + \zeta_+ \right )^{-1} \left \{ - (1 + z_+) +
\beta /3 \ y^{-1} \right .\nn \\ 
&&\left . - z_{R_+} \beta \left [ y^{-1} - (1 + z_+) - \left ( \varepsilon_+ + \zeta_+ \right )
(2y^{-1} - 1) \right ] \right \} \ \ \ .
\label{7.35}
\eea

\noi For the - case we have to work a little bit in order to obtain $d\mu_-/d\bar{\alpha}_-$.
First, we note that because $y^2 = 2 \bar{\alpha}_+/\bar{\alpha}_-$ we have 

\beq
d \bar{\alpha}_+/\bar{\alpha}_+ = d \bar{\alpha}_-/\bar{\alpha}_- + 2 dy/y
\label{7.36}
\eeq

\noi and that (\ref{7.31}) gives us for $\ell = -$

\bea
&&d\mu_- \left ( \varepsilon_- + \zeta_- \right ) + d \bar{\alpha}_- (1 + z_-) = \nn \\
&&d \bar{\alpha}_+/\bar{\alpha}_+ \left \{ (\beta /6) (\mu_- + \bar{\alpha}_-)-z_{R_-} \beta \left
[ 1/2 \ (\mu_- + \bar{\alpha}_-) - (1 + z_-) \bar{\alpha}_- \right . \right . \nn \\
&&- \left . \left . \left ( \varepsilon_- + \zeta_- \right ) \mu_- \right ] \right \} \ \ \ .
 \label{7.37}
\eea

\noi Together with (\ref{7.32}) taken for $\ell_-$ we then have the system of equations

\bminiG{EDh} 
\label{7.38a}
d\mu_- (\varepsilon_- + \zeta_- ) + d \bar{\alpha}_- \left [ (1 + z_-) - \{ \} / \bar{\alpha}_-
\right ] = 2 dy/y \ \{ \}  \eeeq 
\beeq
\label{7.38b}
d\mu_- \ \bar{\alpha}_-/\mu_-^2 - d \bar{\alpha}_- \ \mu_-^{-1} = y^{-2} dy
\emini

\noi where $\{ \}$ denotes the quantity between brackets on the right-hand side of (\ref{7.37}).
Then, solving (7.38) gives 

\bminiG{EDh} 
\label{7.39a}
d\mu_- = (dy/y) \left [ 2 \{ \} /\mu_- + y^{-1} \left (  (1 + z_-) - \{ \} / \bar{\alpha}_- \right )
\right ] / \Delta  
\eeeq 
\beeq
d\bar{\alpha}_- = (dy/y) \left [ 2(\bar{\alpha}_-/ \mu_-^2) \{ \} - (\varepsilon_- + \zeta_-)
y^{-1} \right ] / \Delta \label{7.39b}
\eeeq
\noi with
\beeq
\label{7.39c}
\Delta = (\varepsilon_- + \zeta_-)/\mu_- + (\bar{\alpha}_-/\mu_-^2) \left [ (1 + z_-) - \{ \} /
\bar{\alpha}_- \right ] \emini

\noi and therefore

\bminiG{EDh} 
\label{7.40a}
d\mu_-/d\bar{\alpha}_- &=& \left \{ 2 c_- + y^{-1} \left [ (1 + z_-) - c_- y/(1 - y) \right ]
\right \}/ \nn \\
&&\left [ 2(y^{-1} - 1) c_- - (\varepsilon_- + \zeta_-)y^{-1} \right ]  \eeeq
\noi with 
\beeq
\label{7.40b}
c_- = \{ \} /\mu_- = \beta \left \{ y^{-1}/6 - z_{R_-} \left [ y^{-1}/2 - (1 + z_-) (y^{-1} - 1) -
(\varepsilon_- + \zeta_-) \right ] \right . \ .
 \emini

\noi Now, (\ref{7.32}) gives ($\ell = +$)

\beq
\label{7.41}
\left ( \Lambda_+^R d \bar{\alpha}_+/dy \right )^{-1} = 1/2 \left [ y^2/\left ( \bar{\alpha}_+
\Lambda_+^R \right ) \right ] \left ( 2y^{-1} - 1 - d \mu_+/d \bar{\alpha}_+ \right ) \ \ \ . \eeq

\noi Taking $A_+$ and $B_+$ from (7.24) and putting them into (\ref{7.34}), we get with the
help of (\ref{7.41})

\bea
\label{7.42}
d\varepsilon_+/dy &=& y^{-1} \left \{ (d/2 + 1 + d/2 \bar{z}_+ ) (\varepsilon_+ - 1) + \left [ (d/2
+ 1) (\varepsilon_+ - \eta_+) + d/2 \ \bar{\zeta}_+ \right ] d\mu_+/d \bar{\alpha}_+ \right \} \nn
\\
&&/(2y^{-1} - 1 - d \mu_+/d\bar{\alpha}_+) \ \ \ .
\eea

Again, (\ref{7.32}) for $\ell = -$ gives

\beq
\label{7.43}
\left ( \Lambda_-^R d \bar{\alpha}_-/dy \right )^{-1} = \left [ y^2/ (\mu_- \Lambda_-^R) \right ]
\left [ - 1 + (y^{-1} - 1) d\mu_-/d\bar{\alpha}_- \right ] \eeq

\noi and taking $A_-$ and $B_-$ from (7.24), putting them into (\ref{7.34}), taking into
account (\ref{7.43}), we get 

\bea
\label{7.44}
d\varepsilon_-/dy &=& y^{-1} \left \{ (d/2 + 1 + d/2 \bar{\zeta}_- ) (\varepsilon_- - 1) + \left [
(d/2 + 1)(\varepsilon_- - \eta_-) + d/2 \ \bar{\zeta}_- \right ] d\mu_-/d\bar{\alpha}_- \right \} \nn
\\
&&/ \left [ - 1 + (y^{-1} - 1) d\mu_-/d\bar{\alpha}_- \right ] \eea

\noi which completes our evaluation of quantitites used in the solution of (\ref{6.1a}).

\mysection{Consistency of $\alpha_i = O(h_0/I)$} 
\hspace*{\parindent}
We have concluded in section 2 that $\bar{\alpha}_i$ and $\mu_i$ should be proportional to $h_0/I$
in order to achieve consistency. This was done, however, neglecting the dependence
$\bar{\alpha}_j(\alpha_i)$ which renormalizes $\Lambda$ into $\Lambda_i^R$ and the factors
$H_i(\alpha_i - \bar{\alpha}_i)$. So one can ask what happens to this consistency when all these
changes are taken into account. The effects of taking into account the $\bar{\alpha}_j(\alpha_i)$
dependence in the equation determining $\bar{\alpha}_i$ is translated into \par

a) changing $\Lambda$ into $\Lambda_i^R = \Lambda + d/2 \ L E_i^0$ \par

b) introducing the factors $J_{\ell}^{-1}$, $H_{\ell}(\alpha_{\ell} - \bar{\alpha}_{\ell})$ and
$\widetilde{H}_0(\bar{\alpha}_{\ell})$ in the integrand (see (\ref{6.1a})). \par

Now, if we look at the $d/2 \ LE_i^0$ part of $\Lambda_i^R$, we see that $E_i^0$ is proportional to
$1/h_0$ times a function of the variables $\beta$, $z_{R i}$, $y$, $\mu_i \Lambda_i^R$,
$\bar{\alpha}_i\Lambda_i^R$. (The $\alpha_i \Lambda_i^R$ dependence is transmutated into a
$\bar{\alpha}_i \Lambda_i^R$ dependence by taking the mean-value of $E_i^0$ as a function of
$\alpha_i \Lambda_i^R$. This is the origin of the parameters $x_{\ell}$ which appear when replacing
$(\mu_{\ell} + \alpha_{\ell})^{-1}$ by $(1 - x_{\ell})(\mu_{\ell} + \bar{\alpha}_{\ell})^{-1}$ in
the expression for $\Lambda_{\ell}^R \ \delta \Lambda_{\ell}/\delta \alpha_i$ which in turn appears
in $\Delta_+/\Delta_-$, which itself is appearing in $A_{\ell {\pm}}$. See the equations
(7.18), (6.9) and (6.10)). $\beta$ being proportional to $\bar{\alpha}_+ \Lambda$ and $z_{R_i}$
being the ratio $\Lambda/\Lambda_i^R$, we can therefore consider $E_i^0$ as a constant when $I, L
\to \infty$ and be perfectly consistent with the assumption $\bar{\alpha}_i = O(h_0/I)$. \par

For $\widetilde{H}_0(\bar{\alpha}_{\ell}) = (1 - z_{R_i}) \bar{\alpha}_{\ell} \Lambda_{\ell}^R$ the
above reasoning leads to $\widetilde{H}_0 (\bar{\alpha}_{\ell})$ being constant. Looking at the
expressions (\ref{7.5a}) and (\ref{7.5b}) for $J_+$ and $J_-$ we see that the varying factors are
$\alpha_+ \Lambda_+^R/[(\mu_+ + \alpha_+) \Lambda_+^R]^2$ and $\alpha_- \Lambda_-^R/[(\mu_- +
\alpha_-)\Lambda_-^R ) ]^2$ which have limited variations for any value of $\alpha_+\Lambda_+^R$ or
$\alpha_- \Lambda_-^R$ respectively. Then, we conclude that $J_+$ and $J_-$ have limited variations
too. Of course, for specific values of $\alpha_{\ell}\Lambda_{\ell}^R$, $J_{\ell}^{-1}$ can have a
pole and then the integrand can become infinite. However, for reasonable values of $a_{\ell}$ this
pole does not exist for real values of $\alpha_{\ell} \Lambda_{\ell}^R$ as the numerical equations
resolution show. \par

There are also factors $H_i(\bar{\alpha}_{i_{\pm}} - \bar{\alpha}_{\pm})$ to consider which comes
from the fact that when $\alpha_i \not= \bar{\alpha}_i$, $Q_G (P, \{ \bar{\alpha}_j\}_{j\not=i},
\alpha_i)$ varies. However, looking at (\ref{4.12}) and (\ref{4.22}) we easily conclude that these
factors too have bounded variations as $\alpha_{i_+}$ or $\alpha_{i_-}$ varies. \par

Then, we can write the equation (\ref{6.1a}) determining $\bar{\alpha}_{\ell}$ (or
$\bar{\alpha}_i$)

\bea
\left ( \mu_i + \bar{\alpha}_i \right )^{-d/2} &=& \left [ I/(h_0 \Lambda_i^R) \right ] \kappa
{\Lambda_i^R}^{d/2} \exp (\mu_i \Lambda_i^R) \nn \\ 
&&\int_{\mu_i \Lambda_i^R}^{\infty} dx \exp (-x) \ x^{-d/2}\  B_i(x) 
\label{8.1}
\eea  

\noi where $B_i(x)$ has a bounded variation. This equation is the equation (\ref{2.21}) with
$\Lambda_i^R$ replacing $\Lambda$ and $B_i(x)$ multiplying the integrand. As $B_i(x)$ has a bounded
variation and $\Lambda_i^R$ is porportional to $\Lambda$ everything we have said in section 2
concerning consistency is still true here and therefore the consistency of the assumption
$\bar{\alpha}_j = O(h_0/I)$ is established in the general case where the variations of
$\bar{\alpha}_j(\alpha_i)$ and $Q_G (P, \{ \bar{\alpha}_j\}_{j\not= i}, \alpha_i)$ are taken into
account.

\mysection{The $m \to 0$ limit of the consistency equations} 
\hspace*{\parindent}
For the determination of $\bar{\alpha}_i$ we have to solve the consistency equation (\ref{5.23}) or
(\ref{6.1a}). We are interested in showing that when the mass $m$ tends to zero, this equation
becomes independent of $m$. This is done most easily by considering the form (\ref{5.23}) of the
equation where the variable change $\alpha_i \Lambda_i^R \to x$ has not been done as in
(\ref{6.1a}). So let us rewrite it 

\bminiG{EDh} 
\label{9.1a}
1 &=& (I/h_0) \kappa \int_0^{h_0} d\alpha_i \left [ (\mu_i + \bar{\alpha}_i)/(\mu_i + \alpha_i )
\right ]^{d/2} \cdot \nn \\
&&\exp (- \alpha_i \Lambda_i^R) H_i (\alpha_i - \bar{\alpha}_{\ell}) \widetilde{H}_0
(\bar{\alpha}_i)  
\eeeq
\noi with   
\beeq
\label{9.1b}
\kappa = \exp \left [ - dL/(2I) - (1 - dL/(2I)) \beta \right ]
\eeeq
\beeq
\label{9.1c}
\widetilde{H}_0(\bar{\alpha}_i) = \exp \left ( d/2 \ L \ E_i^0 \bar{\alpha}_i \right )
\eeeq
\beeq
\label{9.1d}
H_{i_+} \left ( \alpha_{i_+} - \bar{\alpha}_+ \right ) = \exp \left ( - \varepsilon_+ \Lambda/m^2
\right ) \eeeq
\beeq
\label{9.1e}
H_{i_-} \left ( \alpha_{i_-} - \bar{\alpha}_- \right ) = \exp \left ( - \varepsilon_- \Lambda /m^2
\right ) \emini

\noi and

\bminiG{EDh}
\varepsilon_+ = t/2 \ L \ \bar{\alpha}_+ <\ell /(2 \ell - 1)> (\alpha_{i_+} - \bar{\alpha}_+)
\mu_+/(\mu_+ + \alpha_{i_+})
  \label{9.2a} 
\eeeq 
\beeq
\varepsilon_- &=& \left [ (\alpha_{\ell_-} - \bar{\alpha}_-) \mu_-/(\mu_- + \alpha_{i_-}) \right ]
\cdot \nn \\
&&(m^2/(\bar{\alpha}_- \Lambda )) y^{-1} \left \{ - (1 - a) \left [ 2/3 \ \ell n \ {\rm
C}^{\rm st} + \ell n (1 - a) \right ] \right \} \label{9.2b} \emini

\noi where the expressions for $H_{i_+}(\alpha_{i_+} - \bar{\alpha}_+)$ and $H_{i_-} (\alpha_{i_-} -
\bar{\alpha}_-)$ have been taken from section 4 and $\widetilde{H}_0(\bar{\alpha}_i)$ from
(\ref{5.22}). We have seen in section 3 that as $m \to 0$

\beq
\label{9.3}
(1 - \beta ) /m^2 \to h_0/Q_G(P, \{\bar{\alpha}\}) 
\eeq

\noi which is independent of $m$ and so 

\beq
\label{9.4}
\Lambda/m^2 = (I - d/2\ L) (1 - \beta )/(h_0m^2)
\eeq

\noi is also independent of $m$ in that limit. Because $\varepsilon_+$ is independent of $m$ we also
conclude that $H_{i_+} (\alpha_{i_+} - \bar{\alpha}_+)$ is independent of $m$ as $m \to 0$.
$\varepsilon_-$ depends on $m$ only through $m^2/\Lambda$ and is also independent of $m$ as $m \to
0$, and so is $H_{i_-}(\alpha_{i_-} - \bar{\alpha}_-)$ for the same reason. As $m \to 0$, $\beta
\to 1$ and therefore $\kappa$ tends to $\exp (- e)$, a constant independent of $m$. Remains
$\widetilde{H}_0(\bar{\alpha}_i)$ which a priori could depend on $m$ through $E_i^0$ and $\exp (-
\alpha_i \Lambda_i^R)$ which also depends on $E_i^0$ because 

\beq
\Lambda_i^R = (I - d/2L) (1 - \beta )/h_0 + d/2 \  L \ E_i^0 \ \ \ .  
\label{9.5}
\eeq
 
\noi $\Lambda_i^R$ can only depend on $m$ through $E_i^0$ because,  as $\beta \to 1$ the first term
in (\ref{9.5}) becomes negligible compared to the second one. This also entails that $z_{R_{i}} \to
0$. \par

We have (see (6.38))

\bminiG{EDh} 
\label{9.6a}
E_+^0 = (1 - y/2) a_{-+}^0 + (y/2 + \eta ) a_{++}^0 
\eeeq 
\beeq
\label{9.6b}
E_-^0 = (1 - y/2) b_{--}^0 + (y/2 + \eta ) b_{+-}^0
\emini

\noi where the coefficients $a_{-+}^0$, $a_{++}^0$, $b_{--}^0$, $b_{+-}^0$ are given by
(6.47). First, as (see (\ref{5.5}))

\beq
\label{9.7}
\eta = (2/d) \ t/2 \ \bar{\alpha}_+ \Lambda/m^2 \ \ \ ,
\eeq

\noi $\eta$ has a limit independent of $m^2$ as $m \to 0$. Now let us rewrite the equations
(6.47)

\bminiG{EDh} 
\label{9.8a}
a_{++}^0 = c_{++} \left ( \beta \ z_{R_+} + A_{++}^{-1} \right )^{-1} 
\eeeq 
\beeq
\label{9.8b}
a_{-+}^0 = A_{-+} \left ( c_{-+} - \beta z_{R_-} \ a_{++}^0 \right )
\eeeq
\beeq
\label{9.8c}
b_{+-}^0 = c_{+-} \left ( \beta z_{R_+} + A_{++}^{-1} \right )^{-1}
\eeeq
\beeq
\label{9.8d}
b_{--}^0 = A_{--} \left ( c_{--} - \beta z_{R_-} b_{+-}^0 \right )
\emini

\noi which lead us to look after the expressions of $c_{i\ell}$ and $A_{i \ell}$. Looking at
(7.2) we see that $c_{i\ell}$ only depends on $m$ through $\Lambda_{\ell}^R$ as $m \to 0$.
Looking at (6.10) defining $A_+$ and $A_-$, we see that these quantities depend on $\delta
\Lambda_{\ell}^R/\Lambda_{\ell}^R$, $\varepsilon_{\ell}$ and $z_{\ell}$ (see (6.9)) and
$\beta$ (see (\ref{6.9a}) and (\ref{6.9c})). \par

However, once more, looking at (7.16) and (7.23) we see that, as $m \to 0$, $\delta
\Lambda_{\ell}^R/\Lambda_{\ell}^R$ only depends on $m$ through $\Lambda_{\ell}^R$. Finally,
$\varepsilon_{\ell}$ and $z_{\ell}$ also have the same property (see (6.3) and the
expressions for $J_{\ell}$ in (7.5)). So $E_+^0$ and $E_-^0$ are determined as a function of
themselves only in the limit $m \to 0$ and therefore do not depend on $m$ in that limit. \par

This completes our verification that the consistency equations determining $\bar{\alpha}_i$ are
indeed independent of $m$ as $m$ tends to zero as they should. 

\mysection{The numerical evaluation of the leading Regge trajectory} 
\hspace*{\parindent}
The consistency equations can be solved numerically when all quantities ap\-pea\-ring in them are
$O(1)$. Here, we deal with the massive case, deferring the massless case to a later study. In
(\ref{6.1a}) we have

\beq
\label{10.1}
(I/h_0) {\Lambda_i^R}^{-1} = \left [ I/(I - d/2 \ L) \right ] (1 - \beta )^{-1} z_{R_i}
\eeq

\noi and therefore the consistency equation for $\bar{\alpha}_i$ takes the following form (for
$\phi^3$ and $d = 4$)

\bea
\label{10.2}
1 &=& z_{R_i} \exp (- 2/3 - \beta / 3) /(1 - \beta ) \int_0^{\infty} \left [ \mu_i \Lambda_i^R +
\bar{\alpha}_i \Lambda_i^R )/(\mu_i \Lambda_i^R + x) \right ]^2 \nn \\
&&\exp (- x) J_i^{-1} H_i(\alpha_i - \bar{\alpha}_i ) \widetilde{H}_0 (\bar{\alpha}_i)
 \eea

\noi where $\mu_i \Lambda_i^R$ and $\bar{\alpha}_i \Lambda_i^R$ are finite unknown quantities. In
the ladder case, there remain two consistency equations and four unknown quantities $\mu_-
\Lambda_-^R$, $\bar{\alpha}_- \Lambda_-^R$, $\mu_+\Lambda_+^R$, $\bar{\alpha}_+\Lambda_+^R$, which,
however are not independent. We have \cite{15r}

\bminiG{EDh} 
\label{10.3a}
\mu_- \Lambda_-^R/\left ( \mu_- \Lambda_-^R + \bar{\alpha}_- \Lambda_-^R \right ) = y 
\eeeq 
\beeq
\label{10.3b}
\bar{\alpha}_+ \Lambda_+^R/ \left ( \mu_+ \Lambda_+^R + \bar{\alpha}_+ \Lambda_+^R \right ) = y/2
\emini

\noi with

\bea
\label{10.4}
y^2 &=&  2 \bar{\alpha}_+/\bar{\alpha}_-  \nn \\
&=& \left ( 2 \bar{\alpha}_+ \Lambda_{R_+} / (\bar{\alpha}_- \Lambda_{R_-}) \right )
\Lambda_{R_-}/\Lambda_{R_+} \nn \\
&=& \left ( 2 \bar{\alpha}_+ \Lambda_{R_+} / ( \bar{\alpha}_- \Lambda_{R_-}) \right )
z_{R_+}/z_{R_-} \eea 

\noi which gives five relations for five unknowns. \par

We recall that (see (6.29))

\beq
z_{R_i} = \Lambda / \Lambda_i^R = (1 - \beta )/\left (1 - \beta + d/2 \ h_0 E_i^0 \right )
\label{10.5}
\eeq
 
\noindent and (see (3.6))

\beq
\label{10.6}
\beta = [t/(2m^2)] \bar{\alpha}_+ \Lambda = [t/(2m^2)] z_{R_+} \bar{\alpha}_+ \Lambda_+^R \ \ \ . 
\eeq

\noindent Moreover, there are parameters appearing in the decomposition of $F_{i\ell}$ in
(6.11) which are 

\beq
x \ , \ \ u \ , \ \ z \ , \ \ a_+ \ , \ \ a_- \ , \ \ x_+ \ , \ \ x_-
\label{10.7}
\eeq

\noindent with the four constraints (\ref{6.23a}), (\ref{6.23b}) and (\ref{6.46a}), (\ref{6.46b}).
Let us also remind that $x_+$ and $x_-$ are introduced because, in order to simplify the
calculations, we replace in the expression of $E_i^0$, i.e. in $c_{i \ell}$, $A_{i\ell}$,
$z_{R_i}$, $(\mu_i + \alpha_i)^{-1}$ by $(1 - x_i)(\mu_i + \bar{\alpha}_i)^{-1}$. Now, another
constraint comes from the relation $\sum\limits_{i=1}^I \bar{\alpha}_i = h_0$ which takes the form
in the ladder case

\beq
\label{10.8}
2L \bar{\alpha}_+ + (L + 1) \bar{\alpha}_- = h_0
\eeq
 
\noi as $L + 1 = \Lambda h_0/(1 - \beta )$, this is converted into (neglecting 1 in front of $L$)

\[
2 \bar{\alpha}_+ \Lambda + \bar{\alpha}_- \Lambda = (1 - \beta )
\]

\noi or

\beq
\label{10.9}
2 \bar{\alpha}_+ \ \Lambda_+^R \ z_{R_+} + \bar{\alpha}_- \ \Lambda_-^R \ z_{R_-} = (1 - \beta ) \ \
\ , \eeq

\noi which we use to obtain $z_{\Lambda_+}$ as a function of $z_{R_-}$. This is the fifth
constraint. \par

The equations (10.3) and (\ref{10.4}) are used to eliminate $\bar{\alpha}_- \Lambda_-^R$,
$\bar{\alpha}_+ \Lambda_+^R$, $\mu_+ \Lambda_+^R$ as free parameters and keep $\mu_-\Lambda_-^R$ and
$y$ as the free ones. Then, we have got nine unknowns

\beq
\label{10.10}
\mu_- \Lambda_-^R \ , \ \ y \ , \ \ x \ , \ \ u \ , \ \ z \ , \ \ a_{+} \ , \ \ a_- \ , \ \ x_+ \
, \ \ x_- \eeq

\noi together with the two consistency conditions (\ref{10.2}) and five constraints (6.23),
(6.46) and (\ref{10.9}), i.e. seven equations. In practice, $\mu_-
\Lambda_-^R/\bar{\alpha}_- \Lambda_-^R$ will be large ($\gsim$ 10) and the mean-value of $(\mu_- +
\alpha_-)^{-1}$ will be very close to $(\mu_- + \bar{\alpha}_-)^{-1}$. So $x_-$ will be close to
zero. So, we take $x_-$ to be zero and we are left with 8 parameters instead of 9. Of course, we
have to have also 

\beq
\label{10.11}
0 < x_+ < 1
\eeq

\noi which can be considered as an eighth constraint. In practice, $x_+$ will be close to 1/2. \par

The procedure we take to solve the systems of equations is to add the absolute values of sides
which have to be zero and minimize their weighted sum. Problems occur because we get a chaotic
behaviour of this sum. This is easily understood because several quantities are expressed as
functions of themselves. So we have to make calculational loops and verify that output values are
the same as input values. This is first done for $z_{R_-}$. Then, the $z_{R_-}$ loop is inserted into
another calculational loop where the value of the left-hand side of (\ref{10.2}) is compared with
one for $i = -$. Again, this loop is contained in a last loop where the left-hand side of
(\ref{10.2}) for $i = +$ is compared with one. This gives a total of three loops in a ``Russian
doll'' configuration. No wonder that we may encounter some chaotic behaviour~! In order to cope
with this phenomenon we have devised a minimization algorithm \cite{17r} which does not use any
gradient approach. It is more in the Monte-Carlo spirit but much more efficient. Its main feature
is the construction of a cube in $n$ dimensions, if there are $n$ parameters, i.e. to calculate two
values of a particular parameter for any other parameter value. So, we have $2^n$ values of the
function to calculate. We take the minimum of these $2^n$ values to construct around it another
cube, but with a side being reduced with respect to the former cube. \par

We found that this algorithm is much more powerful than a well-known minimization program known as
MINUIT \cite{18r}, widely used by experimentalists for instance. \par

The results for the trajectory $\alpha (t/m^2)$ are contained in fig. 4 and fig. 5. In fig. 4 we have
taken two values of the coupling constant $\gamma$ such that $\ell n \gamma_m =
\ell n (\gamma e / (m 4 \pi 3\sqrt{3})$ is equal to $- 0.1$ and $0$. The obtained intercept are,
roughly 0.25 and 0.47 respectively. Calculations \cite{14r} using the Bethe-Salpeter approach
give an intercept (assuming a zero mass for the central-rung fields)

\beq
\label{10.12}
\alpha (0) = - 3/2 + \sqrt{1/4 + \gamma^2/(16 \pi^2 m^2)}
\eeq

\noi corresponding to values $\simeq$ 0.300 and 0.475 for the same values of $\gamma$ as quoted
above. The intercept that we calculated is compared with (\ref{10.12}) in the range\break \noindent
-.4 $\leq \ell n \ \gamma_m <$ .5 in fig. 5. Agreement is obtained for $\alpha (t/m^2) \ \gsim \
0.3$. \par

The fact that for $\ell n \ \gamma_m \ \lsim \ 0.1$ our calculated intercept is lower than that given
by (\ref{10.12}) can easily be explained. We know that when $\gamma \to 0$ the finite ladders give
the dominating contribution to the scattering amplitude. What we see on Fig.~5 is that the finite
ladders still dominate for $\ell n \ \gamma_m \ \lsim \ 0.1$ and when $\gamma_m$ grows larger the
saddle-point contribution of infinite ladders takes over. \par 

In table 1, we report the values of
$\mu$-$\Lambda$, $y$, $x$, $u$, $z$, $a_+$, $x_+$ and $\alpha (t/m^2)$ for $\ell n \ \gamma_m = 0$
and - 3.6 $\leq t/m^2 \ \leq$ 2.0. We remark that $\alpha (t/m^2)$ {\it is compatible with a linear
function of $t/m^2$}. Such linear fit made with the eye are drawn on fig. 4. This result is new and
in an improvement over a previous \cite{11r} determination of $\alpha (t/m^2)$ with the same method
where we could not have a real result for $t/m^2 > 0.8$. This change is due to the correction of some
errors among which was the omission of the term (\ref{7.20}) in $(\Lambda_{\ell}^R)^{-1} \delta
\Lambda_{\ell}^R/\delta \alpha_i$. So, apparently we can go further out in $t/m^2$ range. However,
this takes more computer time and this is the reason why we limited ourselves to the present range.
Let us note that the loop over $z_{R_-}$ is made twice, that for the $-$ consistency equation four
times, but that for the $+$ consistency equation is made fifteen times in order to get a reasonable
safety in convergence. As $|t/m^2|$ grows it becomes more and more difficult to get precise results.
\par

Let us digress a little bit on the linear property. This is what would be expected if the
infinite number of loops part of $\phi^3$ was equivalent to a string theory. Of course, this
argument is not new \cite{19r}. We even found \cite{20r} that a local Polyakov lagrangian could
be deduced (with some weak logarithmic corrections) from the planar $\phi^3$ graphs with an
infinite density of vertices. So we would expect linear Regge trajectories for this sector of
$\phi^3$. Our present work is an indication that this may be true indeed.  

\mysection{Conclusion} 

\hspace*{\parindent} We have shown that the infinite loop limit in scalar field theories can be
accessible to practical calculation. Of course, in general, we have an infinite system of
consistency equations if no symmetry appears in the topology of the considered Feynman graphs.
However, in the ladder case, only two of them survive, allowing us to calculate the leading Regge
trajectory, which, our results show, may be linear. \par

Consistenty equations are obtained by using the mean-value theorem for all $\alpha$-parameters and
for all but one, $\alpha_i$, the one for which usual integration is needed in order to determine
its mean-value $\bar{\alpha}_i$.  \par

So doing, a crucial parameter $\mu_i$ appears to be related to the ratio of the sum of weighted
spanning trees going through $i$ to the sum of weighted spanning trees not going through $i$,
which is in fact, $\mu_i/\bar{\alpha}_i$. This parameter $\mu_i$ represents the {\it local}
topological properties of the graph. In the ladder case the ratio $\mu_i/\bar{\alpha}_i$ is easily
determined for each kind of propagators, belonging to the sides of the ladder or central. As $\mu_i$
is homogeneous to one power of $\alpha$-parameter it should have a priori the same behaviour as a
function of $I$ as $I$ tends to infinity, i.e. it should behave like $O(h_0/I)$. We proved that
$\mu_i$ is $O(h_0/I)$ for all propagators except for an infinitesimal proportion of them. In
fact, once the behaviour of $\mu_i$ is known, that of $\bar{\alpha}_i$ is determined by the
consistency equation and when $\mu_i$ is $O(h_0/I)$, $\bar{\alpha}_i$ has been shown to have the
same behaviour as expected for homogeneity reasons. In that respect the Gaussian propagator
representation is therefore wholly consistent. A scale $\Lambda$ proportional to $I/h_0$ was also
introduced making $\bar{\alpha}_i\Lambda$ a constant as $I$ tends to infinity. Because
$\bar{\alpha}_i\Lambda$ and $\mu_i\Lambda$ are constant, they appear in practical numerical
computations rather than $\bar{\alpha}_i$ and $\mu_i$. When the variation of
$\bar{\alpha}_j(\alpha_i)$ is taken into account a renormalization of $\Lambda$ into $\Lambda_j^R
= \Lambda + d/2 \ E_j^0L$ occurs where $E_j^0$ is some constant, leaving $\Lambda_j^R$ also
proportional to $I/h_0$. In the ladder case, $E_j^0$ has been explicitly determined. In fact,
$E_j^0$ is the result of the interaction of $\alpha_j$ and $\alpha_i$ through terms proportional
to $\alpha_i\alpha_j$ in the consistency equations making $d\bar{\alpha}_j(\alpha_i)/d \alpha_i$
of order $1/I$. A finite renormalization effect occurs because there are $I$ propagators. It
has been shown that this renormalization leaves unchanged the consistency of the scheme. We
expect the renormalization procedure developed for the ladder topology to be only slightly
modified in the general topology case as the procedure used for the ladder topology can be
readily used in the general topology case. What has to be provided in order to have a complete
resolution of the general case is the ratio $\mu_i/\bar{\alpha}_i$, which is of {\it local
nature on the graph}. However, dealing with sums over graphs amplitudes instead of individual
graph amplitude could be the way for treating the general case, $\mu_i/\bar{\alpha}_i$ then
taking an average value for sums of graphs amplitudes. With this averaging procedure only {\it
one} consistency equation would be needed, simplifying somewhat the scheme. Therefore, we
expect the road to be open to a complete solution of massive scalar $\phi^3$ field theory,
using the Gaussian propagator formalism. We have seen that the massless limit, being
independent of the mass, is also tractable in our scheme. This opens the road to QCD if the
reduced kernel can be found for the multi-loop case. Therefore, finding this reduced kernel
will be one of our priorities in the near future.     

\newpage
\appendix
\mysection{- Appendix}

\underbar{\bf Definitions} \\

i) Let us define the contraction of a propagator by the fusion of its two end-vertices. \\

ii) We define the contraction of a loop by the contraction of all propgators belonging to the loop.
\par
If we draw loops on a surface we can define an interior and an exterior for a loop. \\

iii) An elementary loop or mesh contains all the propagators on the lines joining its vertices if
these lines are drawn on the closed interior of the loop. The boun\-da\-ry of the interior of the
loop is then the loop itself. In other words, there are no propagators on the open interior of an
elemenntary loop $\sq$. \\

So, from now on we will consider graphs ordered by a topological expansion as the consideration of
elementary loops will take a primordial importance. We consider first the effect of the contraction
of an elementary loop ${\cal L}$ on a graph $G$ containing the propagator $i$. All spanning trees
on $G$ can be constructed by cutting open all loops of $G$. In particular, if the propagator $i$
is cut we have cut ${\cal L}$ open at $i$. \par

We begin by considering the spanning trees on $G$ with ${\cal L}$ cut {\it once}. Then, {\it all
vertices of ${\cal L}$ are connected}. It follows that if we contract ${\cal L}$, we have, after
this contraction, the spanning trees of $G$ which are constructed from ${\cal L}$ cut once
becoming spanning trees of $G_{{\cal L}_{c}}$ where $G_{{\cal L}_{c}}$ means $G$ with
${\cal L}$ contracted. This is because having contracted some connected piece of a tree, the result
of this contraction is still a tree. So, for ${\cal L}$ cut once, all spanning trees of $G$ can
be cosntructed by constructing first all spanning trees on $G_{{\cal L}_{c}}$ and then, return on
$G$ (that is decontracting ${\cal L}$) and cut ${\cal L}$ at some propagator. The net result is
that we have built the spanning trees on $G$ with ${\cal L}$ cut one in two {\it independent}
steps. Then, the ratio $\mu_i/\bar{\alpha}_i$ for these spanning trees is simply     

\beq
\label{A.1}
\mu_i/\bar{\alpha}_i = \sum_{j \in {\cal L}\atop j \not= i} \bar{\alpha}_j/\bar{\alpha}_i
\eeq

\noi and, therefore

\beq
\label{A.2}
\mu_i + \bar{\alpha}_i = \sum_{j \in {\cal L}} \bar{\alpha}_j \ \ \ . 
\eeq

Looking at the consistency equation (\ref{2.16}), we see that the left-hand side is exactly the
same for all $\bar{\alpha}_j$'s belonging to ${\cal L}$ if we were only considering the spanning
trees built from ${\cal L}$ cut once. So, we have a symmetry between all $\bar{\alpha}_j$'s
belonging to ${\cal L}$ which reduces $n_{\cal L}$ consistency equations to one with the constraint

\beq
\label{A.3}
\mu_i/\bar{\alpha}_i = n_{\cal L} - 1
\eeq

\noi if $n_{\cal L}$ is the total number of propagators of ${\cal L}$. \par

We note that a finite ratio $\mu_i/\bar{\alpha}_i$ implies that $\mu_i I$ cannot tend to zero
because otherwise $\mu_i\Lambda \to 0$ and $\bar{\alpha}_i/\mu_i \to \infty$ as deduced in
(\ref{2.26}). \par

Now, we argue that for most propagators $n_{\cal L}$ is finite. Indeed, let us start the
construction of $G$ with one loop incident with all external lines (this is possible if $G$ is
1-line irreducible). If we keep $G$ planar and add loops to it, each time a loop is added, three
propagators are added. When the number of loops is infinite with respect to the number of external
lines, only an infinitesimal proportion of the propagators will be part of only an elementary loop
with an infinite number of propagators. So, for almost all propagators $n_{\cal L}$ will be finite
if the number of loops of $G$ is sufficiently high. \par

We now examine the effect of taking into account the topologies of the spanning trees on $G$ where
${\cal L}$ is cut more than once. \par

In order to keep having a tree when cutting ${\cal L}$ twice we consider a loop ${\cal L}_1$ having
some propagator $j$ in common with ${\cal L}$. So, cutting ${\cal L}$ at $i$ and $j$ we have a tree
on ${\cal L} \cup {\cal L}_1$. All the spanning trees on $G$ having a spanning tree on ${\cal L} \cup {\cal
L}_1$ are built by taking all the spanning trees on $G_{({\cal L} \cup {\cal L}_1)_c}$ (i.e. $G$
where ${\cal L}_1$ and ${\cal L}_2$ are contracted) and combining them with all spanning trees on
${\cal L} \cup {\cal L}_1$. This can be done because contraction preserves the tree topology. We
can continue the process by considering all spanning trees on ${\cal L} \cup {\cal L}_1 \cup
{\cal L}_2$ where ${\cal L}_2$ is a third elementary loop. Then, a total of three propagators will
be cut on them. Again, the factorization will be at work for all spanning trees on $G$ having a
spanning tree on ${\cal L} \cup {\cal L}_1 \cup {\cal L}_2$. The factorization will continue to
work out in the same way taking an arbitrary number of connected elementary loops ${\cal L}_1$,
${\cal L}_2$, $\cdots$, ${\cal L}_n$ on $G$. This will alow ${\cal L}$ to be cut an arbitrary
number of times. Of course, an arbitrary spanning treee on $G$ can have disconnected sub-trees
on ${\cal L} \cup {\cal L}_1 \cup \cdots \cup {\cal L}_n$. In order to recover all topologies
we have to let $n$ tending to infinity until $G$ is completely covered by ${\cal L} \cup {\cal L}_1 \cup \cdots
\cup {\cal L}_n$. However, we have a systematic way of constructing spanning trees on $G$ starting
from the one loop topology with the important property that {\it factorization will continue to
work out when we add an arbitrary number of loops}. \par

Now, let us find a general expression for the total weight $W_{{\cal L} \cup \cdots \cup {\cal
L}_n}$ of spanning trees on ${\cal L} \cup \cdots \cup {\cal L}_n$, ${\cal L}_1$, ${\cal
L}_2$, $\cdots$, ${\cal L}_n$ being a propagator-connected set of elementary loops. For ${\cal L}$
alone

\beq
\label{A.3a}
W_{\cal L} = \sum_{\ell \in {\cal L}} \bar{\alpha}_{\ell}
\eeq     

\noi and for ${\cal L} \cup {\cal L}_1$

\beq
\label{A.4}
W_{{\cal L} \cup {\cal L}_1} = \left ( \sum_{\ell \in {\cal L}} \bar{\alpha}_{\ell} \right ) \left
( \sum_{k \in {\cal L}_1} \bar{\alpha}_k \right ) - \bar{\alpha}_j^2 \eeq

\noi where $j$ is the propagator common to ${\cal L}$ and ${\cal L}_1$ ($j \in {\cal L} \cap
{\cal L}_1$). We remark that when a loop ${\cal L}_i$ is added we can multiply the weight of the set
of loops to which it is connected by $\sum\limits_{k_i \in {\cal L}_i} \bar{\alpha}_{k_i}$ but we
have to subtract the terms which contains $\bar{\alpha}_j^2$ if $j$ is a propagator common to ${\cal
L}_i$ and the set of other loops. The reason is that we cannot cut twice the same propagator.
However, the first term in (\ref{A.4}) is the manifestation of the factorization property.
Furthermore, adding a loop, we cannot cut that loop more than once on propagators not belonging to
other loops because this would create disconnected sub-trees on the set of connected loops
considered. This is why we only have a first power polynomial for each loop. These remarks help
enormously writing down the weight of spanning trees on any number of loops. For three loops
${\cal L}$, ${\cal L}_1$ and ${\cal L}_2$ we have

\beq
\label{A.5}
W_{{\cal L} \cup {\cal L}_1 \cup {\cal L}_2} = \left ( \sum_{\ell \in {\cal L}} \bar{\alpha}_{\ell}
\right ) W_{{\cal L}_1 \cup {\cal L}_2} - \bar{\alpha}^2_{j_1} W_{{\cal L}_2} -
\bar{\alpha}^2_{j_2} W_{{\cal L}_1} \eeq

\noi where $j_1$ and $j_2$ are propagators of ${\cal L}$ common to ${\cal L}_1$ and ${\cal L}_2$
respectively. If, for isntance, ${\cal L}_2$ has no propagator in common with ${\cal L}$, then
the third term in (\ref{A.5}) disappears. We can generalize to ${\cal L} \cup {\cal L}_1 \cup \cdots \cup {\cal
L}_n$, and taking the notation where ${\cal L}_1 \cup \cdots \widehat{\cal L}_k \cdots \cup {\cal
L}_n$ means that propagators of ${\cal L}_k$ not belonging to ${\cal L}_1 , \cdots , {\cal
L}_{k-1}, {\cal L}_{k+1} , \cdots , {\cal L}_n$ are suppressed.  

has been omitted from ${\cal L}_1 \cup \cdots \cup
{\cal L}_n$, we get

\bminiG{EDh}  
\label{A.6a}
W_{{\cal L}_i \cup {\cal L}_1 \cup \cdots \cup {\cal L}_n} &=& \left ( \sum_{\ell \in {\cal L}}
\bar{\alpha}_{\ell} \right ) W_{{\cal L}_1 \cup \cdots \cup {\cal L}_n} \nn \\
&&- \sum_k \bar{\alpha}_{j_k}^2 \ W_{{\cal L}_1 \cup \cdots \widehat{\cal L}_k \cdots \cup {\cal
L}_n}  \eeeq   \beeq
j_h \in {\cal L} \cap {\cal L}_h \ \ \ . 
\label{A.6b} 
\emini

\noindent This can easily be understood because if there is a spanning tree on ${\cal L}_1 \cup
\cdots \cup {\cal L}_n$ its restriction on ${\cal L}_1 \cup \cdots \widehat{{\cal
L}}_{k} \cdots \cup  {\cal L}_n$ is still a spanning tree if ${\cal L}_1 \cup \cdots \widehat{{\cal
L}}_{k} \cdots \cup  {\cal L}_n$ is connected or a set of spanning trees if ${\cal L}_1 \cup \cdots \widehat{{\cal
L}}_{k} \cdots \cup  {\cal L}_n$ happens to be disconnected. Reciprocally, if there is a
spanning-tree on ${\cal L}_1 \cup \cdots \widehat{{\cal L}}_{k} \cdots \cup  {\cal L}_n$ and if
we add all the propagators of ${\cal L}_k$ not already in ${\cal L}_1 \cup \cdots \widehat{{\cal
L}}_{k} \cdots \cup  {\cal L}_n$, except for $j_k$, we get a spanning tree on ${\cal L}_1
\cup \cdots {\cal L}_n$. \par  

We recall that the propagator $i$ on ${\cal L}$ is never shared with another loop of ${\cal L}_1$,
${\cal L}_2$, $\cdots$ ${\cal L}_n$ in this construction. Then, we can write

\bea
\label{A.7}
&&W_{{\cal L} \cup {\cal L}_1 \cup \cdots \cup {\cal L}_n} = \bar{\alpha}_i \ W_{{\cal L}_1 \cup
\cdots \cup {\cal L}_n} \nn \\
&& + \sum_{\ell_k} \bar{\alpha}_{\ell_k} \left ( W_{{\cal L}_1 \cup \cdots
\cup {\cal L}_n} - \bar{\alpha}_{\ell_k} W_{{\cal L}_1 \cup \cdots \widehat{\cal L}_k \cdots \cup
{\cal L}_n} \right ) \eea

\noi with $\ell_k \in {\cal L}$, $\ell_k \not= i$, taking the convention that 

\beq
\label{A.8}
W_{{\cal L}_1 \cup \cdots \widehat{\cal L}_k \cdots \cup {\cal L}_n} = 0
\eeq

\noi whenever no loop ${\cal L}_k$ shares the propagator $\ell_k$ with ${\cal L}$. It follows from
(\ref{A.7}) that $\mu_i$ can be written (see (\ref{5.9}) for its definition)

\bminiG{EDh} 
\label{A.9a}
\mu_i = \sum_{\ell_k} \bar{\alpha}_{\ell_k} \left ( 1 - \bar{\alpha}_{\ell_k} \ W_{{\cal L}_1 \cup
\cdots \widehat{\cal L}_k \cdots \cup {\cal L}_n}/W_{{\cal L}_1 \cup \cdots \cup {\cal L}_n} \right
) \eeeq  \beeq
\ell_k \not= i \ , &&\ell_k \in {\cal L} \cap {\cal L}_k \quad \hbox{if} \ {\cal L}_k \ \hbox{exists}
\nn \\ &&\ell_k \in {\cal L} \quad \hbox{if} \ {\cal L}_k \ \hbox{does not exists} \ \ \ .
\label{A.9b}
\emini

\noi However, $W_{{\cal L}_1 \cup \cdots \cup {\cal L}_n}$ can also be expressed as

\bminiG{EDh} \label{A.10a}
&&W_{{\cal L}_1 \cup \cdots \cup {\cal L}_n} = \bar{\alpha}_{\ell_k} \ W_{{\cal L}_1 \cup \cdots
\widehat{\cal L}_k \cdots \cup {\cal L}_n} \nn \\
&&+ \sum_{k_m} \bar{\alpha}_{k_m} \left ( W_{{\cal L}_1 \cup \cdots \widehat{\cal L}_k \cdots \cup
{\cal L}_n} - \bar{\alpha}_{k_m} W_{{\cal L}_1 \cup \cdots \widehat{\cal L}_k \cdots \widehat{\cal
L}_m \cdots \cup {\cal L}_n} \right ) \eeeq  
\beeq
\label{A.10b}
k_m \not= \ell_k \ , &&k_m \in {\cal L}_k \cap {\cal L}_m \quad \hbox{if} \ {\cal L}_m \
\hbox{exists} \nn \\ 
&&k_m \in {\cal L}_k \quad \hbox{if} \ {\cal L}_m \ \hbox{does not exist}
\emini

\noi with the same convention that

\[ W_{{\cal L}_1 \cup \cdots \widehat{\cal L}_k \cdots \widehat{\cal L}_m \cdots {\cal L}_n} = 0 \]

\noi whenever $k_m$ is not shared by ${\cal L}_k$ with any other loop of ${\cal L}_1 \cup \cdots
\cup {\cal L}_n$. \par

This leads us to write $\mu_i$ as

\bminiG{EDh}
\label{A.11a}
\mu_i &=& \left . \sum_{\ell_k} \bar{\alpha}_{\ell_k} \Big \{ 1 -
\varepsilon_{\ell_k} \ \bar{\alpha}_{\ell_k}/\Big [ \bar{\alpha}_{\ell_k} + \sum_{k_m}
\bar{\alpha}_{k_m} \Big  ( 1 - \bar{\alpha}_{k_m} W_{{\cal L}_1 \cup \cdots \widehat{\cal L}_k
\cdots \widehat{\cal L}_m \cdots \cup {\cal L}_n} \right / \nn \\
&&W_{{\cal L}_1 \cup \cdots {\cal L}_k \cdots \cup {\cal L}_n} \Big )
\Big ] \Big \}
 \eeeq
\beeq
&&\varepsilon_{\ell_k} = 1 \qquad \hbox{if} \qquad W_{{\cal L}_1 \cup \cdots \widehat{\cal L}_k
\cdots \cup {\cal L}_n \not= 0} \nn \\
&&\varepsilon_{\ell_k} = 0 \qquad \hbox{if} \qquad W_{{\cal L}_1 \cup \cdots \widehat{\cal L}_k
\cdots \cup {\cal L}_n = 0} \label{A.11b} 
\emini 

\noi introducing a {\it continued fraction representation} for $\mu_i$. Let us now see how we can
make $\mu_i$ vanish faster than $1/I$. \\

i) The most immediate way to make $\mu_i$ tends to zero faster than $1/I$ is to assume that every
$\bar{\alpha}_{\ell_k}$ in (A.11) is decreasing faster than 1/I. However note that if
$\bar{\alpha}_{\ell}$ is zero for some propagator $\ell$ it will not contribute either to $P_G(\{
\bar{\alpha}\})$ nor to $Q_G(P, \{ \bar{\alpha} \})$, see ((\ref{2.2}) and (2.3)), so that we
can consider that $\ell$ has been erased from $G$. If we assume that the $\bar{\alpha}_{k_m}$'s in
(A.11) are $O(1/I)$ we then have

\beq
\label{A.12}
\bar{\alpha}_{\ell_k}/\bar{\alpha}_{k_m} \to 0
\eeq 

\noindent and we can proceed as if the propagators $\ell_k$ did not exist, this amounting to
replace the loop ${\cal L}$ by a loop ${\cal L}'$ made of all propagators belonging to the loops
${\cal L}_k$ having a propagator in common with ${\cal L}$ except for the common propagators
$\ell_k$. We then have to write $\mu_i$ with ${\cal L}$ replaced by ${\cal L}'$ which leads to an
expression different from (A.12a) and we have to redo this reasoning again if we want to make
$\mu_i$ vanish faster than $1/I$. \par

ii) The second way to make $\mu_i$ vanish faster than $1/I$ is to assume that for every propagator
$k_m$

\beq
\bar{\alpha}_{k_m}/\bar{\alpha}_{\ell_k} \to 0 \ \ \ .
\label{A.13}
\eeq

\noindent According to what we said in i) this would make the propagators $k_m$ disappear and
every loop ${\cal L}_k$ would fuse with a neighbouring loop ${\cal L}_m$ (erasing their common
propagator) giving a loop ${\cal L}'_k$ instead. Again, (A.12) would be modified by such a
change and $\mu_i$ would be $O(1/I)$ unless the reasoning is repeated. \par

So, we see that in order to make $\mu_i$ vanish faster than $1/I$ we have to repeat the loop
cancellation mechanism forever. This leads to have all $\bar{\alpha}_j$'s to be vanishing faster
than $1/I$ which is forbidden by the constraint (2.9). We therefore conclude that $\mu_i$ must be
$O(1/I)$. \\

\noindent {\bf \underbar{Remarks}} \par

i) In order to be able to use the cancellation mechanism, the sum of the weights of the spanning
trees having a factor $\bar{\alpha}_{\ell_k}$ should be negligible with respect to the sum of the
weights of the spanning trees not having this factor on ${\cal L}_1 \cup \cdots \cup {\cal L}_n$.
So, instead of (A.13), a priori, the real condition would be, see (A.11a),

\beq
\label{A.15} 
\bar{\alpha}_{\ell_k} \ W_{{\cal L}_1 \cup \cdots \widehat{\cal L}_k \cdots \cup {\cal L}_n}/
\sum_{k_m} \bar{\alpha}_{k_m} \left ( W_{{\cal L}_1 \cup \cdots \widehat{\cal L}_k \cdots \cup
{\cal L}_n} - \bar{\alpha}_{k_m} W_{{\cal L}_1 \cup \cdots \widehat{\cal L}_k \cdots
\widehat{\cal L}_m \cdots \cup {\cal L}_n} \right ) \to 0 
\eeq

\noindent which is equivalent to (A.13) provided there is at least one $k_m$ such that 

\beq
\label{A.16}
1 - \bar{\alpha}_{k_m} W_{{\cal L}_1 \cup \cdots \widehat{\cal L}_k \cdots \widehat{\cal L}_m
\cdots \cup {\cal L}_n}/W_{{\cal L}_1 \cup \cdots \widehat{\cal L}_k \cdots \cup {\cal L}_n} \not=
0 \ . 
\eeq

However, breaking (\ref{A.16}) would mean that in 

\[
W_{{\cal L}_1 \cup \cdots \widehat{\cal L}_k \cdots \cup {\cal L}_n} = \bar{\alpha}_{k_m} \ W_{{\cal
L}_1 \cup \cdots \widehat{\cal L}_k \cdots \widehat{\cal L}_m \cdots \cup {\cal L}_n}
\] 
\beq
+ \sum_{m_{\rho}} \bar{\alpha}_{m_{\rho}} \left ( W_{{\cal L}_1 \cup \cdots \widehat{\cal L}_k
\cdots \widehat{\cal L}_m \cdots \cup {\cal L}_n} - \bar{\alpha}_{m_{\rho}} W_{{\cal L}_1 \cup
\cdots \widehat{\cal L}_k \cdots \widehat{\cal L}_m \cdots \widehat{\cal L}_{\rho} \cdots \cup
{\cal L}_n} \right ) \label{A.17} 
\eeq

\noindent we should have for every $m_{\rho}$, either 

\beq
\bar{\alpha}_{m_{\rho}}/\bar{\alpha}_{k_m} \to 0
\label{A.18}
\eeq

\noindent leading to a cancellation of $m_{\rho}$ or an analog of (A.16) leading to a cancellation
at a further step of the reasoning. So, indeed, (A.13) is sufficient to induce the cancellation of
$\ell_k$. \par

ii) Concerning (A.14), we can look at (A.11a) and note that if
$\bar{\alpha}_{k_m}/\bar{\alpha}_{\ell_k} \to 0$, then

\[
\left . \bar{\alpha}_{k_m} \left ( W_{{\cal L}_1 \cup \cdots \widehat{\cal L}_k \cdots \cup {\cal
L}_n} - \bar{\alpha}_{k_m} \ W_{{\cal
L}_1 \cup \cdots \widehat{\cal L}_k \cdots \widehat{\cal L}_m \cdots \cup {\cal L}_n} \right )
\right / \]
\beq
\label{A.19}
\bar{\alpha}_{k_m} \ W_{{\cal L}_1 \cup \cdots \widehat{\cal L}_k \cdots \cup {\cal L}_n} \to 0
\eeq

\noindent because

\beq
W_{{\cal L}_1 \cup \cdots \widehat{\cal L}_k \cdots \cup {\cal L}_n} > W_{{\cal L}_1 \cup \cdots
\widehat{\cal L}_k \cdots \cup {\cal L}_n} - \bar{\alpha}_{k_m} \ W_{{\cal L}_1 \cup \cdots
\widehat{\cal L}_k \cdots \widehat{\cal L}_m \cdots \cup {\cal L}_n} > 0 \ .\label{A.20} \eeq

\noindent So, in the same way, (A.14) is sufficient to have $k_m$ cancelled (of course the same also
apply to (A.18) and the cancellation of $m_p$).
 \newpage

\section*{Table 1} 
\begin{center}
$a_- = x_- = 0$
\vskip 3 truemm
\begin{tabular}{|r|r|r|r|r|r|r|r|r|}
\hline
$t/m^2$ &$\mu_- \Lambda$ &$y \quad$ &$x \quad$ &$u \quad$ &$z \quad$ &$a_+\ $ &$x_+\ $ &$\alpha
(t/m^2)$ \\ \hline
2.0 &11.01 &0.92796 &1.223 &1.269 &0.639 &-0.311 &0.584 &0.50$\ \ $ \\
\hline
1.5 &10.93 &0.92794 &1.210 &1.262 &0.637 &-0.298 &0.588 &0.438 \\
\hline
1.0 &10.82 &0.92779 &1.179 &1.263 &0.639 &-0.276 &0.588 &0.423 \\
\hline
0.5 &10.80 &0.92778 &1.155 &1.262 &0.640 &-0.264 &0.590 &0.321 \\
\hline
0.0 &10.69 &0.92793 &1.140 &1.267 &0.640 &-0.250 &0.590 &0.240 \\
\hline
-0.5 &10.58 &0.92776 &1.116 &1.270 &0.633 &-0.231 &0.590 &0.163 \\
\hline
-1.0 &10.48 &0.92783 &1.082 &1.261 &0.647 &-0.210 &0.585 &0.079 \\
\hline
-1.5 &10.40 &0.92782 &1.064 &1.266 &0.644 &-0.196 &0.586 &0.012 \\
\hline
-2.0 &10.32 &0.92777 &1.041 &1.267 &0.638 &-0.176 &0.586 &-0.020 \\
\hline
-2.5 &10.19 &0.92773 &1.017 &1.266 &0.645 &-0.158 &0.586 &-0.108 \\
\hline
-3.0 &10.05 &0.92772 &0.996 &1.264 &0.638 &-0.143 &0.587 &-0.170 \\
\hline
-3.5 &10.04 &0.92786 &0.963 &1.271 &0.640 &-0.126 &0.586 &-0.195 \\
\hline
\end{tabular}
\end{center}

\newpage
\section*{Figure Captions}

\noindent {\bf \underbar{Fig. 1} :}\par
A ladder graph is shown. Central-propagators are weighted by $\alpha_-$ and side-propagators by
$\alpha_+$. $t$ is the invariant equal to the sum squared of momenta entering at one end of the
ladder. $s$ is the large invariant when Regge behaviour is obtained. It is equal to the sum squared
of momenta entering at one side of the ladder (here, one side is up and the other down). \par \vskip
3 truemm

\noindent {\bf \underbar{Fig. 2} :}\par
We display the three kinds of topology obtained by removing propagators in order to obtain spanning
trees on the ladder. When a central propagator is removed a factor $\bar{\alpha}_-$ is obtained and
when a side propagator is removed a factor $\bar{\alpha}_+$ is obtained. Cells of lengths $\ell_1$,
$\ell_2 , \cdots , \ell_{L-p}$ are formed. Each cell has propagators on its border. Removed central
propagators are shown as dashed lines, removed side-propagators are simply cancelled. In a) the
end-cells are opened on the sides of the ladder. In b) one end-cell is opened at one end of the
ladder. In c) both end-cells are opened at the ends of the ladder. \par \vskip
3 truemm

\noindent {\bf \underbar{Fig. 3} :}\par
A cell of length $\ell$ is displayed as well as a propagator $i_+$ having a weight $\alpha_{i_+}$
(and not $\bar{\alpha}_+$). In this configuration the removed propagator on the down-side of the
cell brings up a factor $\bar{\alpha}_+$. When $i_+$ is on the opposite side of the removed
propagator it can be removed too, bringing up a factor $\alpha_{i_+}$. When $i_+$ is on the same
side as the removed propagator it cannot be removed because some propagators would be isolated
from the rest of the ladder without being attached to an external line. \par \vskip
3 truemm

\noindent {\bf \underbar{Fig. 4} :}\par
We display the Regge trajectory $\alpha (t/m^2)$ for $\ell n \ \gamma_m$ equal to - 0.1 (lower
line and squares) and to 0 (upper line and losanges). Both lines are straight-lines, which are
parallel and give a good fit to the computed data. The error bars show the dispersion given by
repeating the calculations several times with different starting values for the parameters. We
remark that the dispersion grows for uncreasing values of $|t/m^2|$. Also displayed is the axis $t=
0$ and two full circles corresponding to $\alpha (0)$ given by (\protect{\ref{10.12}}).  \par \vskip
3 truemm

\noindent {\bf \underbar{Fig. 5} :}\par
The intercept $\alpha (0)$ is shown for $\ell n \gamma_m$ ranging from - 0.4 to 0.5. The squares
are given by (\protect{\ref{10.12}}) and the crosses are the result of our calculations. Agreement
is observed for $\alpha (0) \ \gsim \ 0.3$. Error bars for crosses give the dispersion of the
calculations.
 
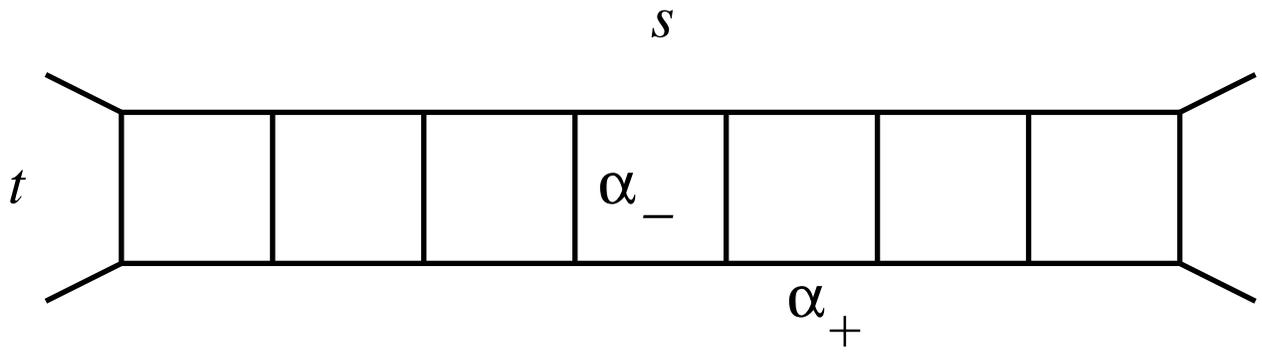
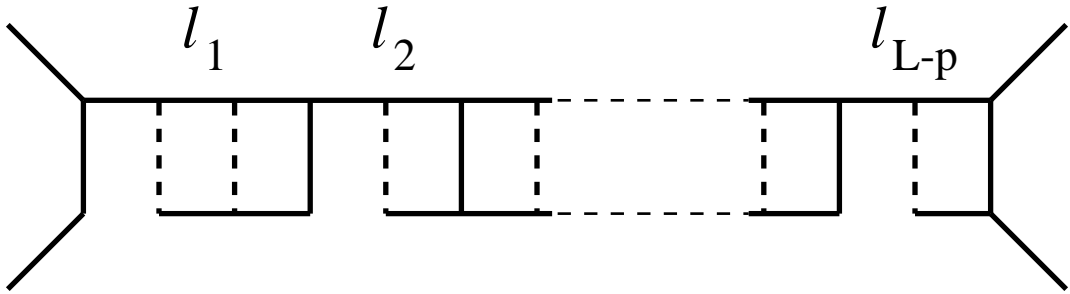
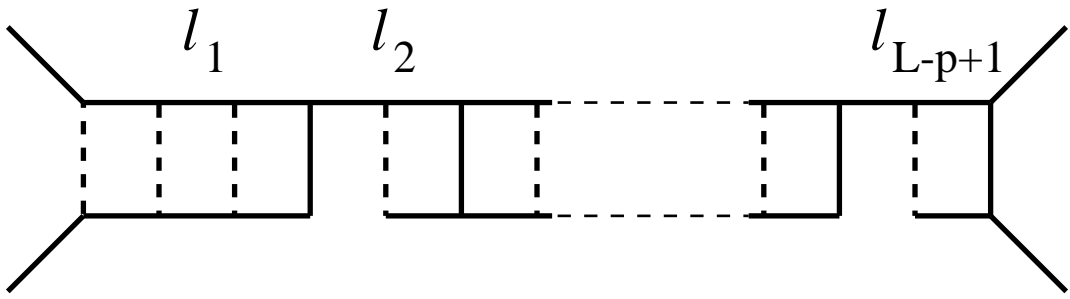
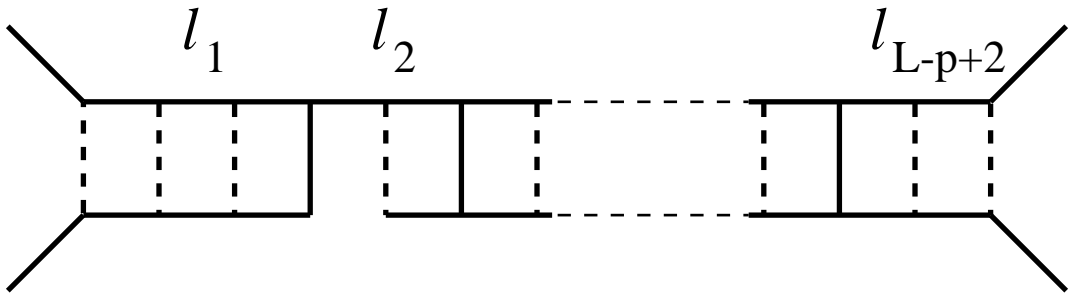
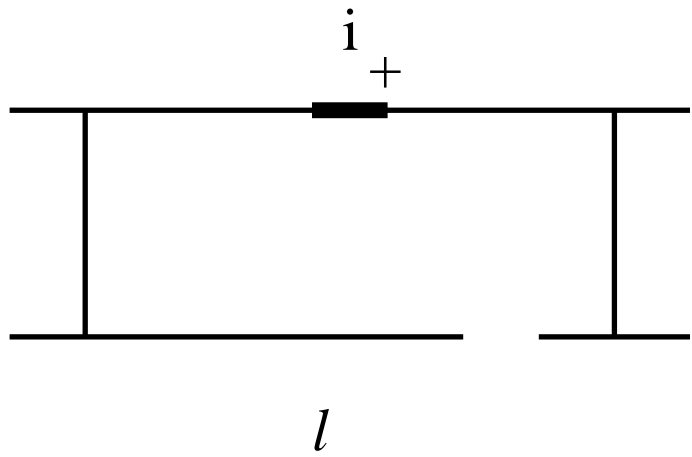
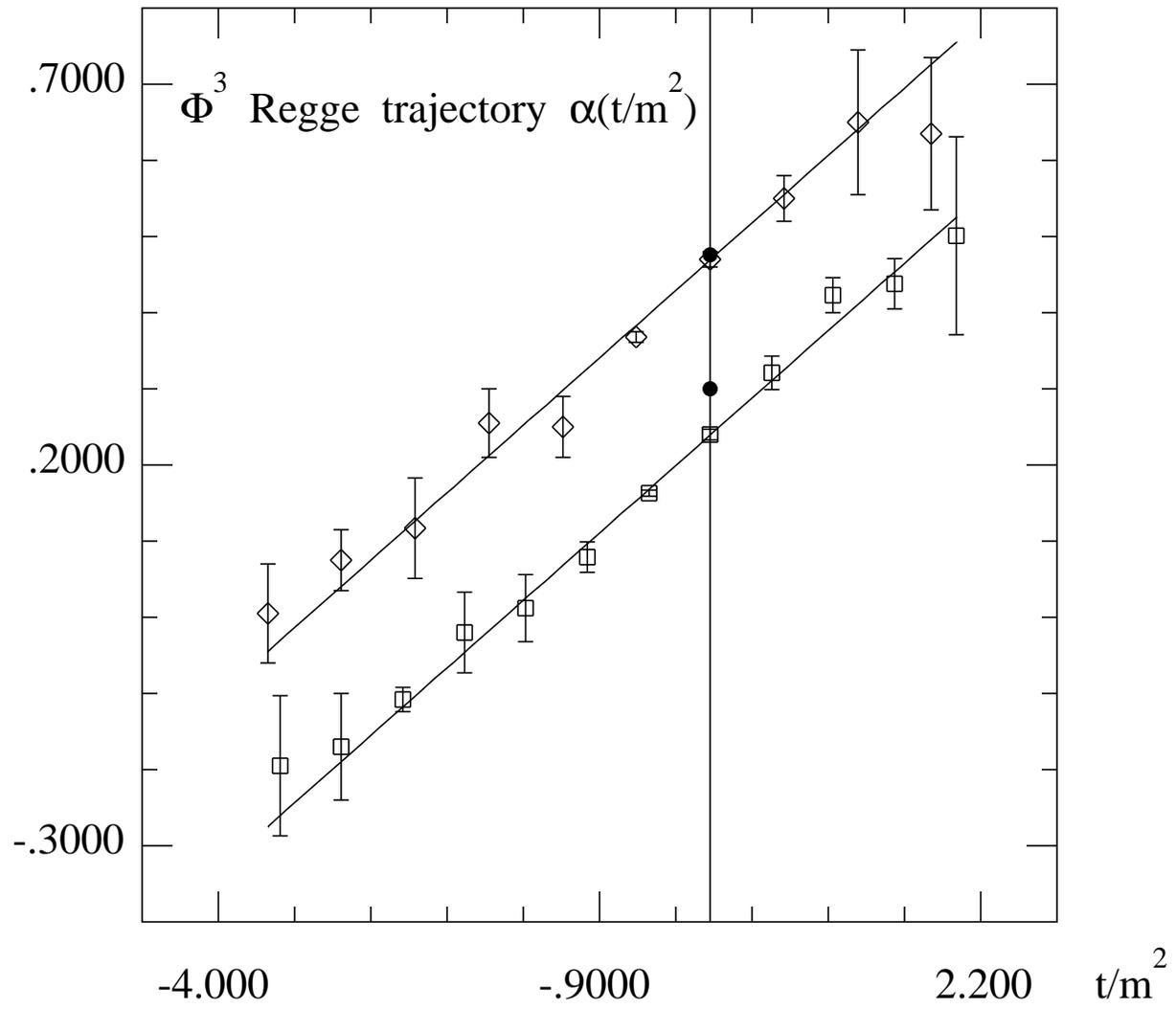
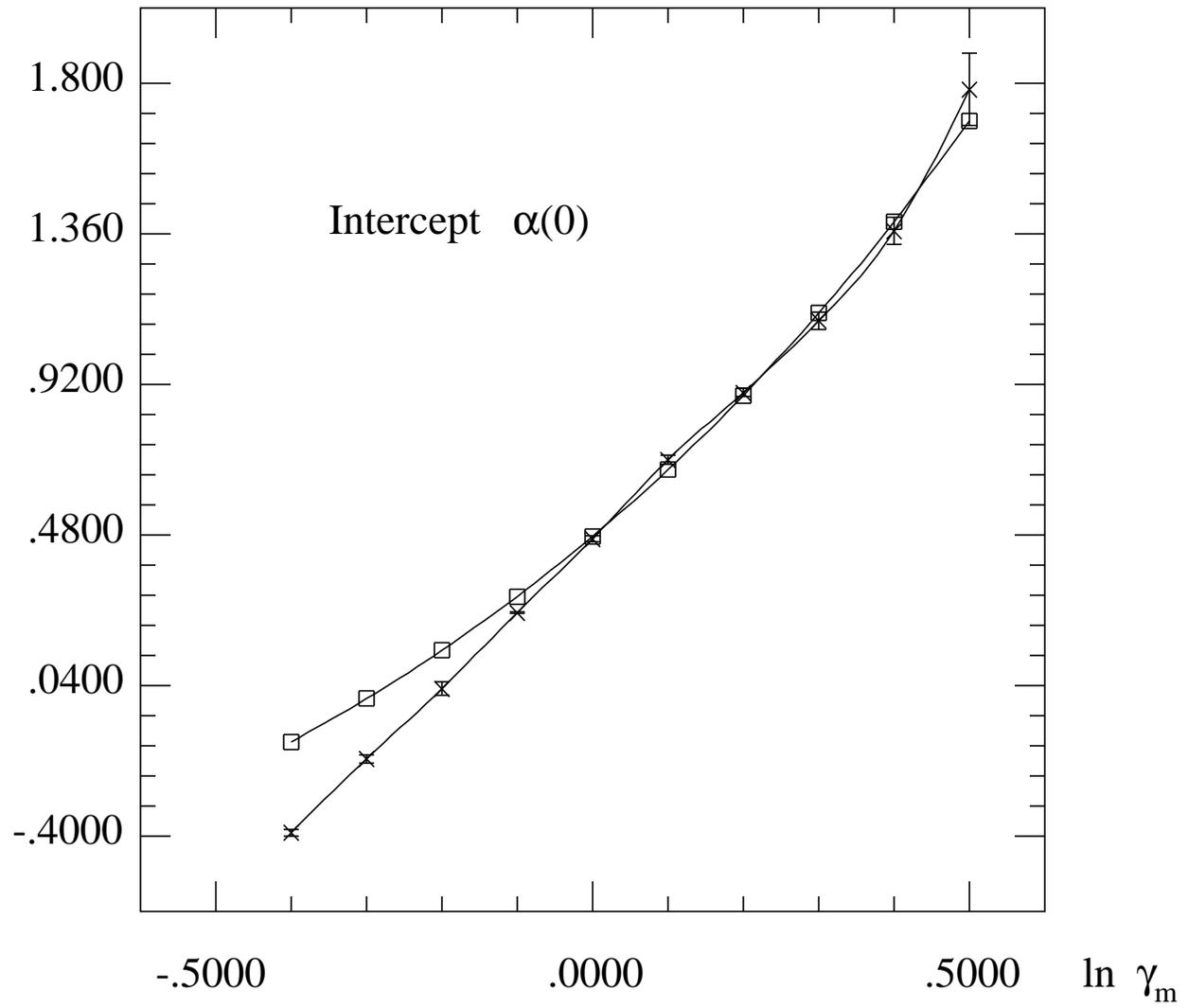
\end{document}